\documentclass[10pt,journal]{IEEEtran}

\usepackage{amsmath,graphicx, booktabs,tabularx,amssymb,bm,svg,tikz,algorithm,algpseudocode,comment,tcolorbox,pgfplots,mwe, color, enumitem, subfig, textcomp, adjustbox, hyperref, overpic}
\usepackage{adjustbox} 
\usepackage{upgreek} 
\usepackage{bbm}

\usepackage{acro}[=v2]
\usetikzlibrary{spy, calc}
\setlist{nosep, leftmargin=14pt}

\usepackage{hyperref}
\definecolor{lightblue}{HTML}{ADD8E6}  
\definecolor{softpink}{HTML}{FFC0CB}     
\definecolor{strongpink}{HTML}{FF69B4}   
\definecolor{vividorange}{HTML}{FF7300}

\usepackage[
	style=ieee,
	dashed=false,
	maxbibnames=20,
	minbibnames=2,
	maxcitenames=1,
	mincitenames=1,
	backend=biber,
	defernumbers=false,
	sorting=none,
	useprefix=false
	]{biblatex}

\algnewcommand\algorithmicparfor{\textbf{parfor}}
\algnewcommand\algorithmicpardo{\textbf{do}}
\algnewcommand\algorithmicendparfor{\textbf{end\ parfor}}
\algrenewtext{ParFor}[1]{\algorithmicparfor\ #1\ \algorithmicpardo}
\algrenewtext{EndParFor}{\algorithmicendparfor}

\DeclareAcronym{1D}{
	short = 1-D,
	long=one-dimensional,}

\DeclareAcronym{2D}{
	short = 2-D,
	long=two-dimensional,
}

\DeclareAcronym{3D}{
	short = 3-D,
	long=three-dimensional,
}

\DeclareAcronym{F18}{
	short = \textsuperscript{18}F,
	long=fluorine-18,
}

\DeclareAcronym{Ga68}{
	short=\textsuperscript{68}Ga,
	long=gallium-68,
}

\DeclareAcronym{Rb82}{
	short=\textsuperscript{82}Rb,
	long=rubidium-82,
}

\DeclareAcronym{FDG}{
	short=FDG,
	long=fluorodeoxyglucose,
}

\DeclareAcronym{PET}{
	short=PET,
	long=positron emission tomography,
}

\DeclareAcronym{SPECT}{
	short=SPECT,
	long= Single Photon Emission Computed Tomography,
}

\DeclareAcronym{CT}{
	short=CT,
	long=computed tomography,
}

\DeclareAcronym{MR}{
	short=MR,
	long= magnetic resonance,
}

\DeclareAcronym{MRI}{
	short=MRI,
	long= magnetic resonance imaging,
}

\DeclareAcronym{GAN}{
	short=GAN,
	long=generative adversarial network,
	long-plural-form = generative adversarial networks,
}

\DeclareAcronym{LS}{
	short=LS,
	long=least squares,
	long-plural-form = least squares,
}

\DeclareAcronym{GATE}{
	short=GATE,
	long=Geant4 Application for Tomography Emission,
}

\DeclareAcronym{MC}{
	short=MC,
	long=Monte Carlo,
}

\DeclareAcronym{PSF}{
	short=PSF,
	long=point spread function,
}

\DeclareAcronym{VAE}{
	short=VAE,
	long= variational autoencoder,
	long-plural-form = variational autoencoders,
}

\DeclareAcronym{LSTM}{
	short=LSTM,
	long=long short-term memory,
}

\DeclareAcronym{ViT}{
	short=ViT,
	long=vision transformer,
}

\DeclareAcronym{NN}{
	short=NN,
	long=neural network,
}

\DeclareAcronym{GNN}{
	short=GNN,
	long=graph neural network,
}

\DeclareAcronym{GRU}{
	short=GRU,
	long=gated recurrent unit,
}

\DeclareAcronym{GCN}{
	short=GCN,
	long=graph convolutional network,
}

\DeclareAcronym{ANN}{
	short=ANN,
	long= artificial neural network,
}

\DeclareAcronym{CNN}{
	short=CNN,
	long=convolutional neural network,
}

\DeclareAcronym{RNN}{
	short=RNN,
	long=recurrent neural network,
	long-plural-form = recurrent neural networks,
}

\DeclareAcronym{FCNN}{
	short=FCNN,
	long=fully-connected neural network,
}

\DeclareAcronym{MLP}{
	short=MLP,
	long=multilayer perceptron,
}	

\DeclareAcronym{LOR}{
	short=LOR,
	long=line of response,
	long-plural-form = lines of response,
}	

\DeclareAcronym{TOF}{
	short=TOF,
	long=time-of-flight
}

\DeclareAcronym{LM}{
	short=LM,
	long=list-mode
}

\DeclareAcronym{AI}{
	short=AI,
	long=artificial intelligence
}

\DeclareAcronym{MPNN}{
	short=MPNN,
	long=message passing neural network
}

\DeclareAcronym{IN}{
	short=IN,
	long=interaction network,
}

\DeclareAcronym{MIN}{
	short=MIN,
	long=modified interaction network,
}

\DeclareAcronym{MDN}{
	short=MDN,
	long=mixture density network,
}

\DeclareAcronym{PDF}{
	short=PDF,
	long=probability density function,
}

\DeclareAcronym{FWHM}{
	short=FWHM,
	long=full width at half maximum ,
}

\DeclareAcronym{RMSE}{
	short=RMSE,
	long=root-mean-square error,
}

\DeclareAcronym{MLEM}{
	short=MLEM,
	long=maximum-likelihood expectation-maximization algorithm,
}

\DeclareAcronym{SNR}{
	short=SNR,
	long=signal-to-noise ratio,
}

\DeclareAcronym{PSNR}{
	short=PSNR,
	long=peak signal-to-noise ratio,
}

\DeclareAcronym{MAE}{
	short=MAE,
	long=mean absolute error,
}

\DeclareAcronym{LHC}{
	short=LHC,
	long=large hadron collider,
}

\DeclareAcronym{XCAT}{
	short=XCAT,
	long=Extended Cardiac-Torso,
}

\DeclareAcronym{EM}{
	short=EM,
	long=expectation maximization,
}

\DeclareAcronym{castor}{
	short=CASToR,
	long=the Customizable and Advanced Software for Tomographic Reconstruction,
}

\DeclareAcronym{GT}{
	short=GT,
	long=ground truth,
}

\DeclareAcronym{ROI}{
	short=ROI,
	long=region of interest,
}

\DeclareAcronym{STD}{
	short=STD,
	long=standard deviation,
}

\DeclareAcronym{LXe}{
	short=LXe,
	long=liquid xenon,
}

\DeclareAcronym{phsp}{
	short=PhSp,
	long=phase space,
}

\DeclareAcronym{MRT}{
	short=MRT,
	long=microbeam radiation therapy,
}

\DeclareAcronym{DL}{
	short=DL,
	long=deep learning,
}

\DeclareAcronym{ML}{
	short=ML,
	long=machine learning,
}

\DeclareAcronym{FOV}{
	short=FOV,
	long=field of view,
}

\DeclareAcronym{MD}{
	short=MD,
	long=multi-discriminator,
}

\DeclareAcronym{PRC}{
	short=PRC,
	long=PR correction,
}

\DeclareAcronym{PR}{
	short=PR,
	long=positron range,
}

\DeclareAcronym{SUV}{
	short=SUV,
	long=standardized uptake value,
}

\DeclareAcronym{DI-DTConv}{
	short=DI-DTConv,
	long=dual-input dynamic transposed convolution,
}

\DeclareAcronym{DDConv}{
	short=DDConv,
	long=Dual-Input Dynamic Convolution,
}

\DeclareAcronym{SVTD}{
	short= SVTD,
	long=spatially-variant and tissue-dependent,
}

\DeclareAcronym{LSO}{
	short= LSO,
	long=lutetium oxyorthosilicate,
}

\DeclareAcronym{GPU}{
    short= GPU,
	long=graphics processing unit,
}

\DeclareAcronym{CPU}{
    short= CPU,
	long=central processing unit,
}

\DeclareAcronym{MAPE}{
    short= MAPE,
	long=mean absolute percentage error,
}

\DeclareAcronym{CNR}{
    short= CNR,
	long=contrast-to-noise ratio,
}

\DeclareAcronym{RC}{
    short= RC,
	long=recovery coefficient,
}

\DeclareAcronym{KL}{
	short=KL,
	long=Kullback--Leibler,
}

\DeclareAcronym{CUDA}{
	short=CUDA,
	long=Compute Unified Device Architecture,
}

\newcommand{\boldd}{\bm{d}}

\newcommand{\boldh}{\bm{h}}

\newcommand{\boldr}{\bm{r}}

\newcommand{\boldu}{\bm{u}}
\newcommand{\boldv}{\bm{v}}
\newcommand{\boldw}{\bm{w}}
\newcommand{\boldx}{\bm{x}}
\newcommand{\boldy}{\bm{y}}
\newcommand{\boldz}{\bm{z}}

\newcommand{\boldmu}{\bm{\mu}}

\newcommand{\boldxhat}{\hat{\boldx}}
\newcommand{\xhat}{\hat{x}}

\newcommand{\boldeta}{\bm{\eta}}

\newcommand{\boldA}{\bm{A}}
\newcommand{\boldB}{\bm{B}}

\newcommand{\boldG}{\bm{G}}
\newcommand{\boldH}{\bm{H}}

\newcommand{\boldP}{\bm{P}}

\newcommand{\calB}{\mathcal{B}}

\newcommand{\calL}{\mathcal{L}}

\newcommand{\calN}{\mathcal{N}}

\newcommand{\calS}{\mathcal{S}}
\newcommand{\calT}{\mathcal{T}}

\newcommand{\boldtheta}{\bm{\theta}}

\newcommand{\boldrho}{\bm{\rho}}

\newcommand{\ybar}{\bar{y}}
\newcommand{\boldybar}{\bar{\boldy}}

\newcommand{\boldzero}{\bm{0}}
\newcommand{\boldone}{\bm{1}}
\newcommand{\R}{\mathbb{R}}

\newcommand{\transp}{^\top}

\title{Dual-Input Dynamic Convolution for Positron Range Correction in PET Image Reconstruction}

\author{Youness Mellak, Alexandre Bousse, Thibaut Merlin, \'Elise \'Emond, Mikko Hakulinen, Dimitris Visvikis
    \thanks{This work received support from the French government, granted to the Labex CominLabs excellence laboratory and managed by the French National Research Agency (ANR) under the ``Investing for the Future'' program, Grant ANR-10-LABX-07-01, and from the ANR under Grant ANR-21-CE19-0036.}
	\thanks{This work did not involve human subjects or animals in its research.}
	\thanks{Youness Mellak, Alexandre Bousse, Thibaut Merlin and Dimitris Visvikis are with Univ. Brest, LaTIM, Inserm, U1101, 29238~Brest, France.}
	\thanks{Élise Émond is with GE HealthCare GmbH, 40468 D\"{u}sseldorf, Germany.}
    \thanks{Mikko Hakulinen is with Diagnostic Imaging Centre, Kuopio University Hospital, Kuopio, Finland.}
	\thanks{Corresponding authors: \href{mailto:bousse@univ-brest.fr}{\texttt{bousse@univ-brest.fr}}}
}

\begin{document}

\maketitle

\begin{abstract}
	\Ac{PR} blurring degrades \ac{PET} image resolution, particularly for high-energy emitters like \ac{Ga68}. We introduce \ac{DDConv}, a novel computationally efficient approach trained with voxel-specific \ac{PR} \acp{PSF} from \ac{MC} simulations and designed to be utilized within an iterative reconstruction algorithm to perform \ac{PRC}. By dynamically inferring local blurring kernels through a trained \ac{CNN}, \ac{DDConv} captures complex tissue interfaces more accurately than prior methods. Additionally, it also computes the transpose  operator, ensuring consistency within iterative \ac{PET} reconstruction. Comparisons with a state-of-the-art, tissue-dependent correction confirm the advantages of \ac{DDConv} in recovering higher-resolution details in heterogeneous regions, including bone--soft tissue and lung--soft tissue boundaries. 
	
	Experiments across digital phantoms and \ac{MC}-simulated data show that \ac{DDConv} offers near-\ac{MC} accuracy and outperforms the state-of-the-art technique, namely \ac{SVTD}, especially in areas with complex material interfaces. 
	
	Results from real phantom experiments further confirm \ac{DDConv}'s robustness and practical applicability: while both \ac{DDConv} and \ac{SVTD} performed similarly in homogeneous soft-tissue regions, \ac{DDConv} provided more accurate activity recovery and sharper delineation at heterogeneous lung--soft tissue interfaces.
	
	Our code  available at \url{https://github.com/mellak/ddconv-prc}.
\end{abstract}


\acresetall

\section{Introduction}\label{sec:intro}

\IEEEPARstart{P}ositron emission tomography (PET) is a nuclear imaging technique that visualizes molecular and metabolic processes by detecting pairs of gamma photons emitted during positron-electron annihilation. During a \acs{PET} scan, a radiopharmaceutical---a biologically active molecule labeled with a positron--emitting radionuclide---is administered to the patient. As the radionuclide decays, it emits positrons, which travel a short distance through tissue before annihilating with electrons. This distance, also referred to as \ac{PR}, displaces the annihilation site from the original tracer location, introducing an inherent blur into the reconstructed image \cite{conti2016physics}. The \ac{PR} is governed by two factors: the radionuclide's positron endpoint energy (the maximum kinetic energy of emitted positrons) and the electron density of the surrounding tissue (e.g., dense bone attenuates positrons more effectively than low-density lung tissue) \cite{sanchez2013comparison}. For widely used radionuclides such as \ac{F18}, which has a low endpoint energy (0.634~MeV), the \ac{PR} is minimal (0.6~mm in water). This blur is negligible compared to the 2--4-mm spatial resolution of modern \ac{PET} scanners, enabling precise imaging of glucose metabolism in oncology. However, clinical demands increasingly require isotopes with higher positron energies. \Ac{Ga68}, used for prostate cancer imaging, exhibits a 1.9~MeV endpoint energy and a \ac{PR} of 2.9~mm in water. Similarly, \ac{Rb82}, employed in cardiac perfusion studies, has a 3.4~MeV endpoint energy and a \ac{PR} of 5.9~mm. These \ac{PR} values exceed the resolution of the scanner, leading to significant blurring that distorts quantitative metrics such as lesion size and \acp{SUV}. This problem is amplified in heterogeneous tissues (e.g., water--lung interfaces), where abrupt changes in electron density further widen the \ac{PR} distribution.

Various \ac{PRC} methods have been developed to mitigate blurring effects caused by \ac{PR} in \ac{PET} imaging, particularly for radionuclides such as \ac{Ga68} \cite{gavriilidis2022positron}. These methods can be broadly categorized into four approaches. 

The first involves reducing the travel distance of the positron by applying strong magnetic fields to confine its trajectory \cite{hammer1994use, wirrwar19974}. While effective, this method requires extremely intense magnetic fields, making it expensive and challenging to implement in clinical \ac{PET} scanners. 

The second approach consists of applying \ac{PRC} before reconstruction (pre-reconstruction) using deconvolution techniques on measured projections \cite{derenzo1986mathematical, haber1990application}. This method assumes a unique  \ac{PR} \ac{PSF}, thus limiting its accuracy in heterogeneous tissues where  \ac{PR} effects are spatially variant. 

The third approach applies corrections directly to reconstructed \ac{PET} images, offering a practical solution when incorporating corrections during acquisition or reconstruction is not feasible. For example, Deep-PRC \cite{herraiz2020deep, encina2024deep} uses a \ac{CNN} to map \ac{Ga68}-blurred images to \ac{F18}-like images, and is trained on images reconstructed from \ac{MC}-simulated data, effectively reducing blurring. However, this method is highly dependent on the quality of the training data, reconstruction parameters, and detected counts. 
Furthermore, self-supervised models have been proposed \cite{xie2025noise}, simulating \ac{Rb82} \ac{PR} kernels using \ac{MC} methods and employing pseudo-labels from \ac{F18}-\ac{FDG} images to approximate the inverse kernel. While promising, these models are limited to isotropic kernels, restricting their applicability in heterogeneous tissues. 

The fourth approach integrates \ac{PRC} directly into the iterative reconstruction process by modeling spatially-variant \ac{PR}  effects in the forward model using voxel-specific convolution kernels. High-precision methods which use \ac{MC} simulations with tissue-specific kernels are capable of achieving accurate \ac{PR} blurring, but they do not incorporate \ac{PR} in the transposed system matrix and are  computationally expensive \cite{autret2015amelioration}, even with \ac{GAN}-based acceleration \cite{mellak2024fast}. Various kernel-based approaches have been developed to address the computational and accuracy challenges of \ac{PRC}. \citeauthor{cal2015tissue}~\cite{cal2015tissue} introduced tissue-dependent and spatially variant kernels derived from \ac{MC} simulations.
However, the computational intensity of \ac{MC} simulations limits their clinical practicality. \citeauthor{bertolli2016pet}~\cite{bertolli2016pet} proposed isotropic and material-specific kernels as a computationally efficient alternative. 
Although efficient, this approach struggles to accurately capture \ac{PR} effects at complex tissue interfaces. 
\citeauthor{kraus2012simulation}~\cite{kraus2012simulation} addressed the challenge of \ac{PR} blurring in heterogeneous environments by precomputing tissue-specific kernels, such as those for lung--soft tissue boundaries. This method improved spatial resolution and reduced artifacts, but lacked adaptability to finer-scale variations within tissues. \citeauthor{kertesz2022implementation}~\cite{kertesz2022implementation} refined this approach by dynamically combining precomputed isotropic kernels using the attenuation maps to approximate the \ac{PR} \ac{PSF}. This allowed for better adaptability in complex anatomies. However, the composition of kernels could still deviate from the true \ac{PR} blurring spatial distribution,  especially near tissue interfaces.  

In addition to kernel-based techniques, new methods based on \ac{DL} architectures have emerged as promising alternatives. \citeauthor{merlin2024deep}~\cite{merlin2024deep} proposed an image translation \ac{GAN} integrated into an \ac{EM} reconstruction framework to dynamically correct \ac{PR} effects during forward projection. This approach demonstrated improved contrast recovery, particularly in low-attenuation tissues, although it operates with an unmatched projector. In contrast, \citeauthor{mellak2024one}~\cite{mellak2024one} introduced a \ac{GNN}-based method that locally predicts the weights of the linear operator responsible for \ac{PR} blurring. This design inherently allows for straightforward computation of the transpose, thus facilitating integration in iterative reconstruction algorithms.

In this study, we expand on previous work and propose a novel \ac{DL}-based method for \ac{PRC}, namely \ac{DDConv}, which can be plugged into iterative \ac{PET} image reconstruction, leveraging a dynamic \ac{CNN} to address accuracy and computational time. Our method is trained on \ac{MC}-simulated data using the \ac{GATE}  \cite{jan2004gate} in order to accurately model \ac{PR} blurring while significantly reducing computational demands. The method inherently computes the transpose of the blurring operator, ensuring consistency between forward and backward projections within iterative reconstruction algorithms. Additionally, \ac{DDConv} depends solely on the tracer and voxel size, making it applicable to any \ac{PET} system, independently of the scanner geometry or detector configuration. 

Section~\ref{sec:method} provides a background on \ac{PR} in \ac{PET} iterative reconstruction, and presents \ac{DDConv}, including the forward blurring and its transposed version, as well as the \ac{MC}-trained \ac{PR} \ac{PSF} predictor. Section~\ref{sec:results} describes our experiments to compare \ac{DDConv} with a state-of-the-art method from the literature, the \ac{SVTD} \ac{PRC} method by \citeauthor{kertesz2022implementation}~\cite{kertesz2022implementation}. The results of this research are summarized in Section~\ref{sec:discussion} and Section~\ref{sec:conclusion} concludes this paper. A method to reduce the  computational time of \ac{DDConv} is proposed in Appendix~\ref{sec:acceleration}. Runtime evaluation and the kernel size analysis are provided in Appendix~\ref{sec:runtime} and Appendix~\ref{sec:kernelsize}, respectively.

\subsection*{Nomenclature}

In the following, `$\transp$' denotes the matrix transposition.
For a given real-valued matrix $\boldA = \{a_{n,m}\}_{n,m=1}^{N,M}\in \R^{N \times M}$, $[\boldA]_{n \times m}$ refers to the entry at position $(n,m)$ in $\boldA$, i.e., $[\boldA]_{n,m} = a_{n,m}$.

The \ac{3D} image is composed of $J$ voxels listed in the set $\calS = \{1,\dots,J\}$. An image defined on $\calS$ takes the form of a real-valued column vector $\boldx = [x_1,\dots,x_J]\transp\in \R^J$ such that for all $j$ the value $x_j$ is the image intensity at voxel $j$. Given a subset of voxels $\calT \subset \calS$, $\boldx_{\calT}$ denotes the restriction of $\boldx$ to $\calT$, i.e., $\boldx_{\calT} = \{x_j\}_{j\in \calT} \subset \R^m$, with $m = \mathrm{card}(\calT)$. 

For each voxel $j$, $\calN_j$ denotes the closed neighborhood of $j$, i.e., $k\in\calN_j\Leftrightarrow j\in\calN_k$ for all $(j,k)$ and $j\in\calN_j$ for all $j$. In this work, we defined $\calN_j$  as the 11\texttimes{}11\texttimes{}11 box centered on $j$ for all $j = 1,\dots,J$ (omitting boundary constraints), and we define by $m = \mathrm{card}(\calN_j) = 11^3$ the number of voxels in each neighborhood. The choice of this neighborhood size is justified in Appendix~\ref{sec:runtime} and Appendix~\ref{sec:kernelsize}. 

$\boldzero$ and $\boldone$ respectively denote the zero vector and the vector consisting entirely of ones, with dimensions determined by the context.


\section{Materials and Methods}\label{sec:method}

\subsection{Problem Formulation}

\subsubsection{PET Reconstruction}

The objective of \ac{PET} reconstruction is to retrieve an activity image $\boldx = [x_1,\dots,x_J]\transp \in \mathbb{R}^J$ from a measurement $\boldy = [y_1,\dots,y_I]\transp \in \R^I$,  $I$ being the number of detector pairs in the \ac{PET} system, by matching the expected measurement $\boldybar(\boldx) = [\ybar_1(\boldx),\dots,\ybar_I(\boldx)]\transp \in \R^I $, given by the linear relation 
\begin{equation}
	\boldybar(\boldx) = \boldH \boldx + \boldr
\end{equation}
where $\boldH \in \R^{I \times J}$ represents the \ac{PET} system matrix, such that $[\boldH]_{i,j}$ denotes the probability that an emission originating from voxel $j$ leads to an annihilation event producing a pair of $\upgamma$-photons detected by detector pair $i$, and $\boldr \in \R^I$ is a background vector representing expected scatter and randoms.  The reconstruction is performed via an optimization problem of the form
\begin{equation}\label{eq:opti}
	\min_{\boldx} \, \ell (\boldy,\boldybar(\boldx))
\end{equation}  
where $\ell$ is a loss function that evaluates the goodness of the fit between $\boldy$ and $\boldybar(\boldx)$, generally defined as the negative Poisson log-likelihood, i.e.,  $\ell(\boldy,\boldybar) = \sum_i  - y_i \log \ybar_i + \ybar_i$ (up to constants), in which case solving \eqref{eq:opti} is achieved via an \ac{EM} algorithm \cite{shepp1982maximum} which computes the estimate $\boldx^{(q+1)}$ at iteration $q+1$ from the estimate $\boldx^{(q)}$ at iteration $q$ with the updating rule 
\begin{equation} \label{eq:mlem}
	\boldx^{(q+1)} = \frac{\boldx^{(q)}}{\boldH\transp \bm{1}}  \boldH\transp \left( \frac{\boldy}{\boldH \boldx^{(q)} +  \boldr} \right)   \, .
\end{equation}
where all vector operations are to be understood element-wise. This algorithm can be generalized for parametric imaging \cite{wang2013direct}.

\subsubsection{Incorporating Positron Range}

The \ac{PET} system matrix $\boldH$ depends on the system's geometry, the \ac{3D} linear attenuation image $\boldmu \in \R^J$---usually derived from an anatomical image such as \acf{CT} or \acf{MR}---and \ac{PR} which depends on the \ac{3D} electronic density image $\boldrho \in \R^J$. In the context of \ac{PET} imaging, $\boldmu$ and $\boldrho$ are strongly correlated \cite{manohara2008effective}, and therefore we assume that \ac{PR} is determined by $\boldmu$.

The matrix $\boldH$ can be decomposed as \cite{reader2002one}
\begin{equation}\label{eq:factorisation}
	\boldH = \boldA(\boldmu) \boldP  \boldB(\boldmu)  
\end{equation} 
where $\boldA(\boldmu) \in \R^{I\times I}$ is a diagonal matrix representing the attenuation factors along the \acp{LOR} for each detector pair, $\boldP\in \R^{I\times J}$ is the \ac{PET} geometric projector defined such that $[\boldP]_{i,j}$ is the probability that an annihilation taking place at voxel $j$ is detected on $i$ in absence of attenuation (taking into account sensitivity and detector resolution), and $\boldB(\boldmu)$ is the \ac{PR} blurring operator defined such that $[\boldB(\boldmu)]_{j',j}$ is the probability that a positron emitted in $j$ interacts with an electron in $j'$. 

The geometric projector $\boldP$ is known from the system's manufacturer, while $\boldA(\boldmu)$ can be computed by integrating $\boldmu$ along each \ac{LOR}. The \ac{PR} blurring operator $\boldB(\boldmu)$ is more challenging, as it performs position-dependent blurring. Consequently, it is often replaced by the identity matrix or a position-independent blurring operator \cite{derenzo1986mathematical}, which may underestimate \ac{PR} in regions with low electron density, such as the lungs.

A \ac{CNN} can be trained to approximate $\boldB(\boldmu) \boldx$ by taking $\boldx$ and $\boldmu$ as inputs and directly producing an image with \ac{PR}  blurring applied \cite{merlin2024deep}. While computationally efficient, this approach cannot be used to compute the transpose of the  \ac{PR} operator $\mathbf{B}(\boldmu)^\top$, leading to the use of an unmatched forward model in the iterative scheme \eqref{eq:mlem}.

\subsection{Dual-Input Dynamic Convolution for Positron Range Modeling}

This section describes our \ac{DDConv} implementation  of the \ac{PR} blurring   $\boldx\mapsto\boldB(\boldmu) \boldx$ and its transposed version $\boldz\mapsto\boldB(\boldmu)\transp \boldz$ which are involved in the \ac{EM} algorithm  \eqref{eq:mlem} through $\boldH$ and $\boldH\transp$.

\subsubsection{Matrix Formulation}

The blurring operator $\boldB(\boldmu) \in \R^{J\times J}$ models the \ac{PR}-induced spatial blurring, transforming an activity distribution image $\boldx \in \R^J$ into an annihilation distribution image $\boldz = [z_1,\dots,z_J]\transp \in \R^J$ defined as 
\begin{equation}\label{eq:blur}
	\boldz = \boldB(\boldmu) \boldx \, ,
\end{equation} 
which represents the spatial locations where positrons undergo annihilation. The attenuation image $\boldmu$ governs this process by defining the local electron density and tissue composition, which influence positron propagation before annihilation. In the following, we assume that \ac{PR} is bounded. More precisely, we assume that a positron emission at voxel $j$ results in an annihilation in an 11\texttimes{}11\texttimes{}11 closed neighborhood of $j$, denoted $\calN_j$, with $m\triangleq \mathrm{card}(\calN_j) = 11^3$. 
 
For all $j=1,\dots,J$, the probability that a positron emitted from $j$ annihilates with an electron located in voxel $k\in \calN_j$ is denoted $w_{j\to k} \in [0,1]$ and is entirely determined by  $\boldmu_{\calN_j} \in \R^m$ for a given radiotracer, and we assume that annihilation is certain, i.e.,
\begin{equation}\label{eq:sum_prob}
	\sum_{k\in\calN_j} w_{j\to k} = 1 \, .
\end{equation}	
In other words, the vector $\boldw_j = \{w_{j\to k}\}_{k\in\calN_j}\in \R^m$ is the \ac{PSF} at voxel $j$. The annihilation distribution image $\boldz$ is obtained at each voxel $k$ by performing a sum of the activity values of $\boldx_{\calN_k}$ weighted by the $w_{j\to k}$'s, $j\in\calN_k$, 
\begin{align}\label{eq:blurring}
	z_k = \sum_{j\in\calN_k}  w_{j\to k} \cdot x_j 
\end{align}		
and thus we have defined blurring operator $\boldB(\boldmu)$ as
\begin{equation} 
	[\boldB(\boldmu)]_{k,j} = 
	\begin{cases}
		w_{j\to k} & \text{if}\, j \in \calN_k \, ,\\
		0 & \text{otherwise.}
	\end{cases}
\end{equation}

\subsubsection{PR Prediction using a CNN}

The position-dependent \ac{PSF} $\{\boldw_j\}_{j\in\calS}$ cannot be stored and therefore we opted for an on-the-fly implementation of the blurring operator $\boldB(\boldmu)$. 

We used a \ac{CNN} $\boldG_{\boldtheta} \colon \R^m \times \R^m \to \R^m$ with trainable parameter $\boldtheta$  to predict $\boldw_j$ from $\boldmu_{\calN_j}$. Additionally, $\boldG_{\boldtheta}$ takes as input a constant vector $\boldd = \{ d_{j,k} \}_{k \in \calN_j}$ with $d_{j,k} = \mathrm{dist}(j,k)$---included as a second channel---to provide spatial information to the \ac{CNN}; similar location-augmentation strategies have been used in prior work \cite{wang2018location,liu2018intriguing,hu2024learning}.

\begin{figure}[t]
	\centering
	\includegraphics[width=\columnwidth]{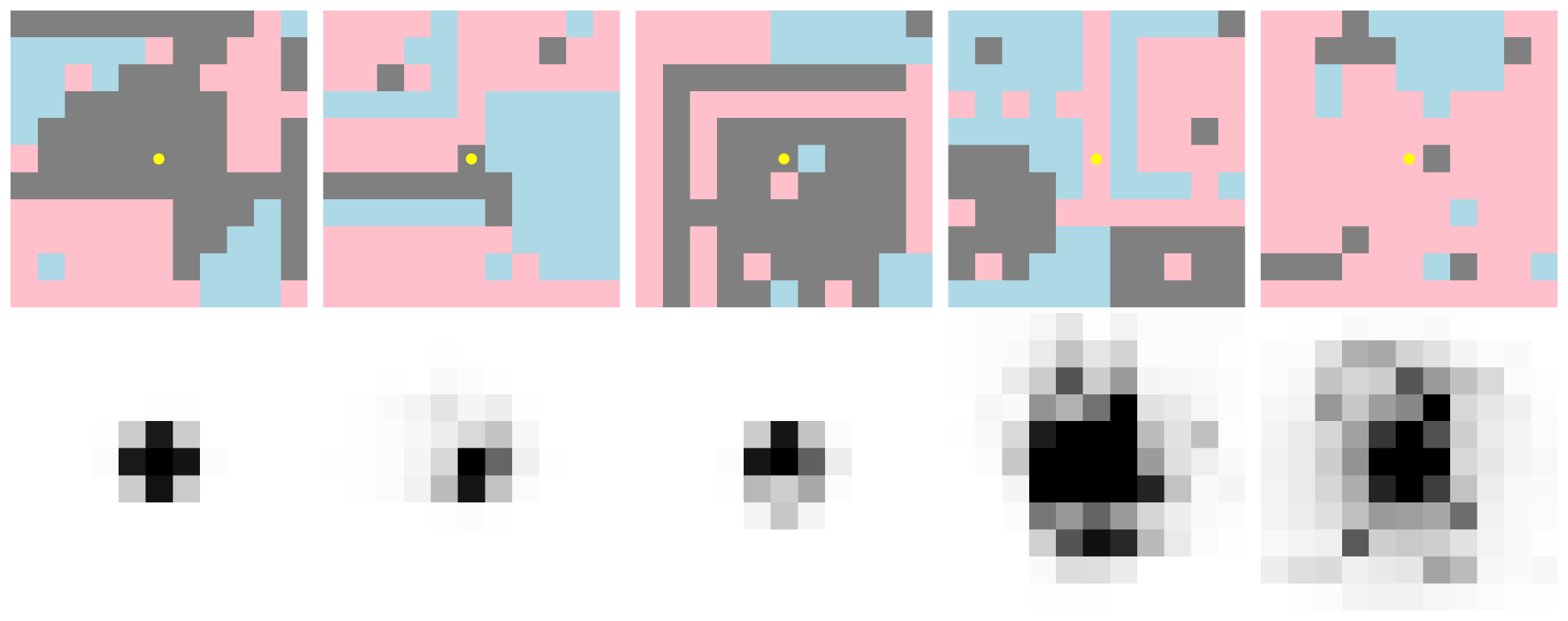}
	\caption{Random material images $\boldeta$ (upper row) with tissue-specific color coding---pink for lung, light blue for water, and gray for bone---and their corresponding \ac{MC}-generated \ac{PR} \ac{PSF} $\boldw_{\boldeta}$ (annihilation image). The yellow spot represents the \ac{Ga68} positron-emitting point source. We used a 11\texttimes{}11\texttimes{}11 window with 2-mm cubic voxels}
	\label{fig:KernelsUsed}
\end{figure}

The training of $\boldG_{\boldtheta}$ is performed using 1,000 small random synthetic 11\texttimes{}11\texttimes{}11  $L$-material images $\boldeta \in \{1,2,\dots,L\}^m$, $m=11^3$. In this work, we considered the lung, rib bone, and water materials ($L=3$) although additional material may be considered for other applications. The synthetic material images---see Figure~\ref{fig:KernelsUsed} (top) for examples---were generated using a custom \ac{3D} shape generator. Starting from an empty volume,  we sequentially placed randomly sized and oriented geometric primitives (cylinders and boxes) with randomly sampled positions and material labels, and superposed them to form extended regions and structured interfaces between materials. To introduce additional local heterogeneity, we then randomly flipped the material label in 10\% of the voxels, which produces isolated voxels and small disconnected fragments. While this procedure yields a diverse and relatively challenging set of phantoms with many sharp transitions and small features, we observed that these shapes were sufficient to train our model.

For each material image $\boldeta$, a \ac{MC} simulation is performed using \ac{GATE} \cite{jan2004gate} with a \ac{Ga68} positron-emitting point source at the center of $\boldeta$ to generate a \ac{PSF} $\boldw_{\boldeta} \in \R^m$. Each simulation generates 10\textsuperscript{6} positron emission events in order to generate a single noise-free \ac{PR} \ac{PSF} $\boldw_{\boldeta}$---see Figure~\ref{fig:KernelsUsed} (bottom).

Supervised training of the \ac{CNN} $\boldG_{\boldtheta}$ is achieved by solving the optimization problem
\begin{equation}
	\min_{\boldtheta} \,  \mathbb{E}_{ \boldeta }   \left[ \calL\left( \boldG_{\boldtheta} (\boldmu_{\boldeta} , \boldd) , \boldw_{\boldeta} \right)  \right] \label{eq:training}
\end{equation}
where $\boldmu_{\boldeta} \in \R^m$  is the attenuation map corresponding to  $\boldeta$ and $\calL$ is a loss function. The complete architecture of $\boldG_{\boldtheta}$ is illustrated in Figure~\ref{fig:fullarchitecture} (right). To compute \eqref{eq:training}, we employed a \ac{KL} divergence for $\calL$ averaged over the 1,000 realizations of $\boldeta$. We observed that, for this kernel size, 1,000 realizations were sufficient to train the model $\boldG_{\boldtheta}$ to accurately predict the \ac{PR} \ac{PSF} $\boldw_{\boldeta}$, although we have not investigated if that number could be reduced. The training was performed for 5,000~epochs using the Adam optimizer ($\text{lr}=10^{-4}$, batch size of~4) on an NVIDIA GeForce~RTX~3060~\ac{GPU} with PyTorch~2.5 and \ac{CUDA} acceleration. The total training time was approximately 3~hours for the 11\texttimes{}11\texttimes{}11 kernel, 3.5~hours for the 21\texttimes{}21\texttimes{}21 kernel, and 4~hours for the 31\texttimes{}31\texttimes{}31 kernel.

\subsubsection{Implementation of the Blurring}

At each voxel $j$, the \ac{PSF} $\boldw_j$ is computed from the local attenuation image $\boldmu_{\calN_j}$ using $\boldG_{\boldtheta}$  to redistribute the activity value $x_j$ in $\calN_j$, using a $\mathtt{spread}$ operation defined as 
\begin{equation}\label{eq:spread}
	\mathtt{spread}(x_j,\boldw_j) = \{w_{j\to k} \cdot x_j  \}_{k\in\calN_j}
\end{equation}
In our implementation, this operation is achieved using the $\mathtt{torch.nn.ConvTranspose3d}$ module provided by PyTorch~\cite{paszke2019pytorch, zeiler2010deconvolutional}.
Starting from an initial annihilation image $\boldz = \boldzero$, the final annihilation image is obtained by summing up the spread activity for each neighborhood $\calN_j$:
\begin{equation}
	\boldz_{\calN_j} \gets \boldz_{\calN_j} + \mathtt{spread}(x_j,\boldw_j)  \, . 
\end{equation}
Conversely, the transposed blurring operator $\boldB(\boldmu)\transp$ is achieved  by summing the annihilation image over $\calN_j$ with weights $w_{j\to k}$, i.e., 
\begin{equation}\label{eq:conv} 
	x_j \gets \sum_{k\in\calN_j}  w_{j\to k} \cdot z_k  \, .
\end{equation}

All these operations can be performed in parallel and in pairwise disjoint batches of voxels $\calB_q$ with $\calS = \cup_{q=1}^Q\calB_q$, $\calB_q \cap \calB_{p} = \varnothing$.

\begin{figure*}[htbp]
    \centering
    \includegraphics[width=0.9\textwidth]{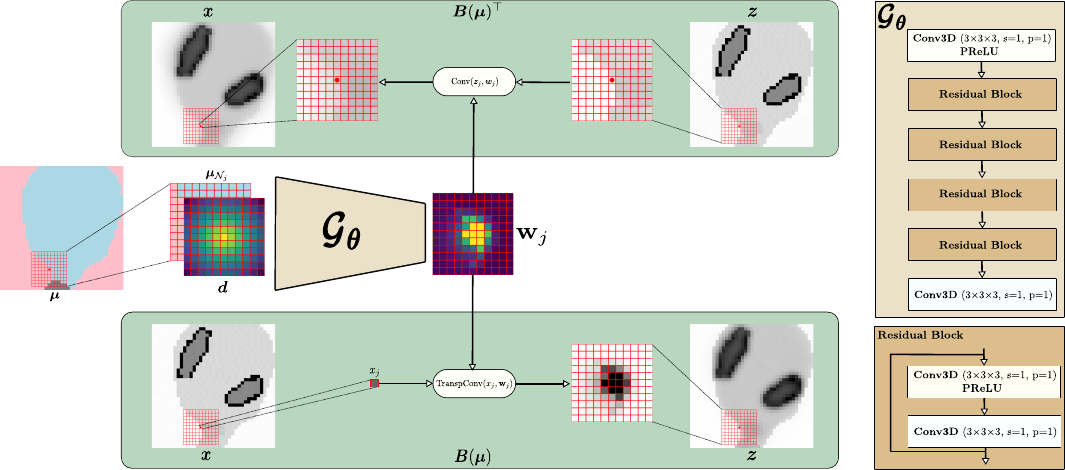}
    \caption{Illustration of the \ac{PR} blurring operators. The top section represents the transposed operator $\boldB(\boldmu)^\top$, while the bottom section shows the forward operator $\boldB(\boldmu)$. Both operations use spatially varying PSFs $\boldw_j$ predicted by the same model $\boldG_{\boldtheta}$, based on the local attenuation image $\boldmu_{\calN_j}$. The right side details the architecture of $\boldG_{\boldtheta}$.}
    \label{fig:fullarchitecture}
\end{figure*}

The overall \ac{DDConv} methodology to compute $\boldB(\boldmu) \boldx$ and $\boldB(\boldmu)\transp \boldz$ is summarized in Figure~\ref{fig:fullarchitecture}, Algorithm~\ref{algo:pr} and Algorithm~\ref{algo:pr_t}.

\begin{algorithm}
	\caption{\Ac{PR} blurring}  \label{algo:pr}
	\begin{algorithmic}[1]
		\Require $\boldx$ (activity), $\boldmu$ (attenuation image), $\boldG_{\boldtheta}$ (\ac{PSF} predictor), $\calS = \cup_{q=1}^Q\calB_q$ with $\calB_q \cap \calB_{p} = \varnothing$ for all $q\ne p$ (batch decomposition).
		\State $\boldz \gets \boldzero$ \Comment{initialization}
		\For{$q=1,\dots,Q$}
			
			\For{$j\in \mathcal{B}_q$}
				\State $\boldw_j \gets  \boldG_{\boldtheta}\left(\boldmu_{\calN_j},\boldd\right)$
				\State $\boldz_{\calN_j}  \gets \boldz_{\calN_j} +  \mathtt{spread}(x_j,\boldw_j)$
			\EndFor
		\EndFor
		\State \Return $\boldz$
	\end{algorithmic}
    \label{alg:AlgorithmForward}
\end{algorithm}

\begin{algorithm}
	\caption{\Ac{PR} transposed blurring}  \label{algo:pr_t}
	\begin{algorithmic}[1]
		\Require $\boldz$ (annihilation image), $\boldmu$ (attenuation image), $\boldG_{\boldtheta}$ (\ac{PSF} predictor), $\calS = \cup_{q=1}^Q\calB_q$ with $\calB_q \cap \calB_{p} = \varnothing$ for all $q\ne p$ (batch decomposition).
		\State $\boldx \gets \boldzero$ \Comment{initialization}
		\For{$q=1,\dots,Q$}
			
			\For{$j\in \mathcal{B}_q$}
				\State $\boldw_j \gets  \boldG_{\boldtheta}\left(\boldmu_{\calN_j},\boldd\right)$ 
		
				\State $x_j \gets \sum_{k\in\calN_j}  w_{j\to k} \cdot z_k $
			\EndFor
		\EndFor
		\State \Return $\boldx$
	\end{algorithmic}

    \label{alg:AlgorithmBackward}
\end{algorithm}

\section{Experiments and Results}\label{sec:results}


\subsection{Experimental Setup and Dataset for Positron Range Correction Evaluation}\label{sec:setup}

The performance of the proposed method was benchmarked against the \ac{SVTD} \ac{PRC} method by \citeauthor{kertesz2022implementation}~\cite{kertesz2022implementation}. This approach utilizes a tissue-dependent anisotropic \ac{PSF}. \Ac{SVTD} approximates the \ac{PR} \ac{PSF} by choosing \ac{MC}-derived \acp{PSF} corresponding to different tissue types (e.g., lung, soft tissue, bone), and then cutting and assembling these according to the tissue boundaries. This approximation is reasonable in homogeneous regions but can be inaccurate at the interface between several tissue types (cf. Figure~5 in \cite{kertesz2022implementation}). Therefore, we opted to focus our evaluation by investigating the accuracy of \ac{SVTD} and \ac{DDConv} in such scenarios.

All experiments were carried out on a workstation equipped with an Intel Xeon E5-1650 v4 CPU (3.6~GHz), 62~GB RAM, and an NVIDIA GeForce RTX 3060 GPU (12 GB VRAM) using PyTorch 2.5 with CUDA acceleration.

We first evaluated the accuracy of the \ac{PR} blurring on digital phantoms (Experiment~1), then in image reconstruction on \ac{MC}-simulated data (Experiment~2) and real phantom data (Experiment~3).

We used a 2\texttimes{}2\texttimes{}2 mm\textsuperscript{3} voxel size for all experiments.

For data acquisition, simulated data were generated using a Siemens mMR \ac{PET} scanner, which has a 60~cm inner diameter, a 90~cm outer diameter, and \ac{LSO} crystals measuring 4\texttimes{}4\texttimes{}20 mm\textsuperscript{3}.  Clinical data were acquired using a Siemens Biograph Vision \ac{PET}/\ac{CT} system, which features a 78-cm bore diameter and \ac{LSO} crystals measuring 3.2\texttimes{}3.2\texttimes{}20 mm\textsuperscript{3}. 

Image reconstructions were performed by \ac{EM} using \acs{castor}~\cite{merlin2018castor}, with incorporation of \ac{DDConv} (i.e.,~$\boldB(\boldmu)$ and $\boldB(\boldmu)\transp$).  We performed reconstruction from \ac{MC}-simulated data generated from a digital phantom and the \ac{XCAT} phantom~\cite{segars20104d} (male, no respiratory or cardiac motion), as well as from real phantom data acquired on the Siemens Biograph Vision system at Kuopio University Hospital (Kuopio, Finland).  Raw \ac{PET} data were simulated with 200~ps \ac{TOF} resolution for the synthetic datasets (no \ac{TOF} for real phantom data).  The intrinsic spatial resolution of the systems was incorporated in~$\boldP$, 
with 4.4\texttimes{}4.4\texttimes{}4.4 mm\textsuperscript{3} \ac{FWHM} for the Siemens mMR and 3.6\texttimes{}3.6\texttimes{}3.6 mm\textsuperscript{3} \ac{FWHM} for the Siemens Biograph Vision. No post-reconstruction filtering was applied.

Three quantitative metrics were used in Experiment~2 to assess lesion quantification performance across iterations: the \ac{RC}, the \ac{SUV}\textsubscript{max} error, and the \ac{MAPE}. Denoting $\boldxhat = [\xhat_1,\dots,\xhat_J]\transp$ and $\boldx^\star = [x^\star_1,\dots,x^\star_J]\transp$ the reconstructed image and the \ac{GT} image respectively, the \ac{RC} quantifies the average recovery of lesion intensity with respect to the \ac{GT} and is defined as
\begin{equation}
	\mathrm{RC} = \frac{\sum_{j\in \mathrm{lesion}  } \xhat_j   }{\sum_{j\in \mathrm{lesion}  } x^\star_j},
\end{equation}
The \ac{SUV}\textsubscript{max} error measures the relative difference between the reconstructed and \ac{GT} maximum voxel value inside the lesion:
\begin{equation}
	\mathrm{SUV_{max}\,Err}~(\%) = 100 \times 
	\frac{\left| \mathrm{SUV}_\mathrm{max}^\mathrm{rec} - \mathrm{SUV}_\mathrm{max}^\star \right|}
	{\mathrm{SUV}_\mathrm{max}^\star},
\end{equation}
where $\mathrm{SUV}_\mathrm{max}^\mathrm{rec} =  \max \{ \xhat_j , \, j\in\mathrm{lesion}  \}  $ and $\mathrm{SUV}_\mathrm{max}^\star =  \max \{ x^\star _j , \, j\in\mathrm{lesion}  \}$.  Finally, the \ac{MAPE} evaluates the average voxelwise deviation within the lesion and is defined as
\begin{equation}
	\mathrm{MAPE}~(\%) = 
	\frac{100}{\operatorname{card}(\mathrm{lesion})} 
	\sum_{j\in\mathrm{lesion}} 
	\left| 
	\frac{\xhat_j - x^\star_j}{x^\star_j} 
	\right|,
\end{equation}
These metrics quantify, respectively, the overall lesion contrast recovery, the local bias of the hottest voxel, and the voxelwise quantitative accuracy.

\subsection{Experiment 1: Blurring Accuracy}\label{sec:exp1}

\subsubsection{Geometric Phantom}

\begin{figure*}[!ht]
	\centering
	\includegraphics[width=0.8\textwidth]{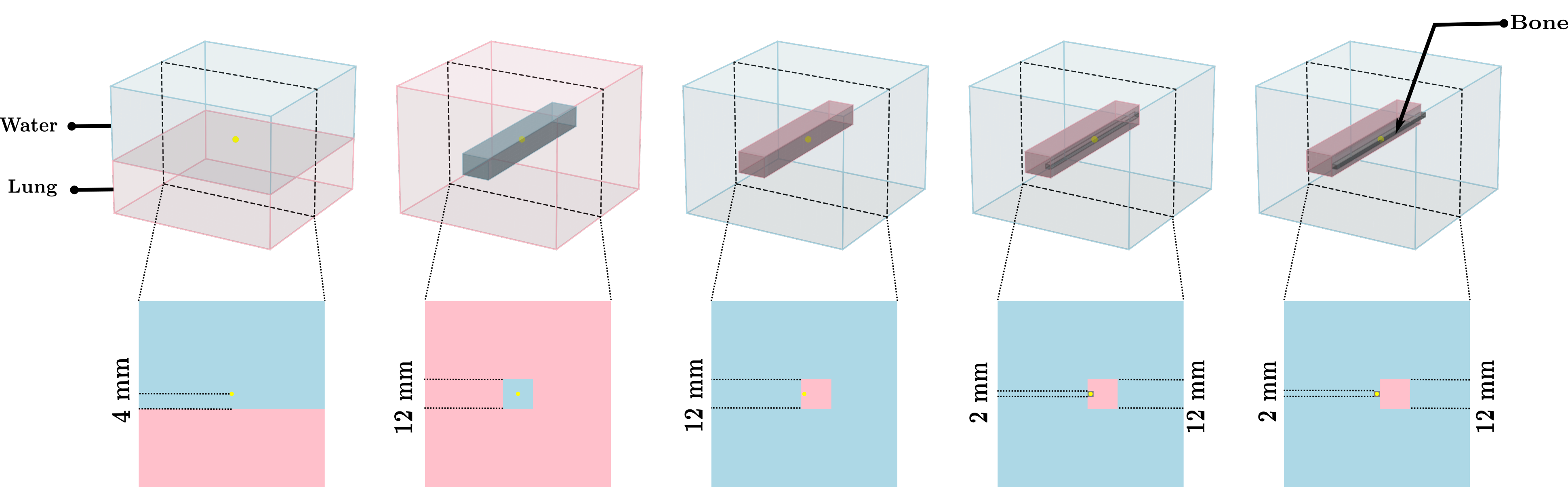}
	\caption{
        Experiment 1---Digital phantoms used to assess \ac{PR} blurring accuracy (pink for lung, light blue for water).
	}
	\label{fig:phantoms}
\end{figure*}

To investigate the spatial variation of \ac{PR} distributions in heterogeneous tissue environments, we designed a series of controlled digital phantoms that simulate distinct biological compositions relevant to \ac{PET} imaging, following the approach of \citeauthor{kertesz2022implementation}~\cite{kertesz2022implementation}. Each phantom is represented as a \ac{3D} volume of 62\texttimes{}62\texttimes{}62 mm\textsuperscript{3}, with a \ac{Ga68} point source placed at the center. We considered five distinct configurations (Figure~\ref{fig:phantoms}):
(i) a lung--water interface, where lung tissue occupies the anterior 26~mm along the $z$-axis, while the remaining 36~mm is filled with water;
(ii) a lung background with a centrally embedded 12\texttimes{}12 mm\textsuperscript{2} water inclusion spanning the full 62~mm in the $x$-dimension;
(iii) a water matrix containing a 12\texttimes{}12 mm\textsuperscript{2} lung region, offset by 4 mm along the $y$-axis;
(iv) a water background embedding a 12\texttimes{}12 mm\textsuperscript{2} lung inclusion that contains a 2-mm bone column extending along the entire $x$-dimension;
(v) the same as (iv), except the lung inclusion is shifted an additional 2~mm (one voxel) along the $y$-axis, while the bone column remains fixed.

Figure~\ref{fig:overview_axes} shows the results of the \ac{PR} blurring from \ac{MC} simulation (reference), \ac{SVTD} and the proposed \ac{DDConv}. The proposed method \ac{DDConv} produces positron annihilation distributions that closely match those obtained from the reference \ac{GATE} \ac{MC} simulations across all phantom configurations, 
highlighting its accuracy in heterogeneous tissue environments. In contrast, the \ac{SVTD} method exhibits significant deviations from the \ac{GATE} distributions, indicating that it is less reliable for accurately modeling complex spatial variations in \ac{PR}. These results are consistent with those of \citeauthor{kertesz2022implementation}~\cite{kertesz2022implementation} (Figure 5).

\newcommand{\mynewroundedbox}[1]{%
	\tikz[baseline=(box.base)]{
		\node[draw, rounded corners=5pt, inner sep=2mm] (box) {#1};%
	}%
}

\begin{figure*}[htbp]
	\centering
	
	\mynewroundedbox{
		\begin{minipage}[t]{0.3\textwidth}
			\centering
			\textbf{Axis 1 (axial)}\\[1mm]
			
			\begin{minipage}{0.03\linewidth}
				\centering
				~
			\end{minipage}
			\begin{minipage}{0.95\linewidth}
				\centering
				\begin{minipage}{0.23\linewidth}\centering\tiny Phantom\end{minipage}%
				\begin{minipage}{0.23\linewidth}\centering\tiny MC (reference)\end{minipage}%
				\begin{minipage}{0.23\linewidth}\centering\tiny SVTD\end{minipage}%
				\begin{minipage}{0.23\linewidth}\centering\tiny DDConv\end{minipage}%
			\end{minipage}\\[1mm]
			
			\begin{minipage}{0.03\linewidth}
				\centering
				\rotatebox{90}{\tiny Phantom 1}
			\end{minipage}
			\begin{minipage}{0.95\linewidth}
				\centering
				\includegraphics[width=0.23\linewidth,angle=90]{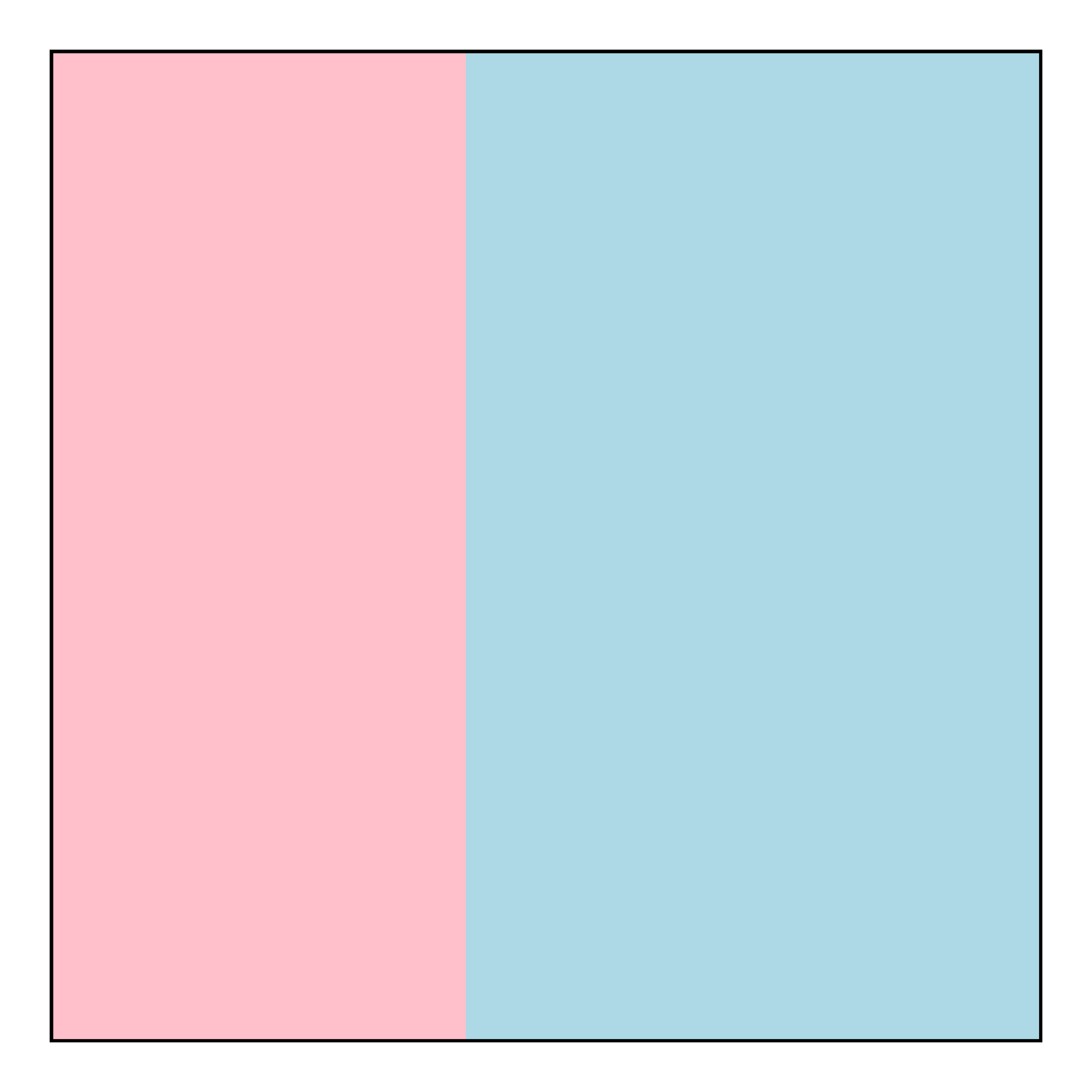}%
				\includegraphics[width=0.23\linewidth,angle=90]{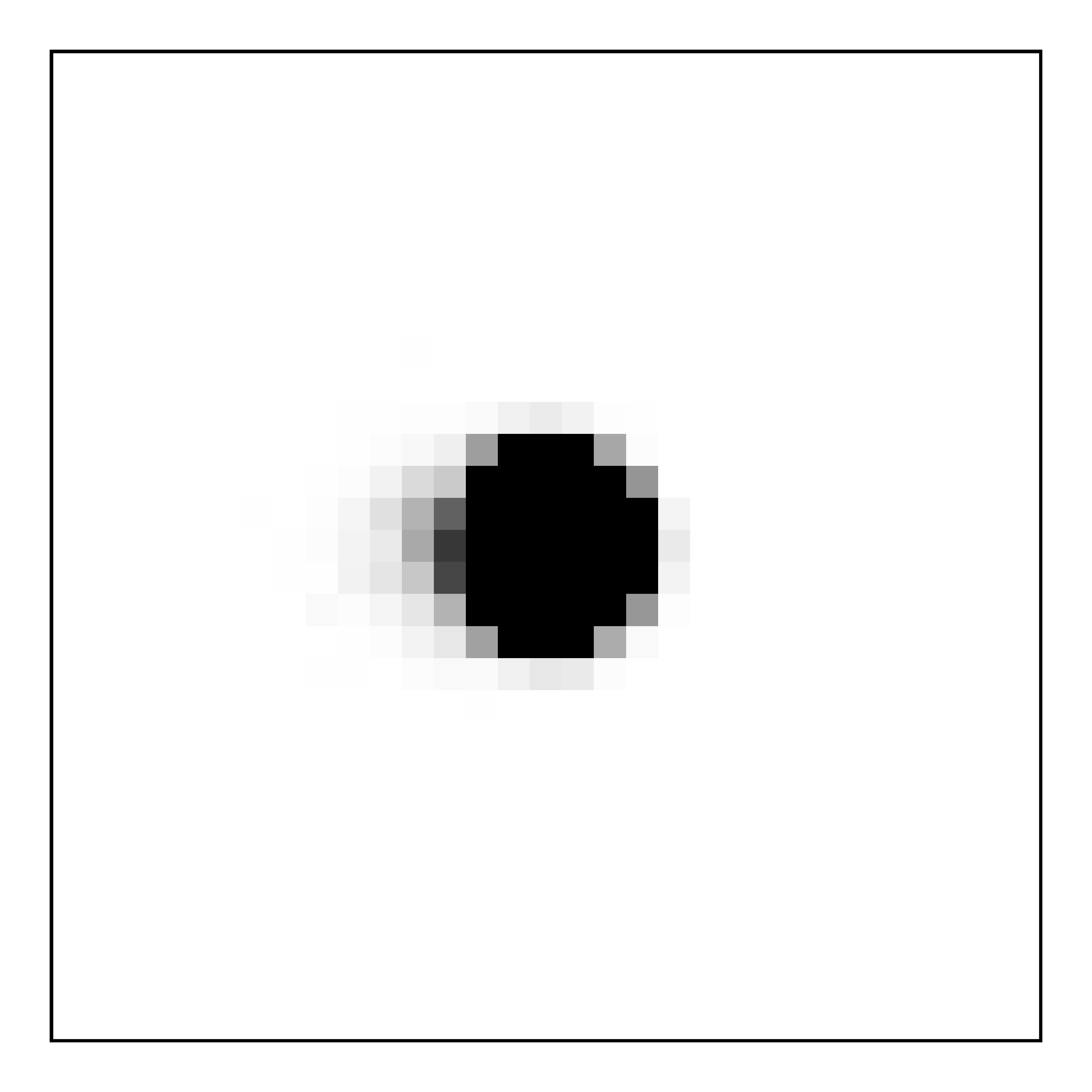}%
				\includegraphics[width=0.23\linewidth,angle=90]{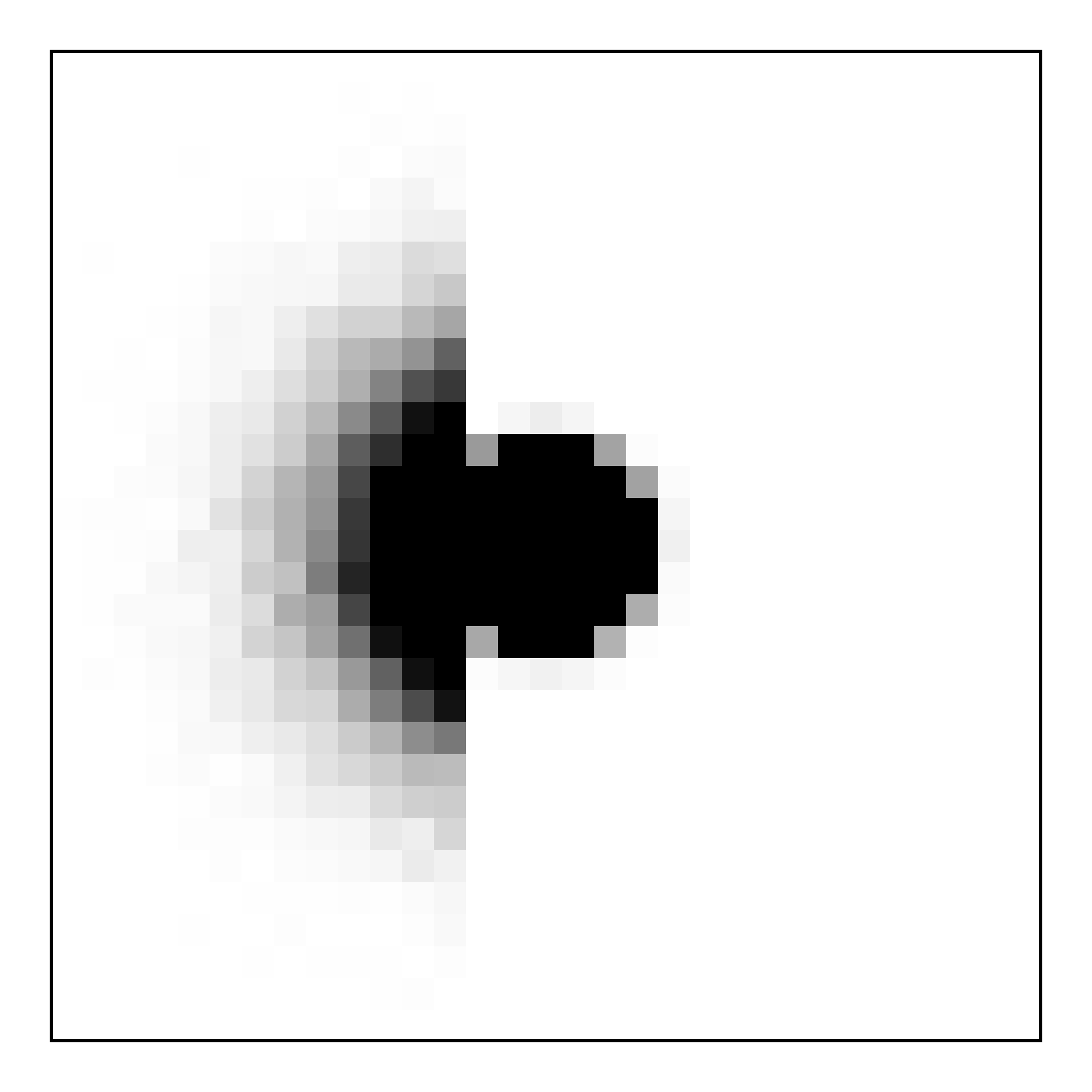}%
				\includegraphics[width=0.23\linewidth,angle=90]{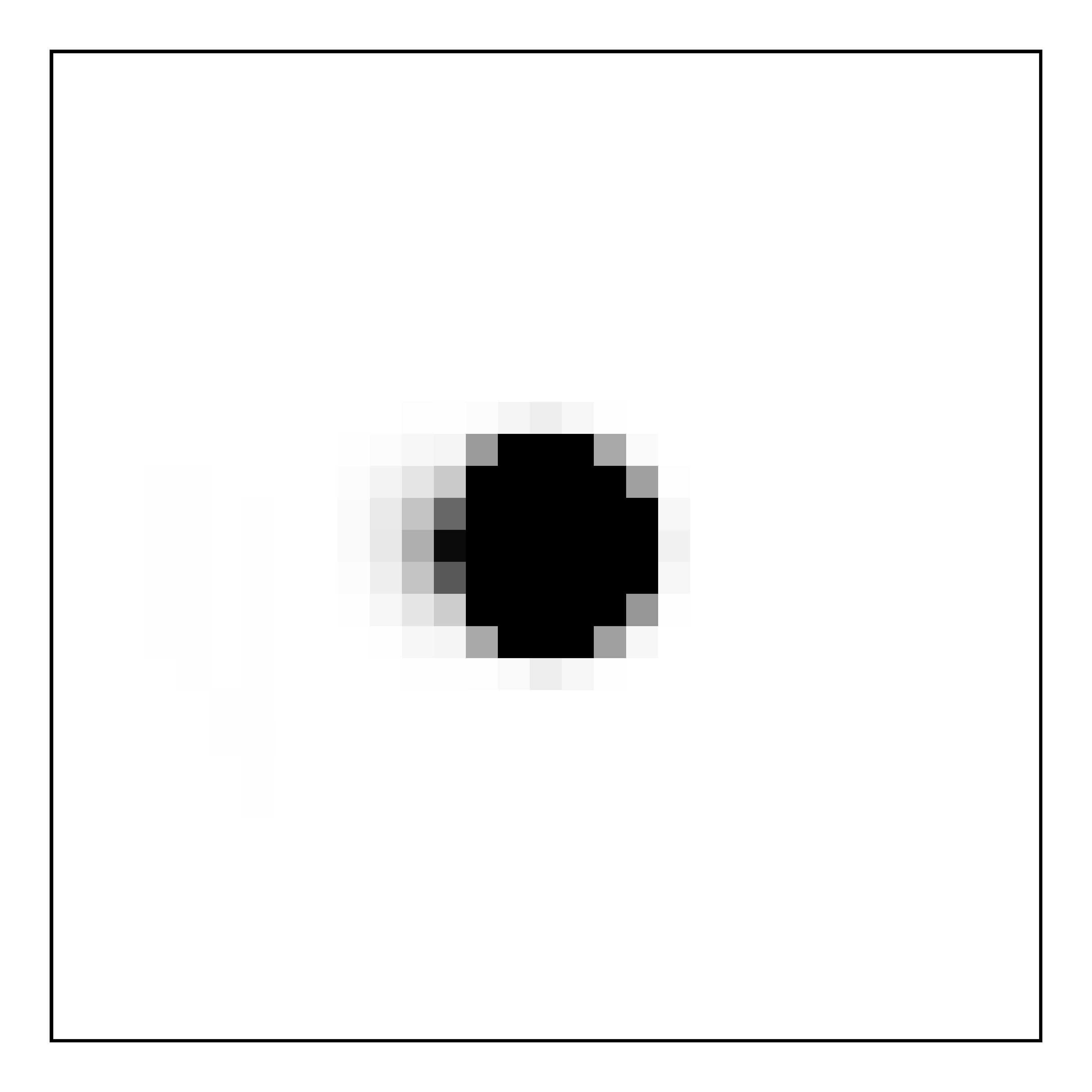}%
			\end{minipage}\\[0.5mm]
			
			\begin{minipage}{0.03\linewidth}
				\centering
				\rotatebox{90}{\tiny Phantom 2}
			\end{minipage}
			\begin{minipage}{0.95\linewidth}
				\centering
				\includegraphics[width=0.23\linewidth,angle=90]{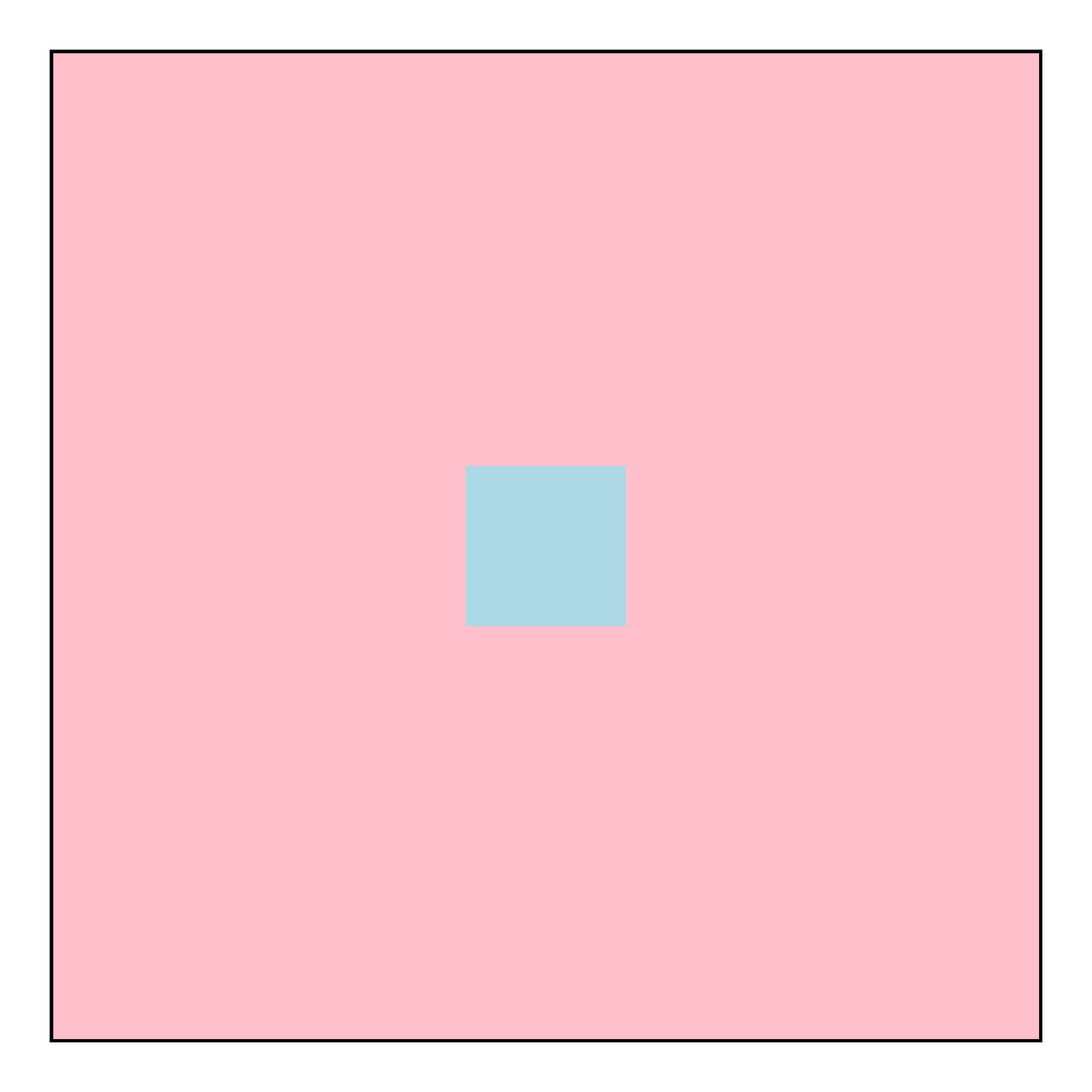}%
				\includegraphics[width=0.23\linewidth,angle=90]{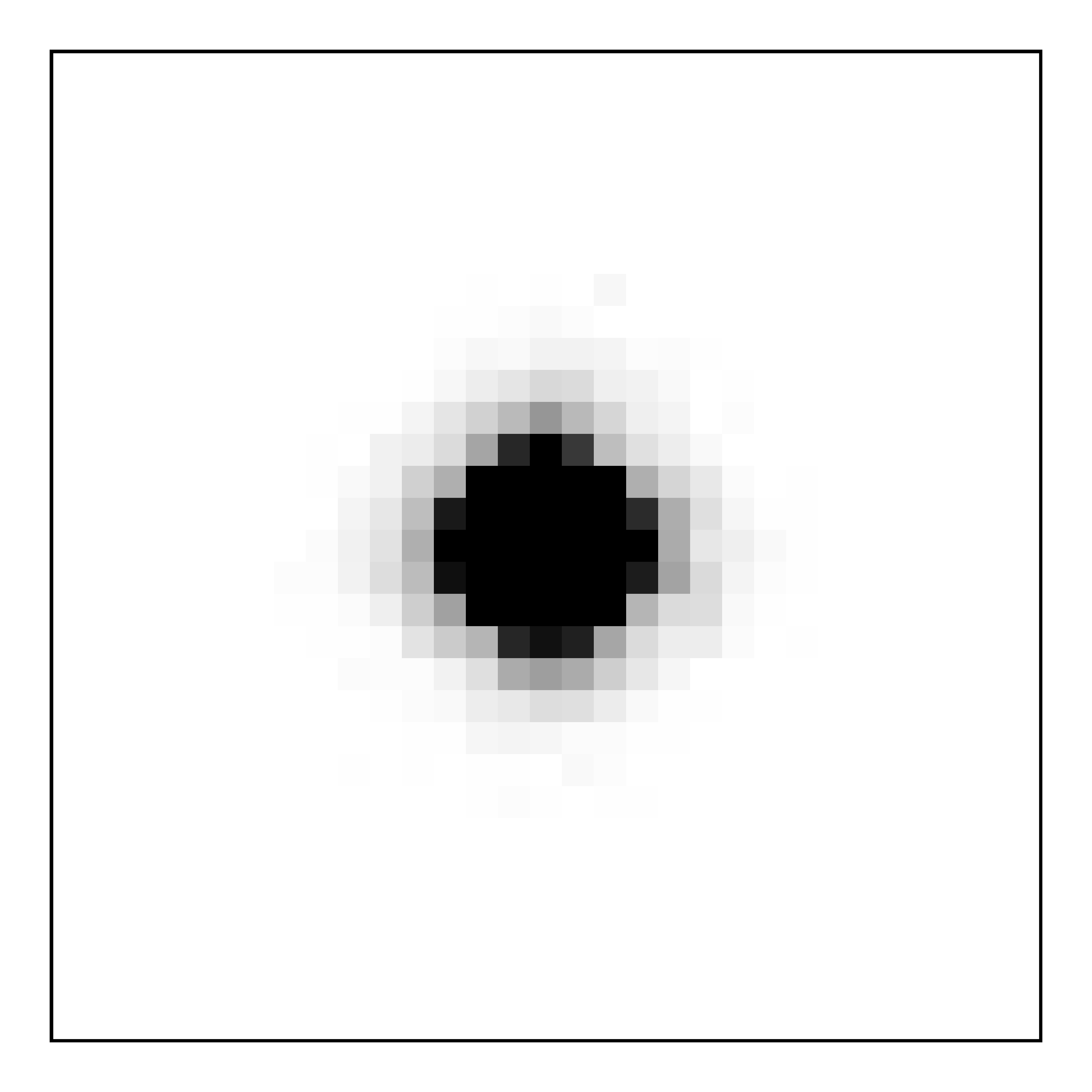}%
				\includegraphics[width=0.23\linewidth,angle=90]{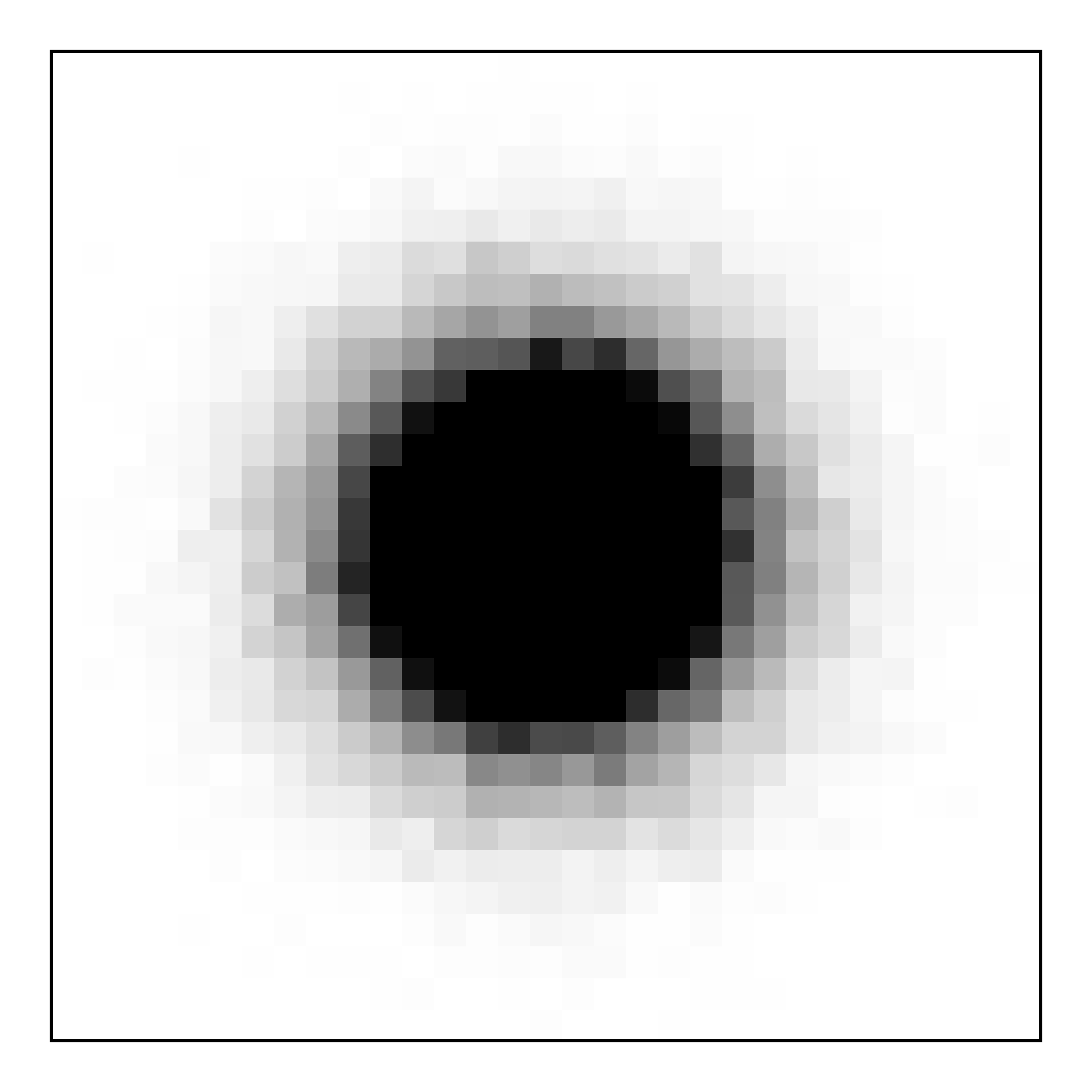}%
				\includegraphics[width=0.23\linewidth,angle=90]{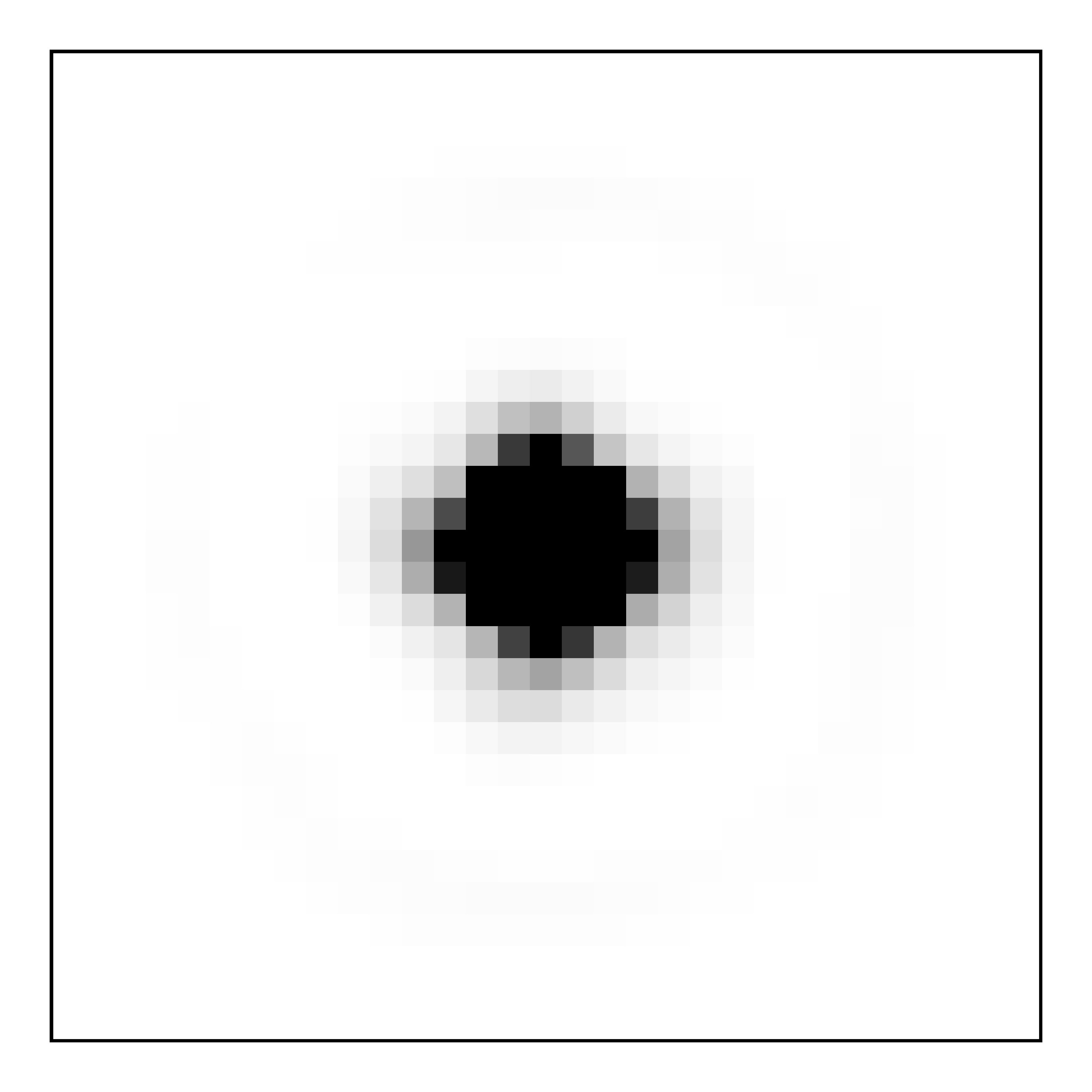}%
			\end{minipage}\\[0.5mm]
			
			\begin{minipage}{0.03\linewidth}
				\centering
				\rotatebox{90}{\tiny Phantom 3}
			\end{minipage}
			\begin{minipage}{0.95\linewidth}
				\centering
				\includegraphics[width=0.23\linewidth,angle=90]{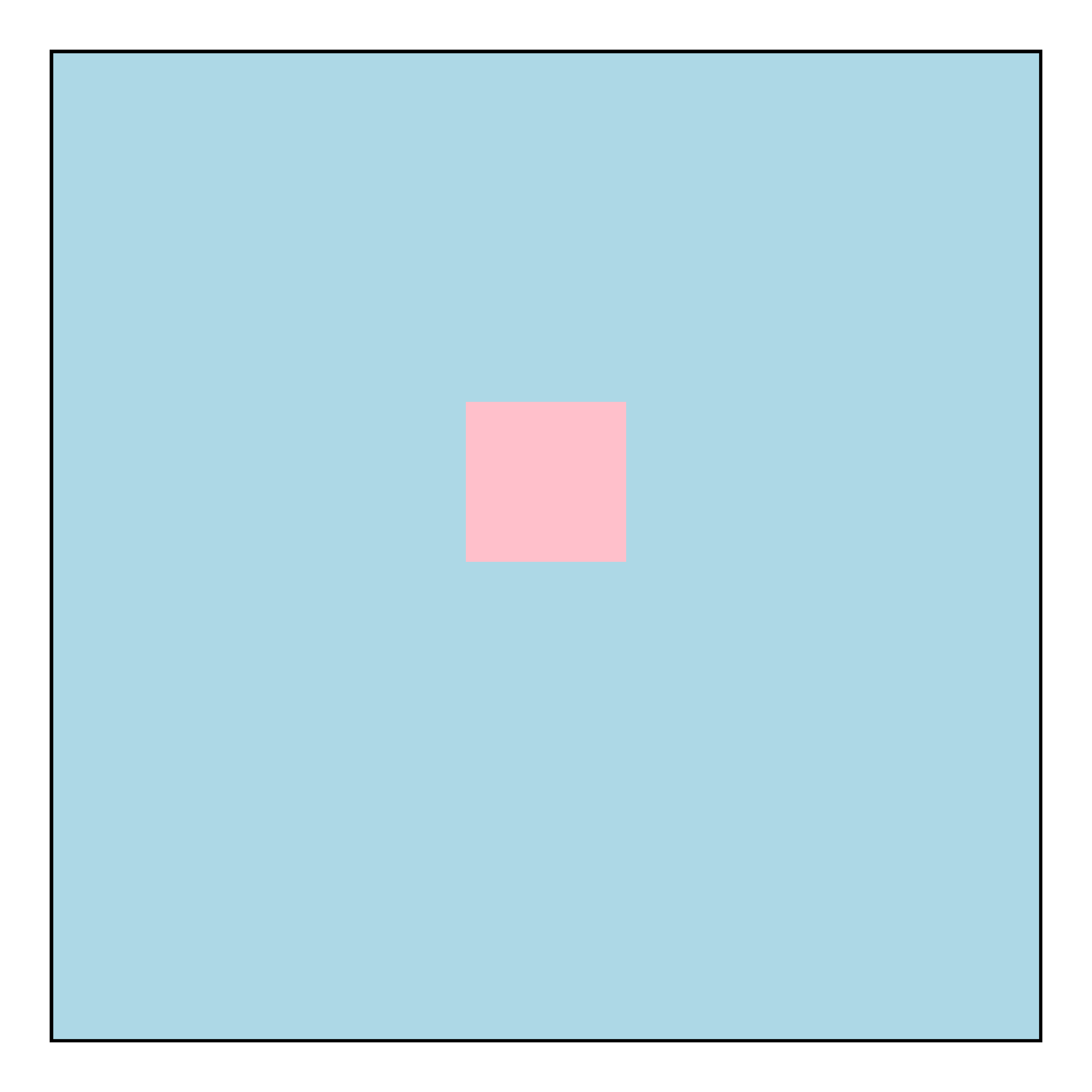}%
				\includegraphics[width=0.23\linewidth,angle=90]{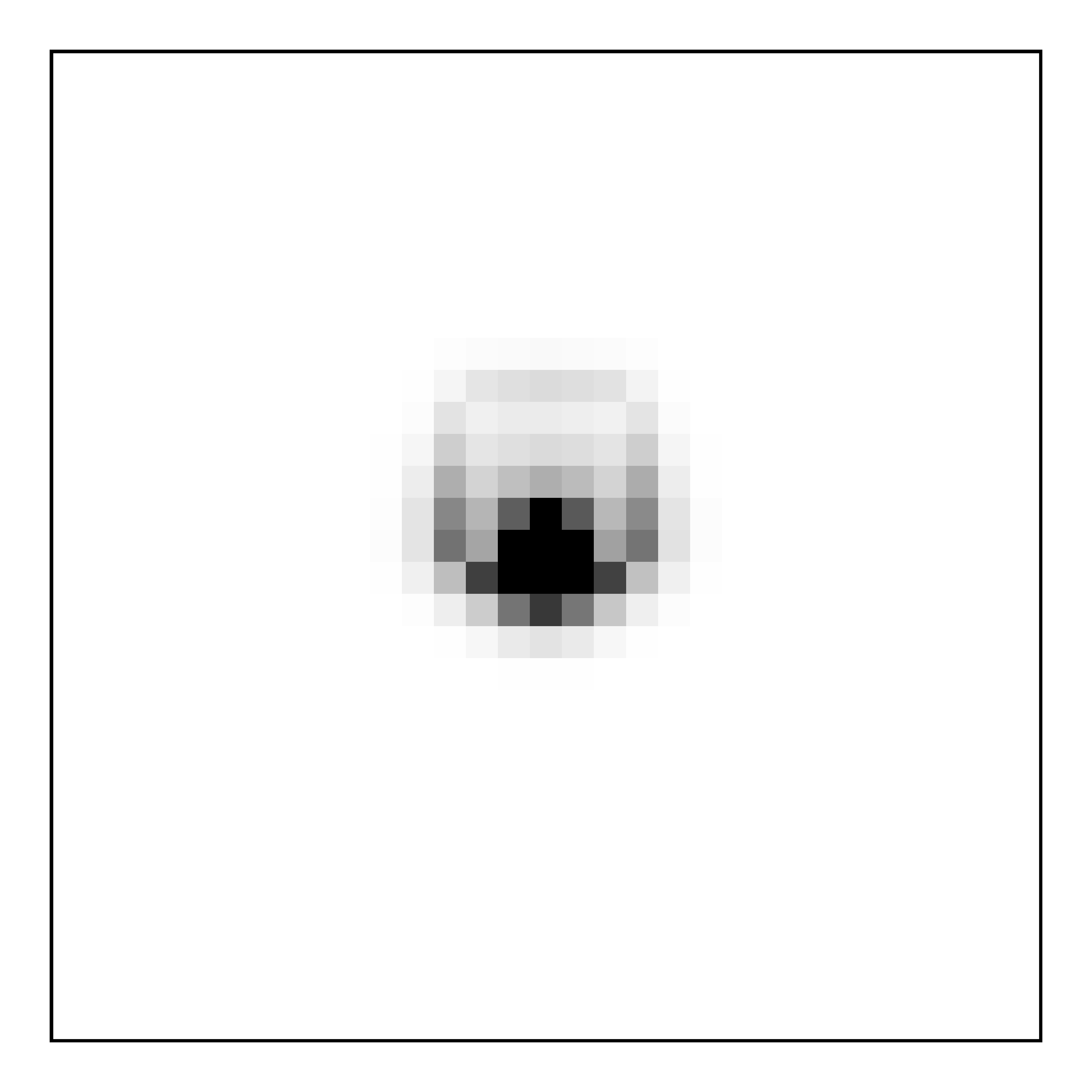}%
				\includegraphics[width=0.23\linewidth,angle=90]{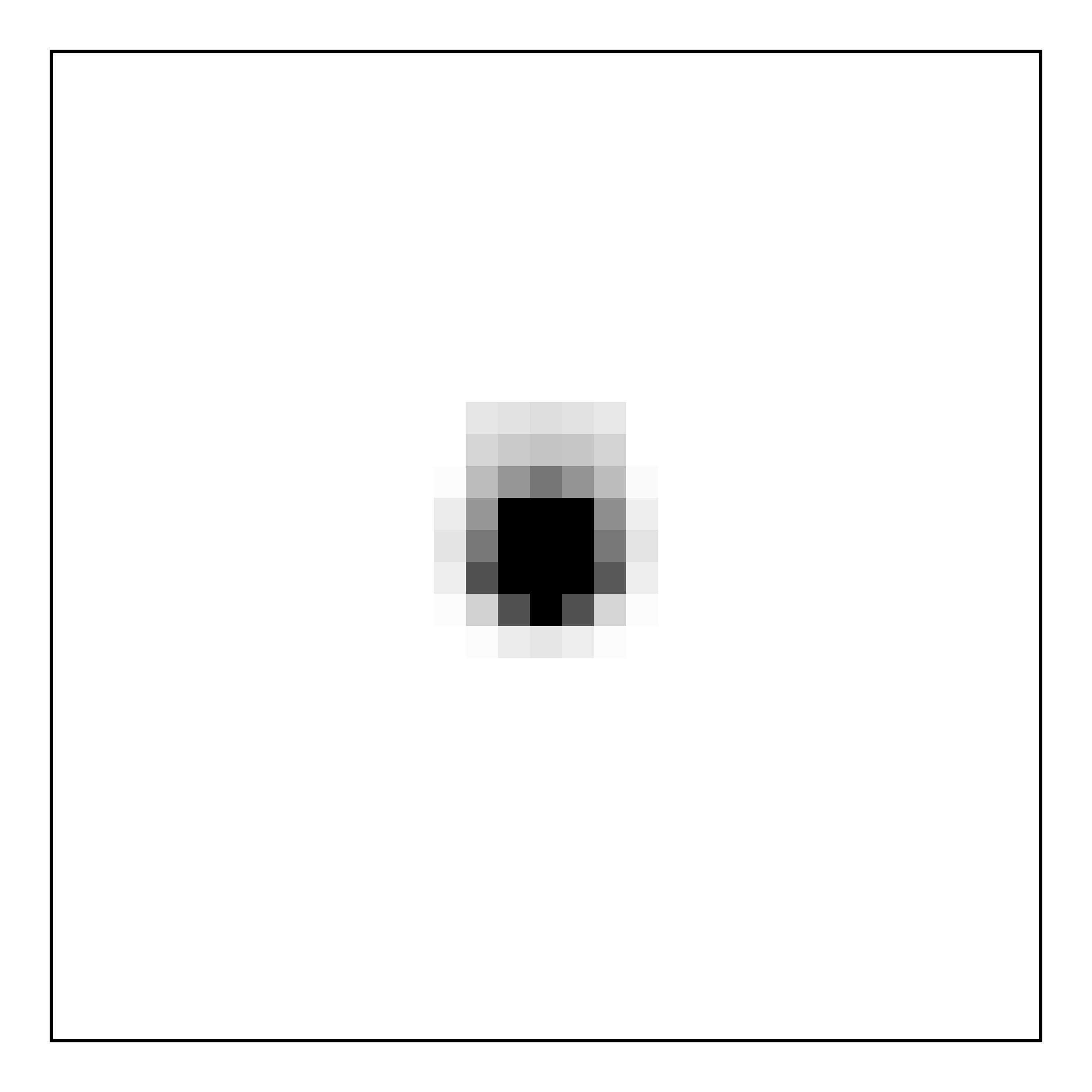}%
				\includegraphics[width=0.23\linewidth,angle=90]{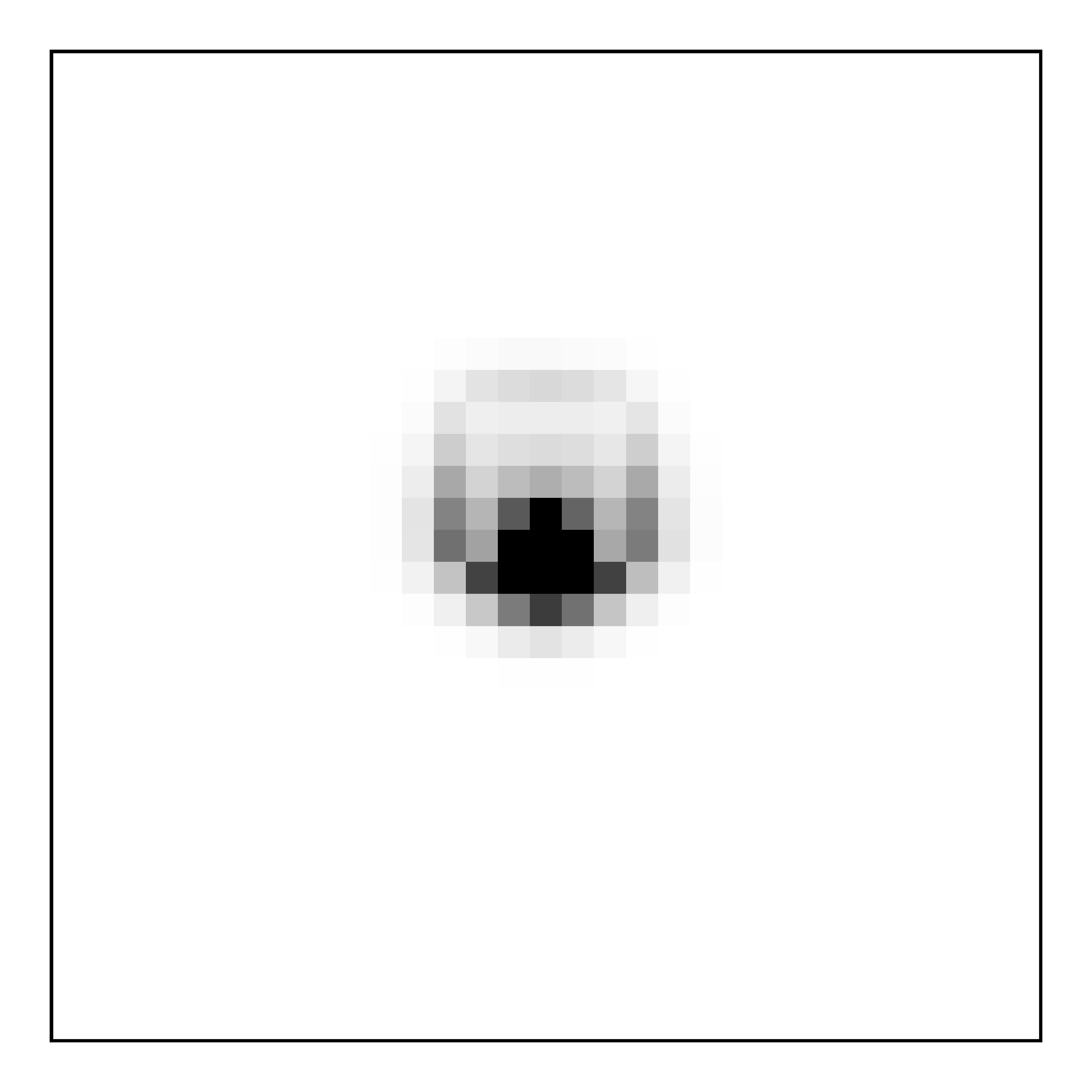}%
			\end{minipage}\\[0.5mm]
			
			\begin{minipage}{0.03\linewidth}
				\centering
				\rotatebox{90}{\tiny Phantom 4}
			\end{minipage}
			\begin{minipage}{0.95\linewidth}
				\centering
				\includegraphics[width=0.23\linewidth,angle=90]{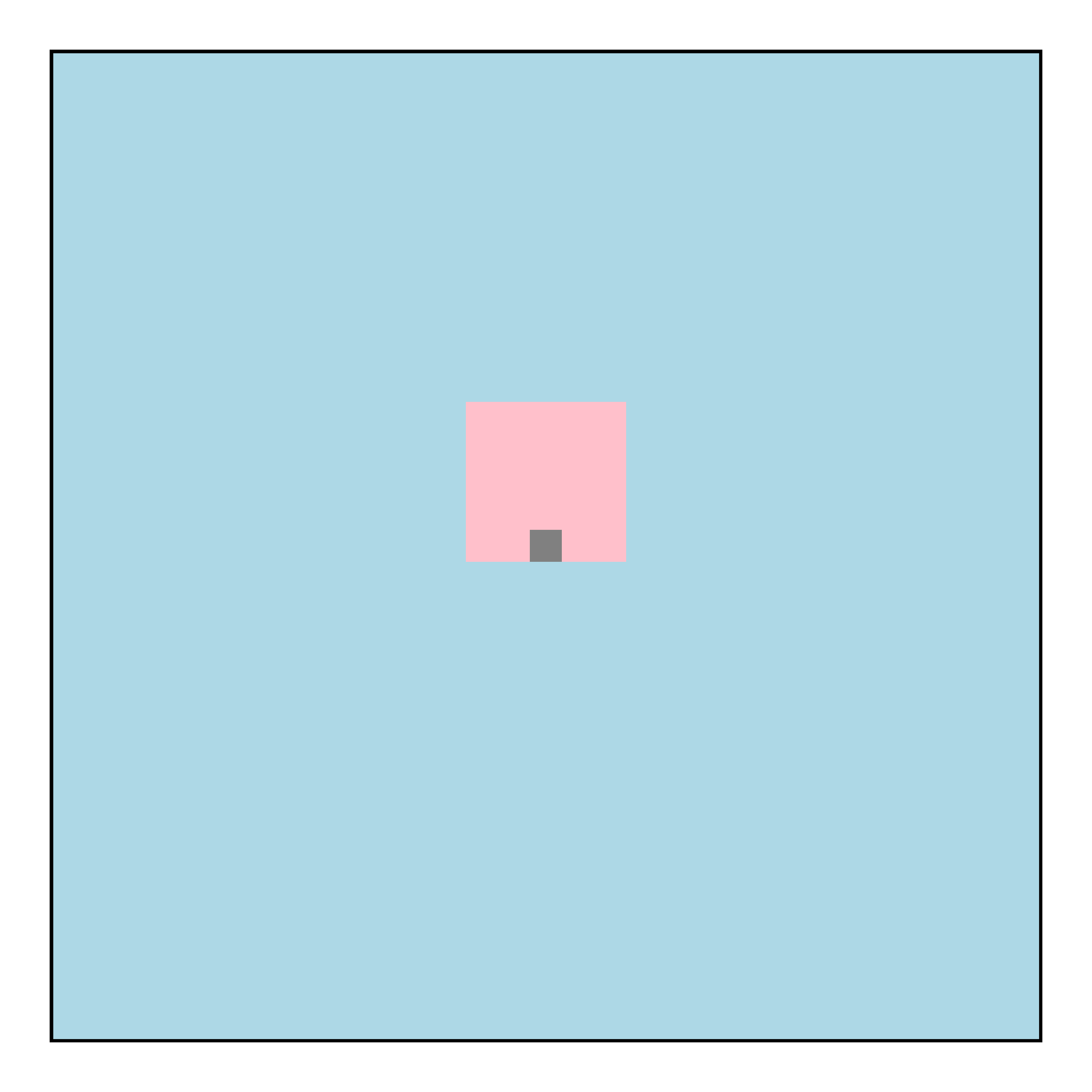}%
				\includegraphics[width=0.23\linewidth,angle=90]{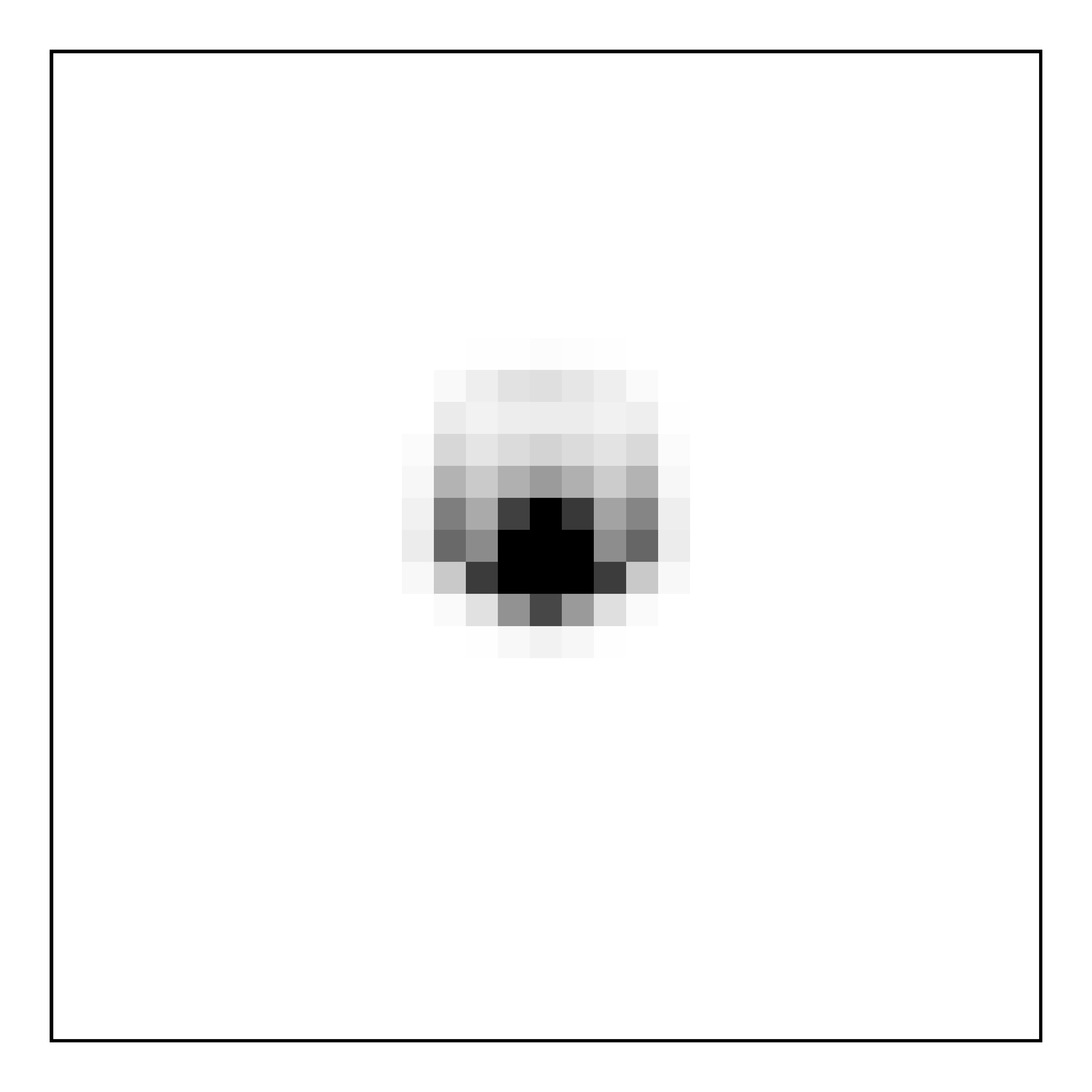}%
				\includegraphics[width=0.23\linewidth,angle=90]{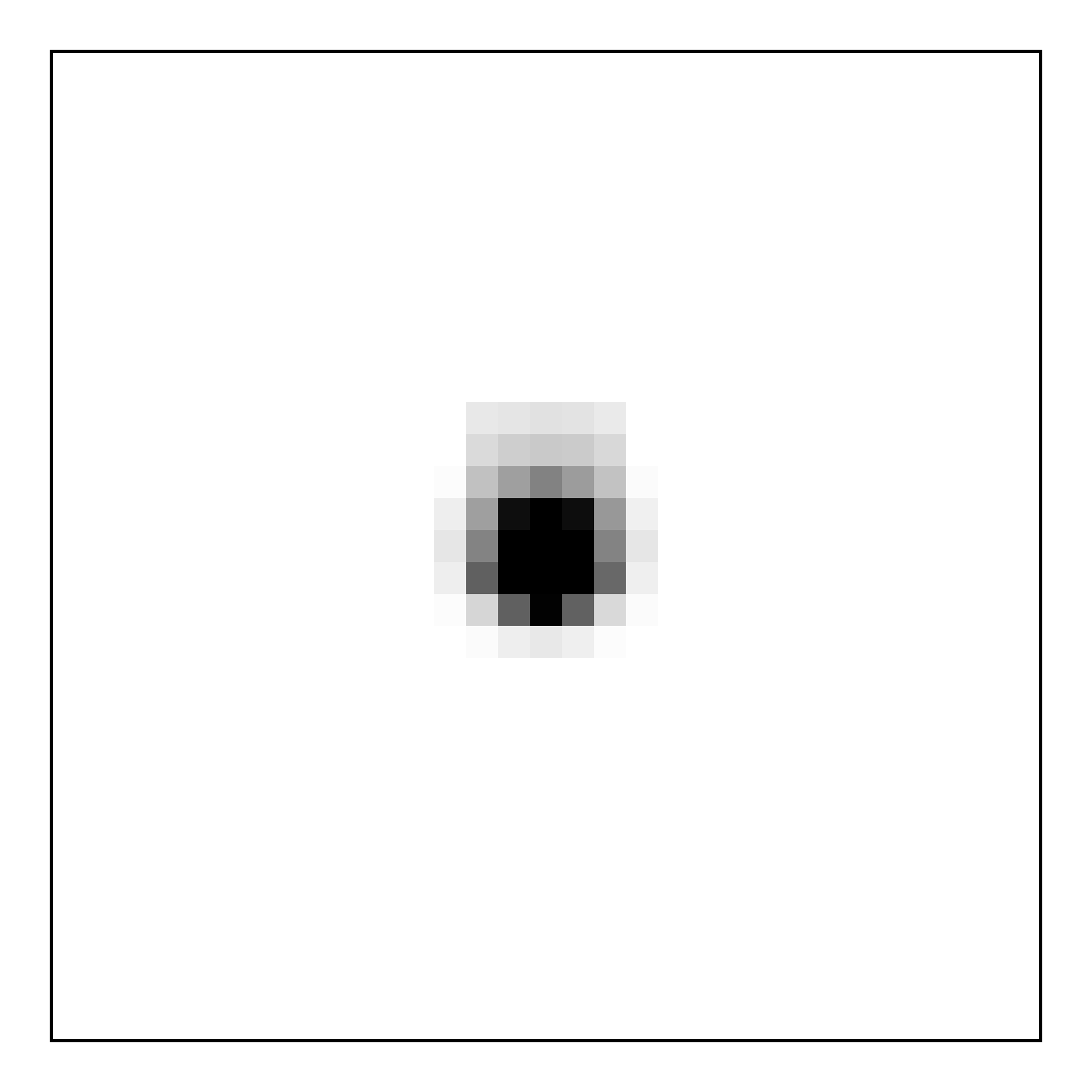}%
				\includegraphics[width=0.23\linewidth,angle=90]{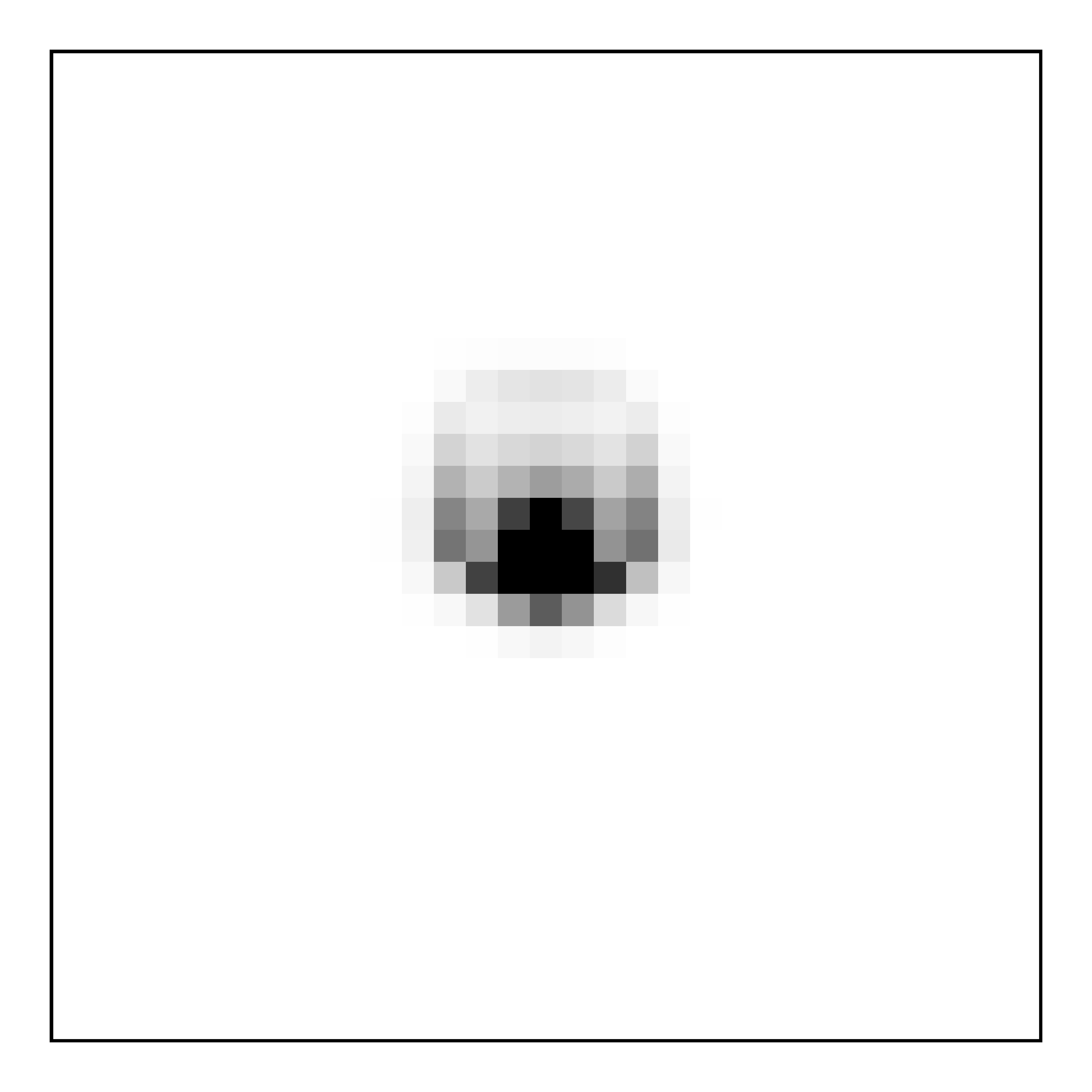}%
			\end{minipage}\\[0.5mm]
			
			\begin{minipage}{0.03\linewidth}
				\centering
				\rotatebox{90}{\tiny Phantom 5}
			\end{minipage}
			\begin{minipage}{0.95\linewidth}
				\centering
				\includegraphics[width=0.23\linewidth,angle=90]{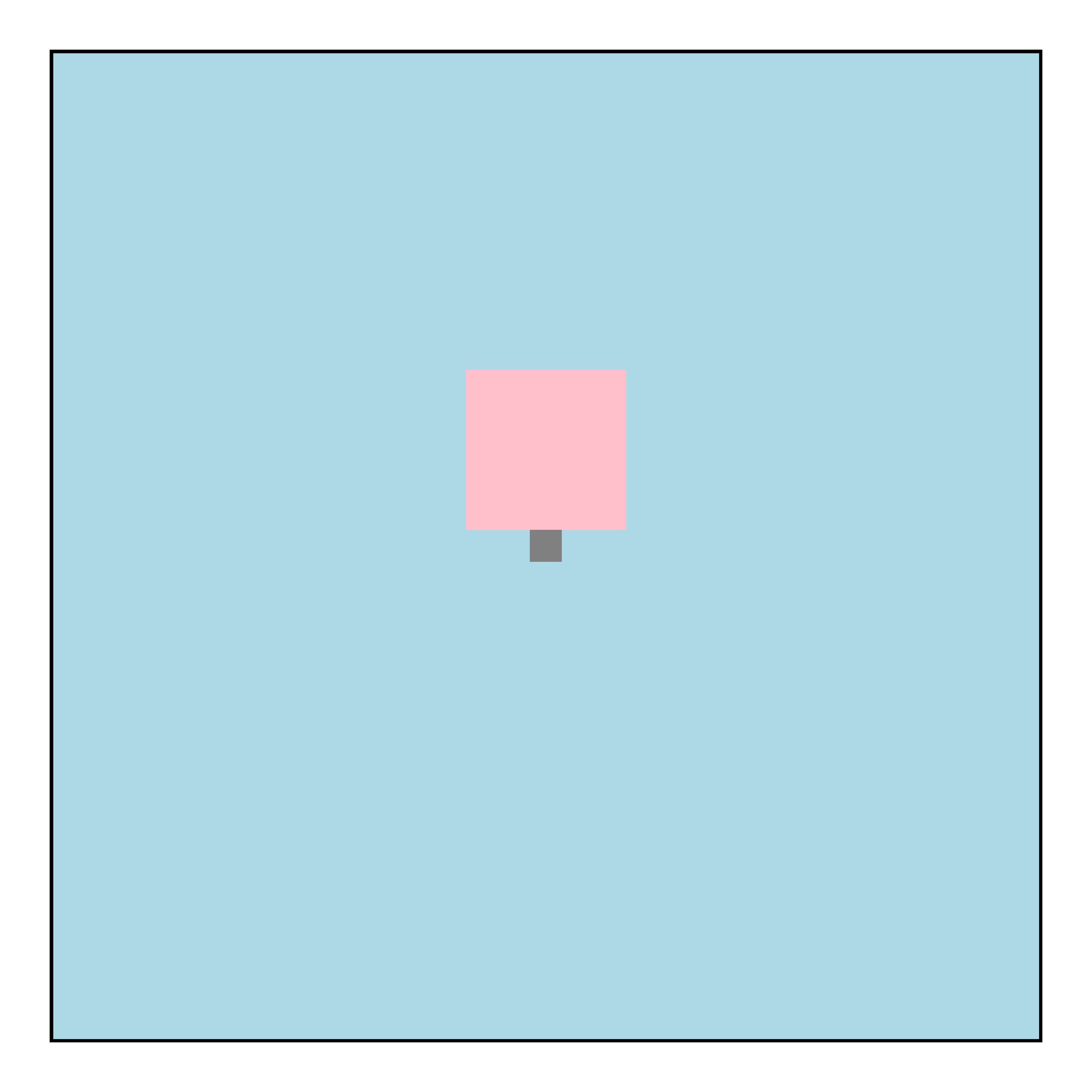}%
				\includegraphics[width=0.23\linewidth,angle=90]{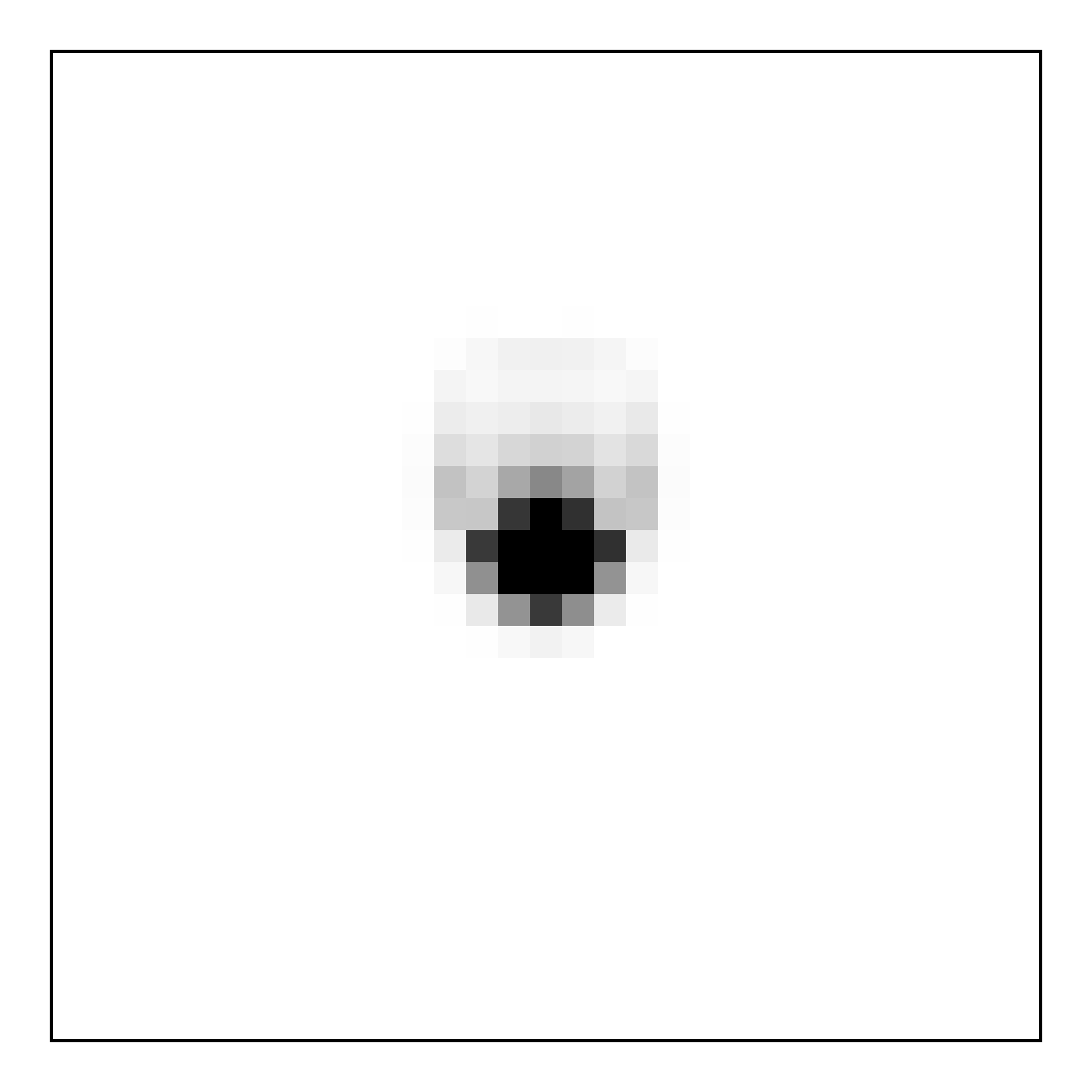}%
				\includegraphics[width=0.23\linewidth,angle=90]{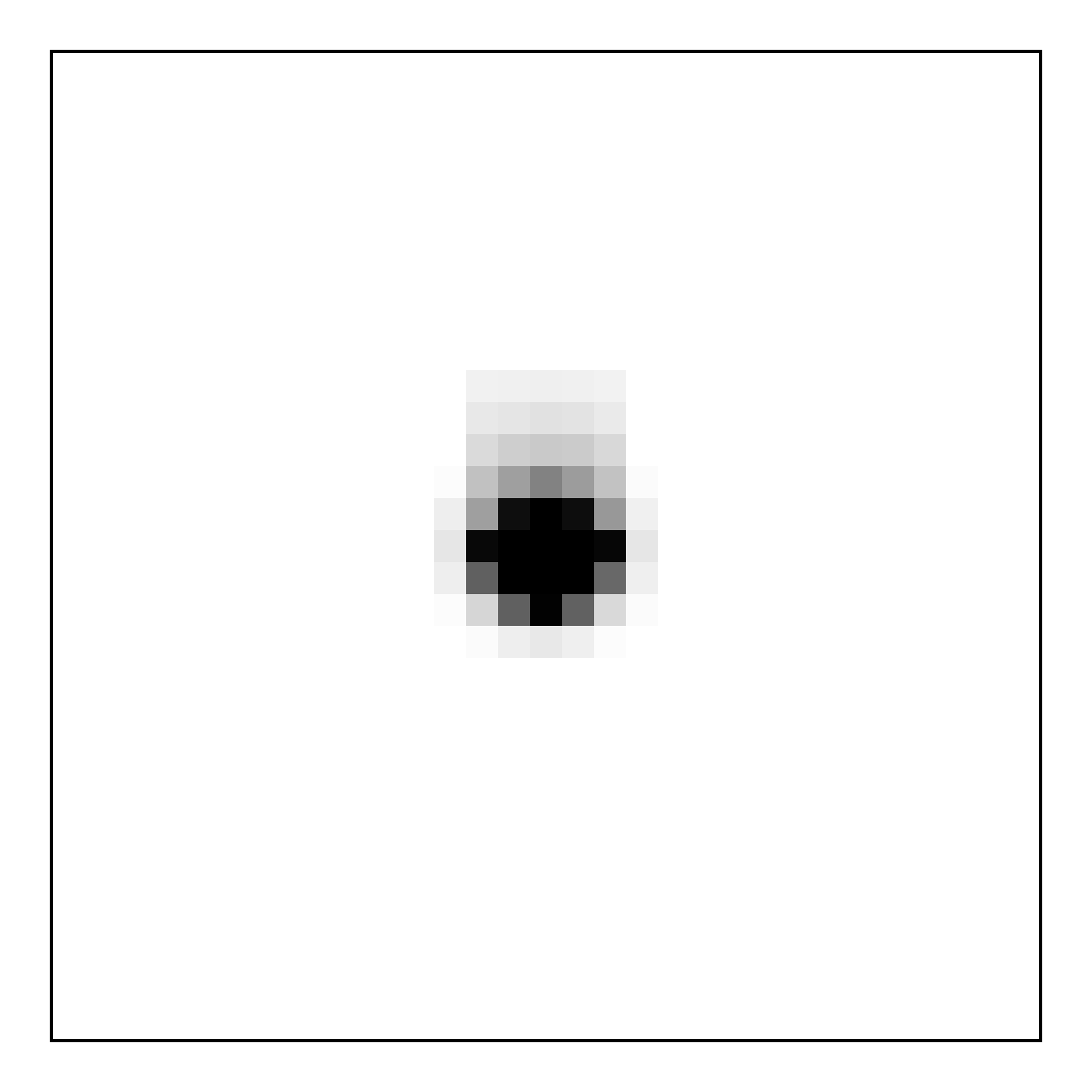}%
				\includegraphics[width=0.23\linewidth,angle=90]{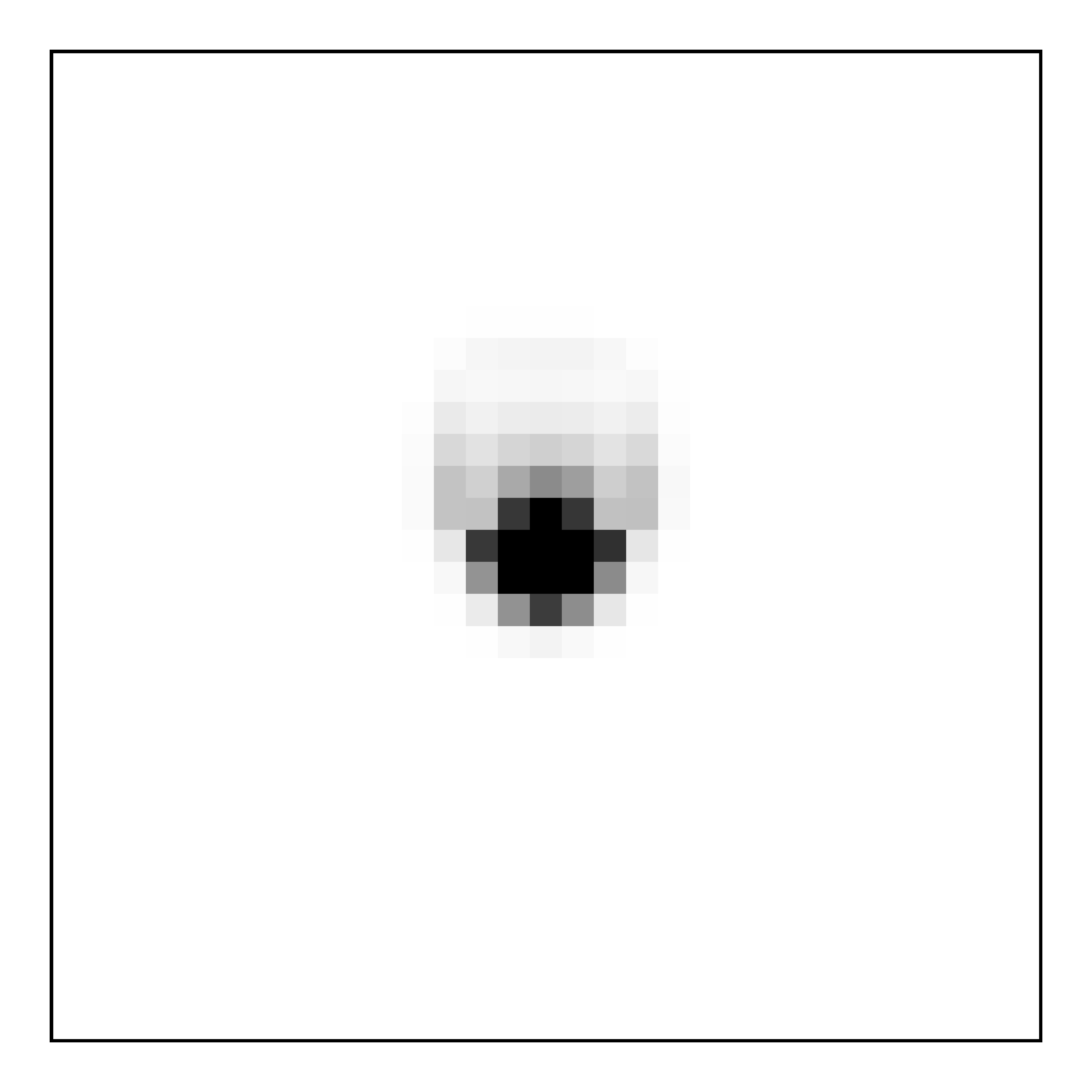}%
			\end{minipage}
		\end{minipage}%
	}
	\hfill
	\mynewroundedbox{
		\begin{minipage}[t]{0.3\textwidth}
			\centering
			\textbf{Axis 2 (coronal)}\\[1mm]
			
			\begin{minipage}{0.03\linewidth}
				\centering
				~
			\end{minipage}
			\begin{minipage}{0.95\linewidth}
				\centering
				\begin{minipage}{0.23\linewidth}\centering\tiny Phantom\end{minipage}%
				\begin{minipage}{0.23\linewidth}\centering\tiny MC (reference)\end{minipage}%
				\begin{minipage}{0.23\linewidth}\centering\tiny SVTD\end{minipage}%
				\begin{minipage}{0.23\linewidth}\centering\tiny DDConv\end{minipage}%
			\end{minipage}\\[1mm]
			
			\begin{minipage}{0.03\linewidth}
				\centering
				\rotatebox{90}{\tiny Phantom 1}
			\end{minipage}
			\begin{minipage}{0.95\linewidth}
				\centering
				\includegraphics[width=0.23\linewidth,angle=90]{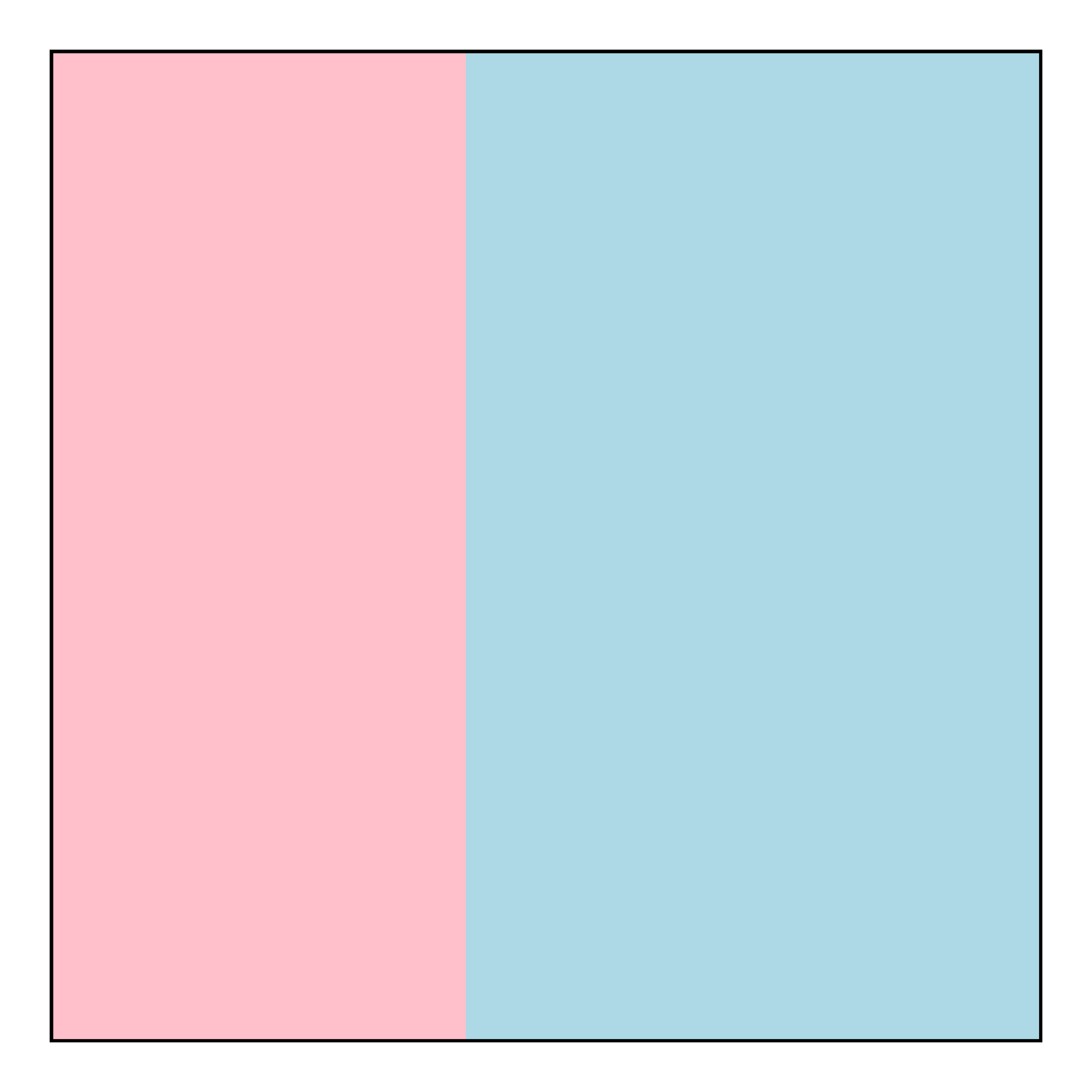}%
				\includegraphics[width=0.23\linewidth,angle=90]{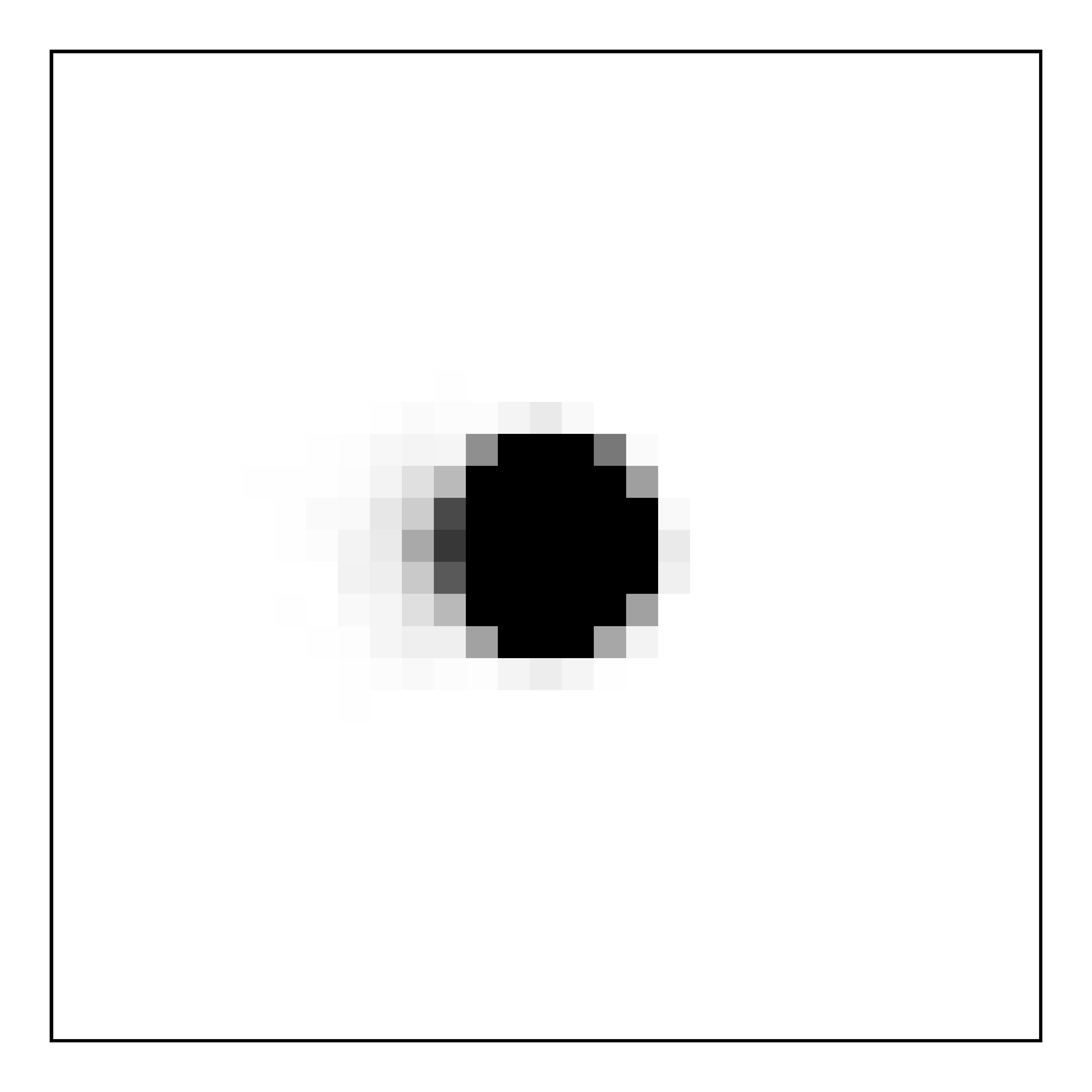}%
				\includegraphics[width=0.23\linewidth,angle=90]{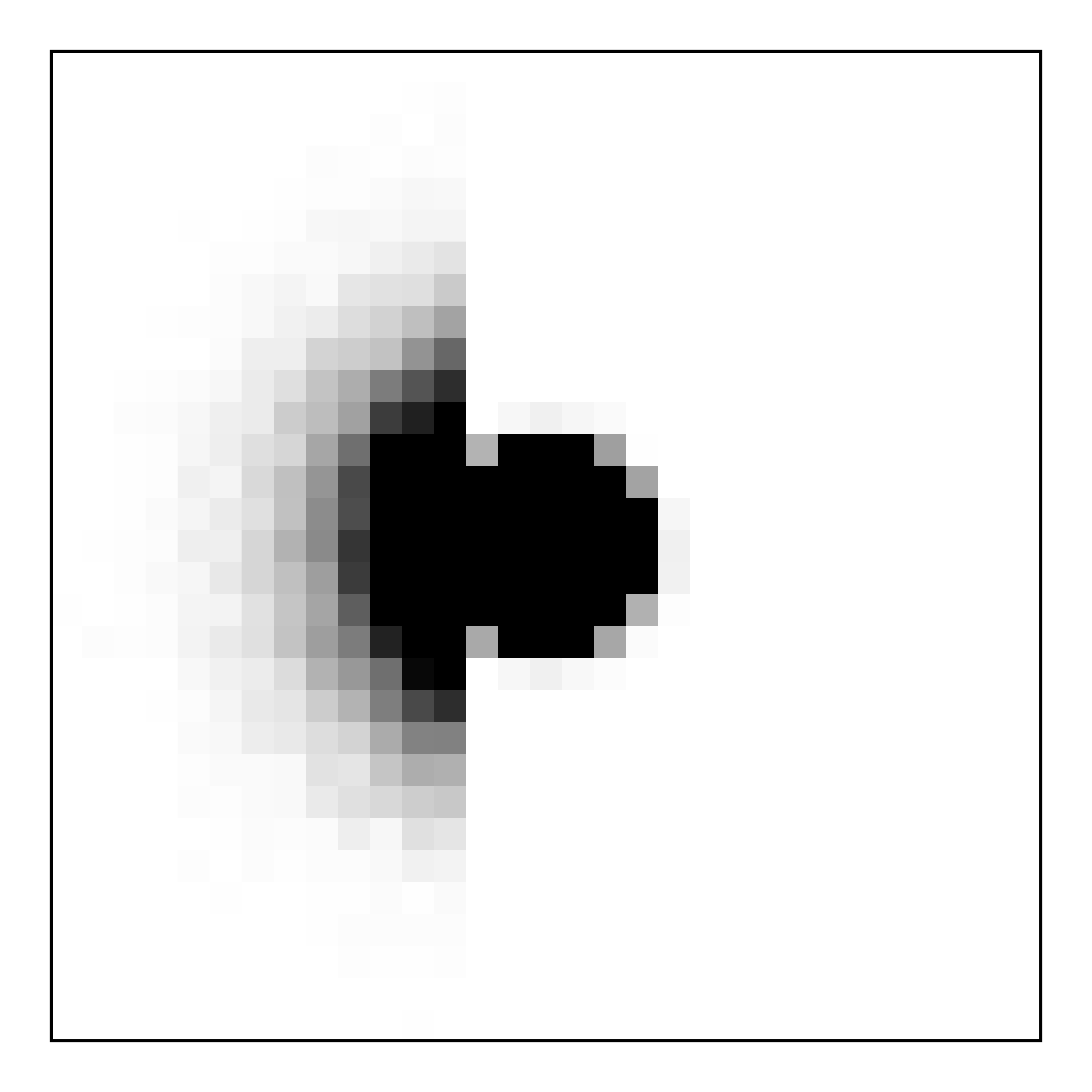}%
				\includegraphics[width=0.23\linewidth,angle=90]{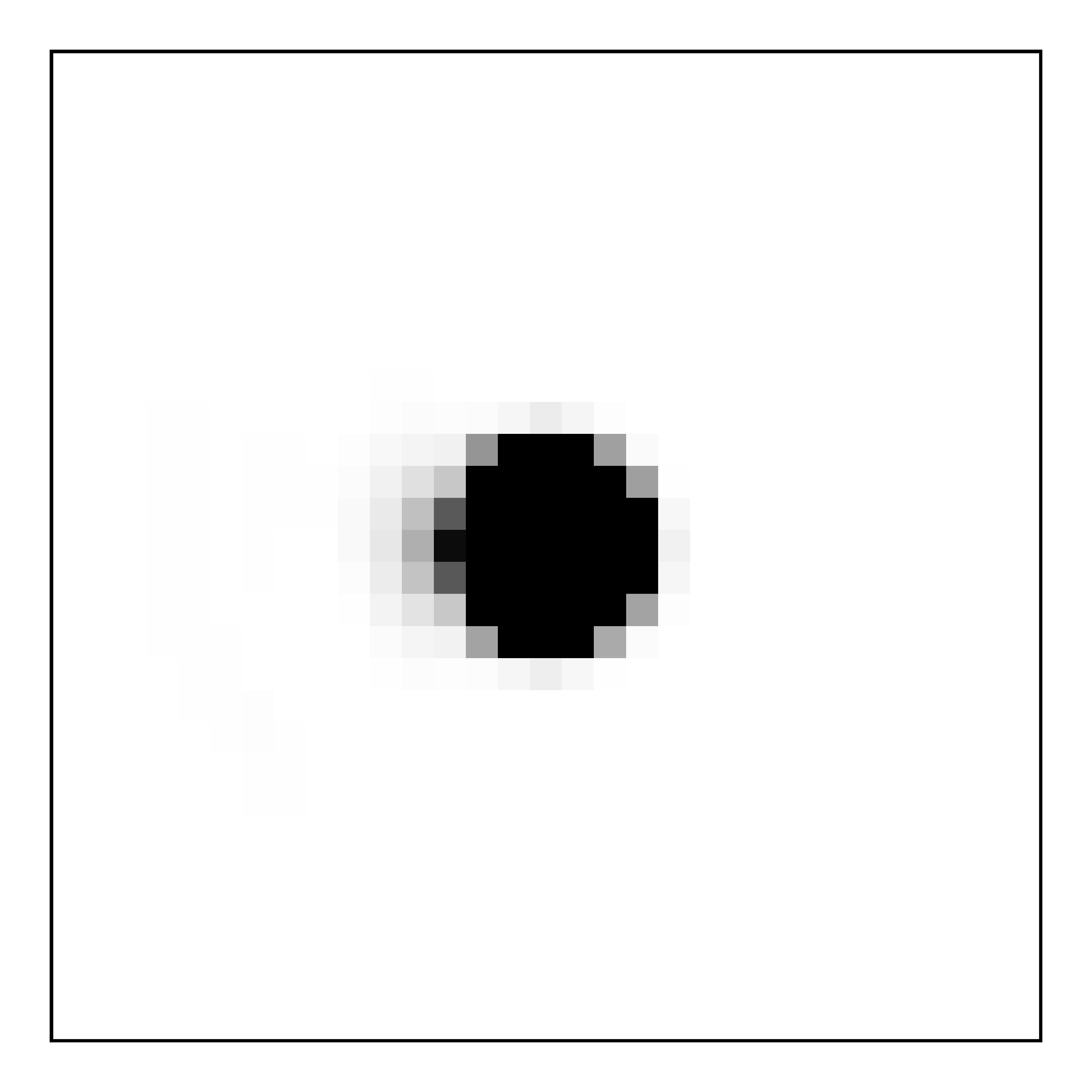}%
			\end{minipage}\\[0.5mm]
			
			\begin{minipage}{0.03\linewidth}
				\centering
				\rotatebox{90}{\tiny Phantom 2}
			\end{minipage}
			\begin{minipage}{0.95\linewidth}
				\centering
				\includegraphics[width=0.23\linewidth,angle=90]{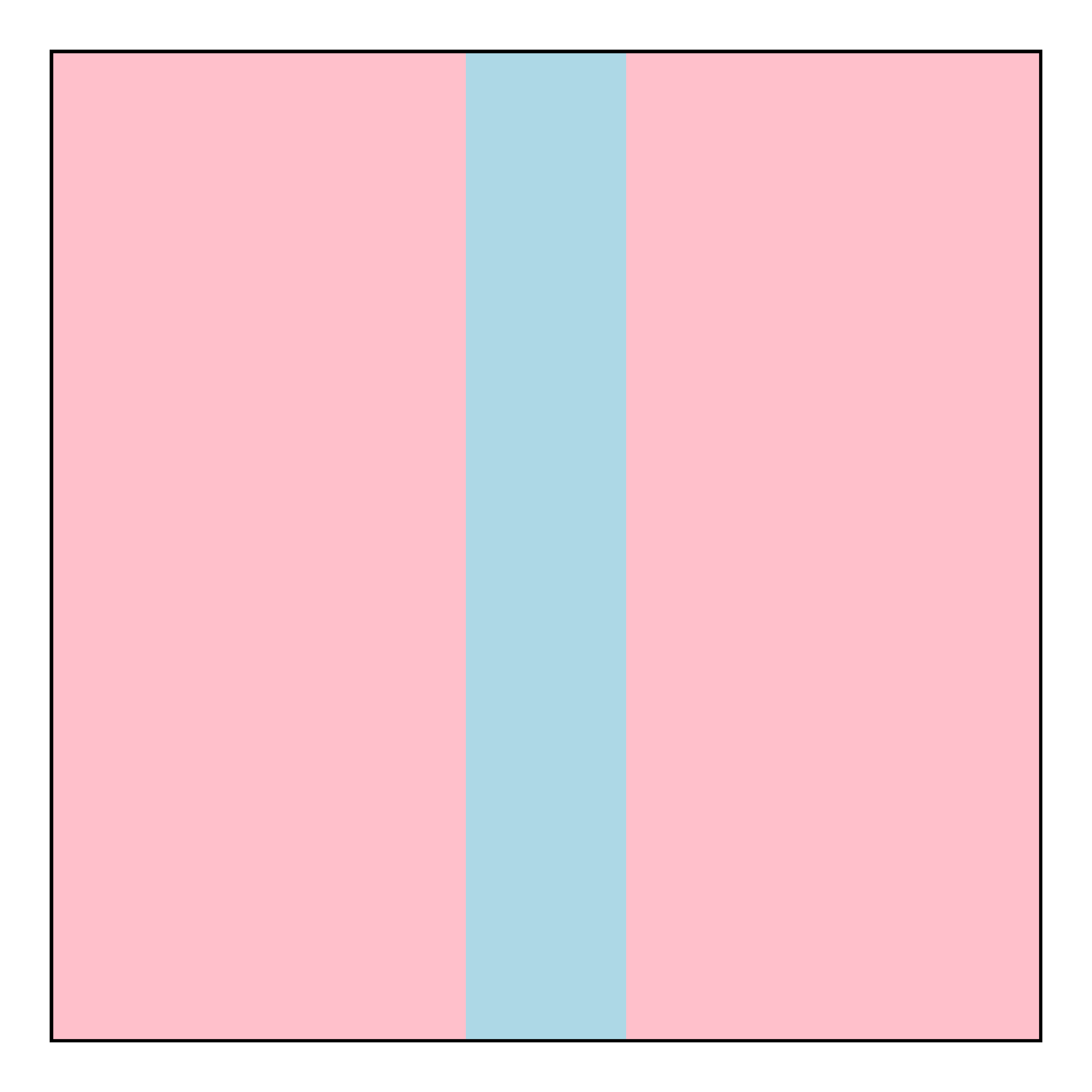}%
				\includegraphics[width=0.23\linewidth,angle=90]{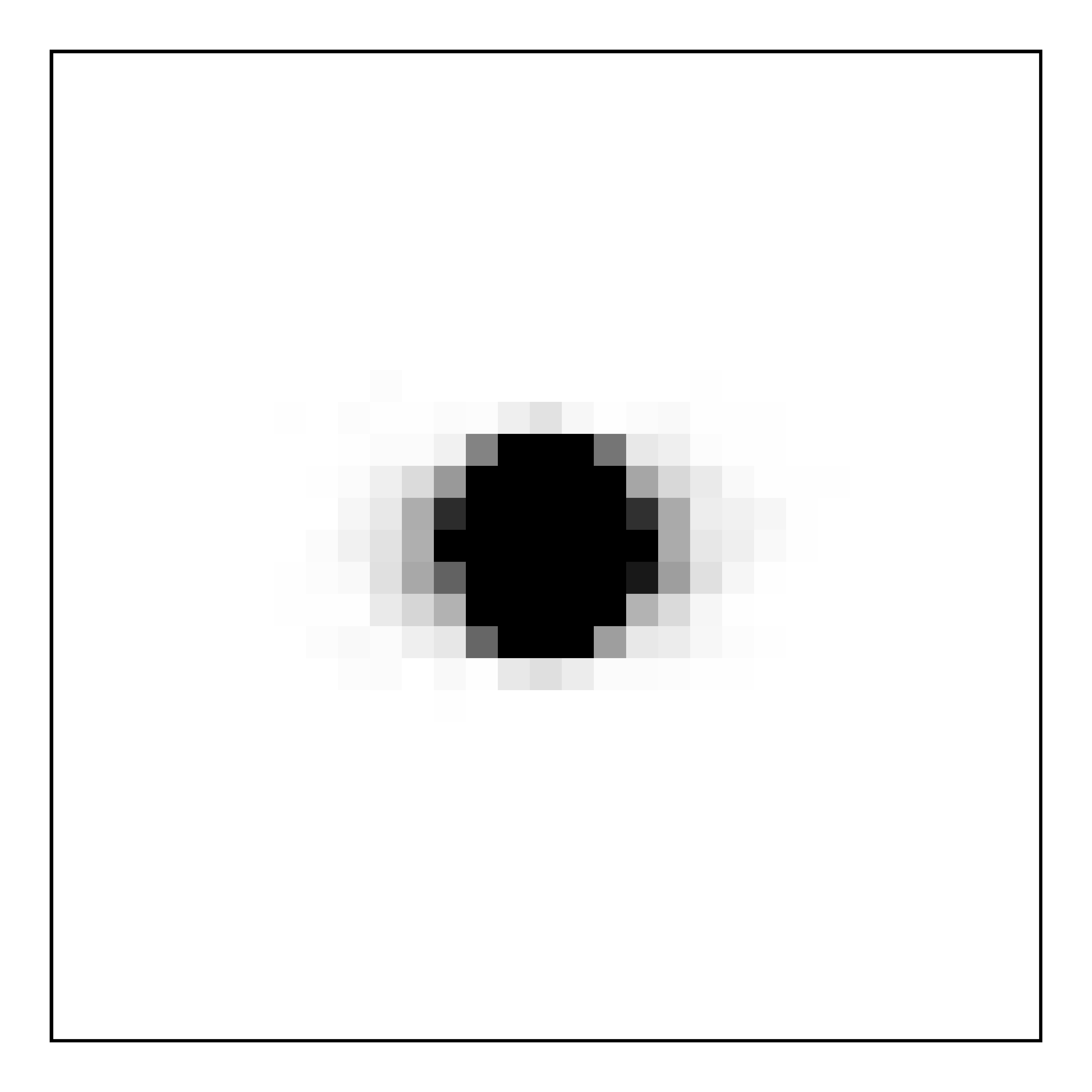}%
				\includegraphics[width=0.23\linewidth,angle=90]{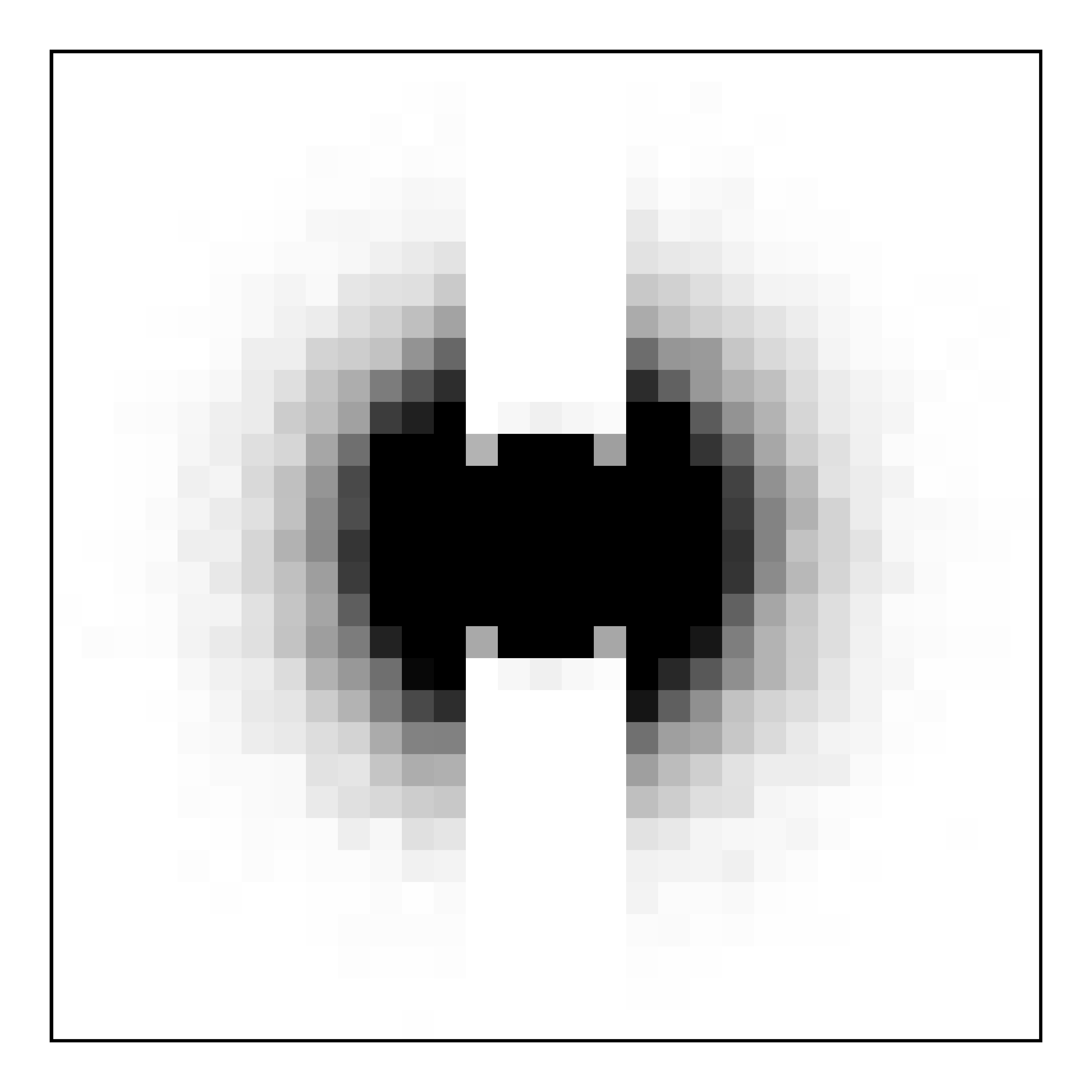}%
				\includegraphics[width=0.23\linewidth,angle=90]{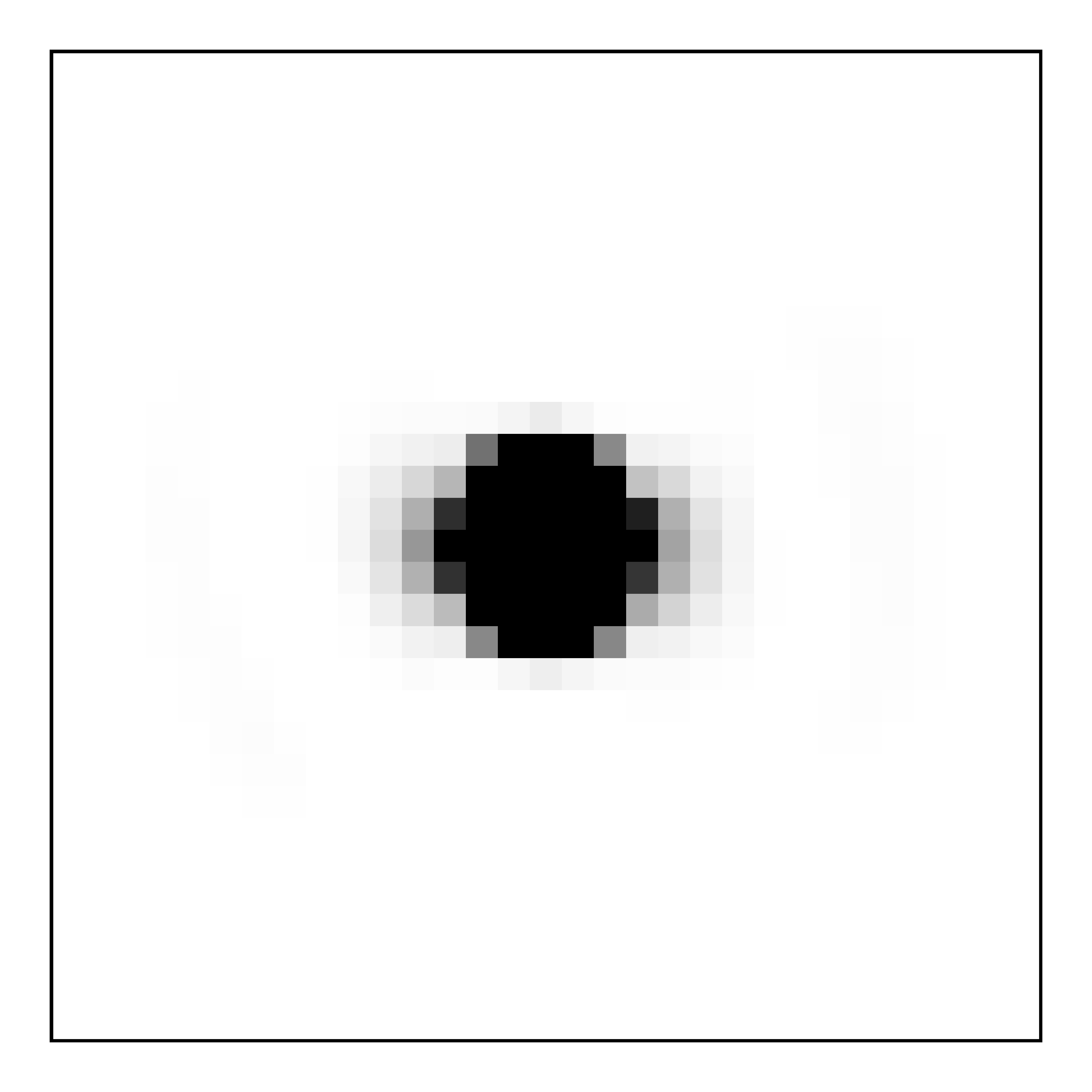}%
			\end{minipage}\\[0.5mm]
			
			\begin{minipage}{0.03\linewidth}
				\centering
				\rotatebox{90}{\tiny Phantom 3}
			\end{minipage}
			\begin{minipage}{0.95\linewidth}
				\centering
				\includegraphics[width=0.23\linewidth,angle=90]{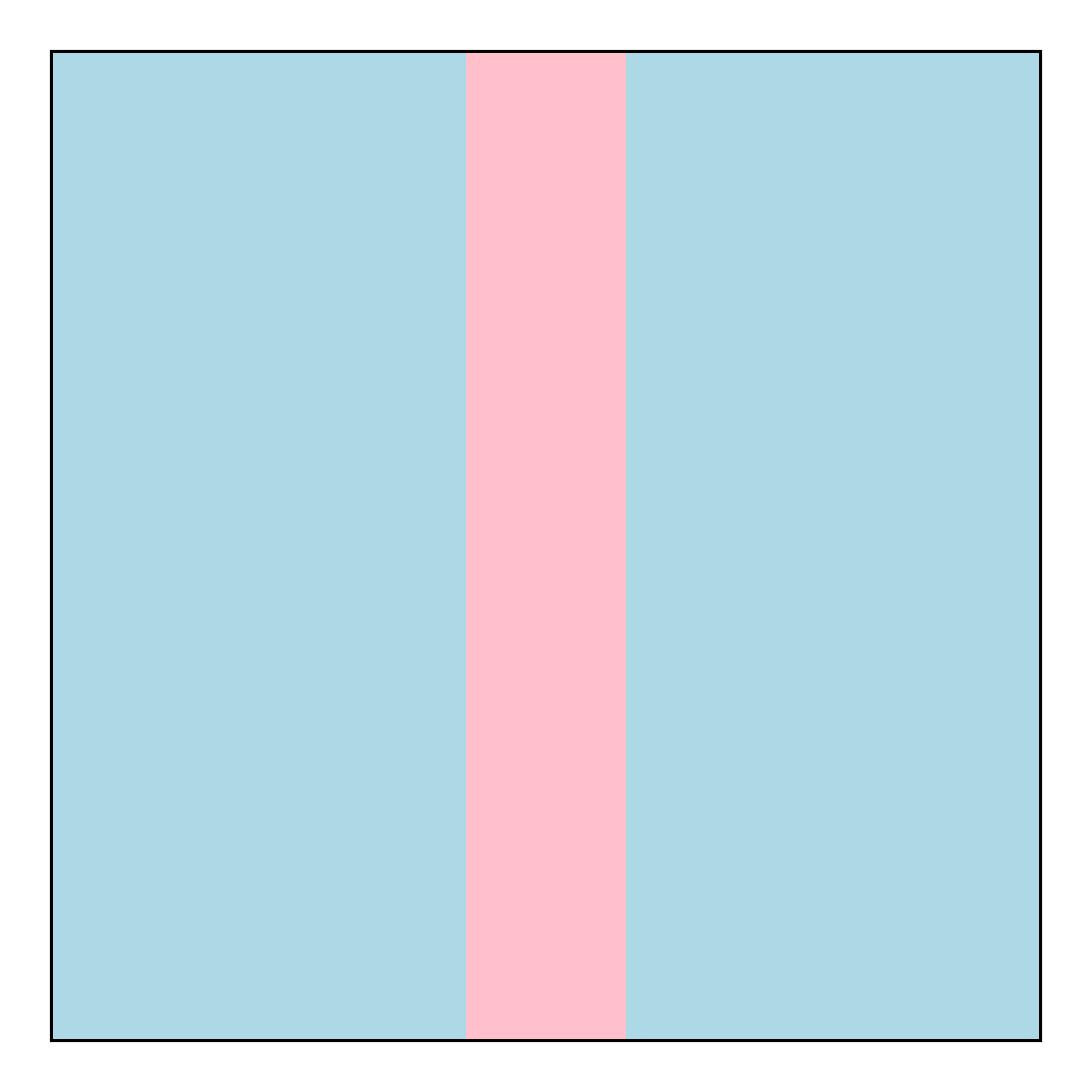}%
				\includegraphics[width=0.23\linewidth,angle=90]{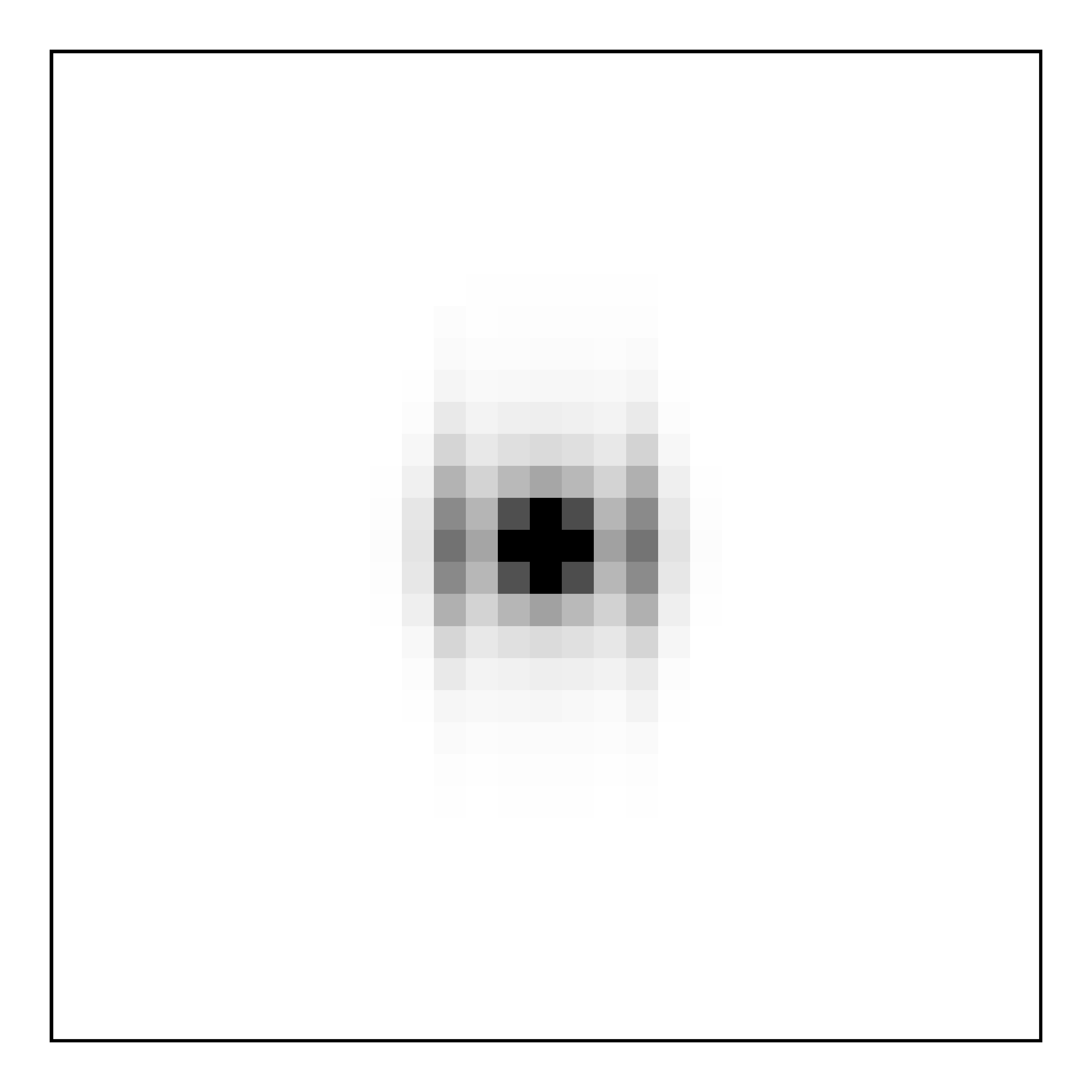}%
				\includegraphics[width=0.23\linewidth,angle=90]{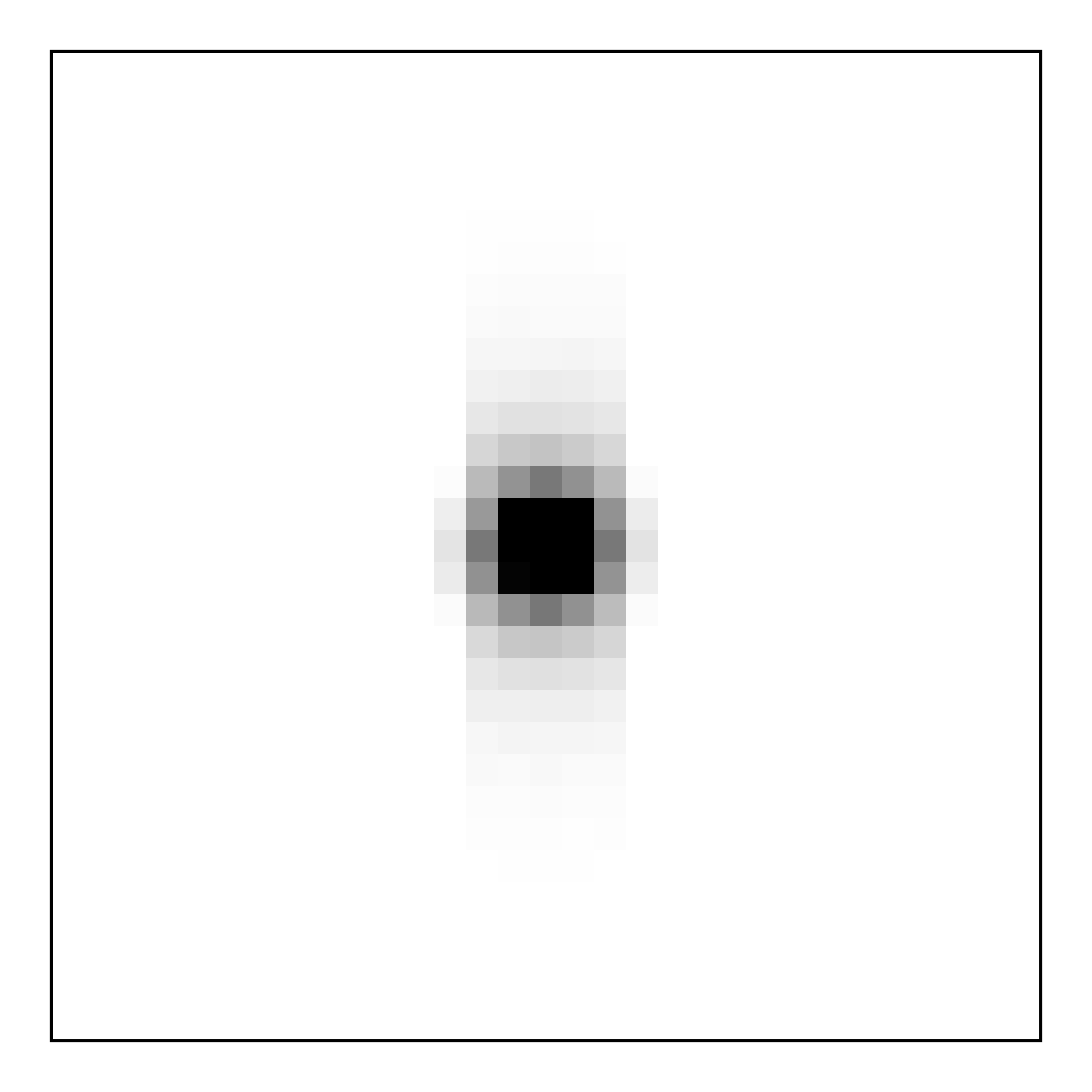}%
				\includegraphics[width=0.23\linewidth,angle=90]{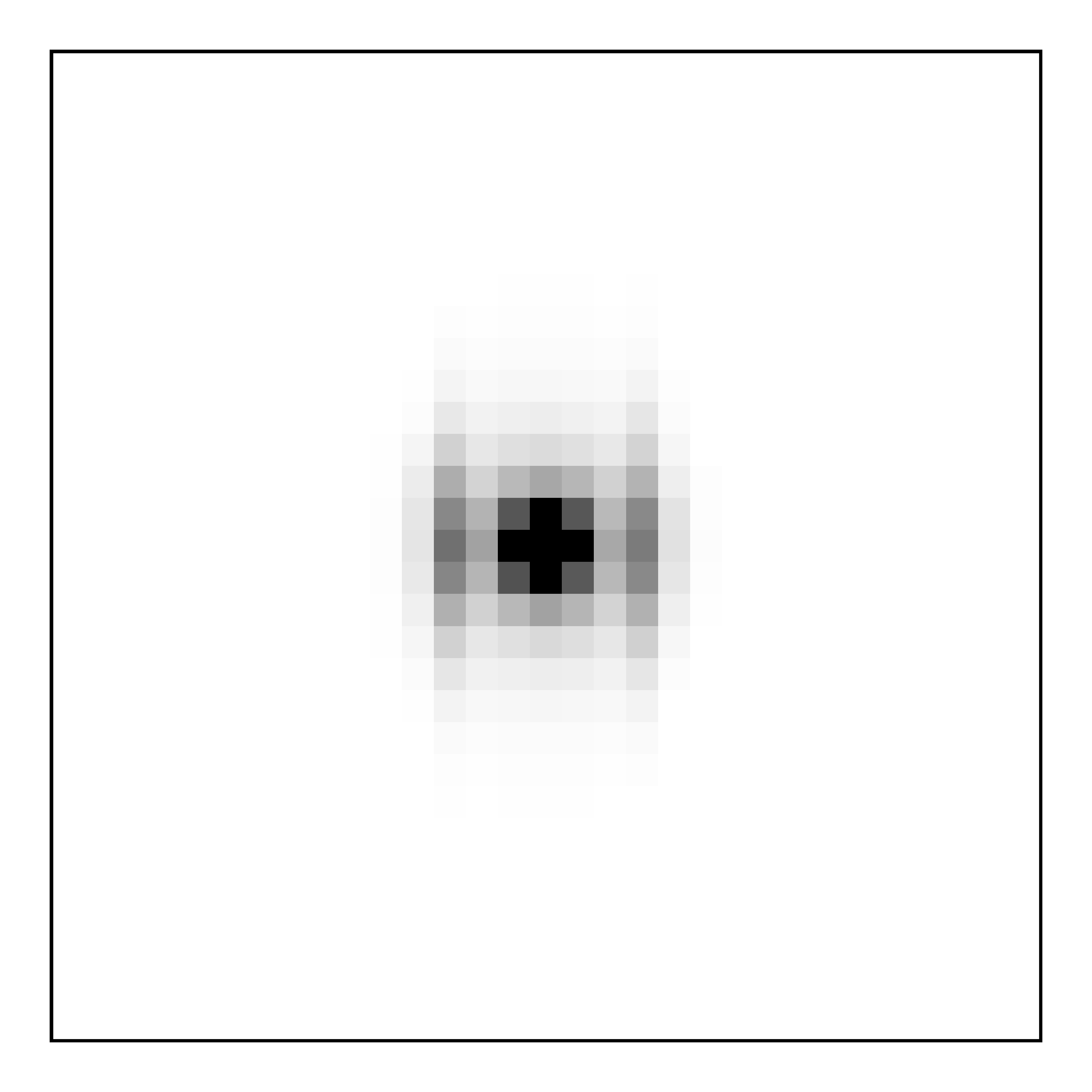}%
			\end{minipage}\\[0.5mm]
			
			\begin{minipage}{0.03\linewidth}
				\centering
				\rotatebox{90}{\tiny Phantom 4}
			\end{minipage}
			\begin{minipage}{0.95\linewidth}
				\centering
				\includegraphics[width=0.23\linewidth,angle=90]{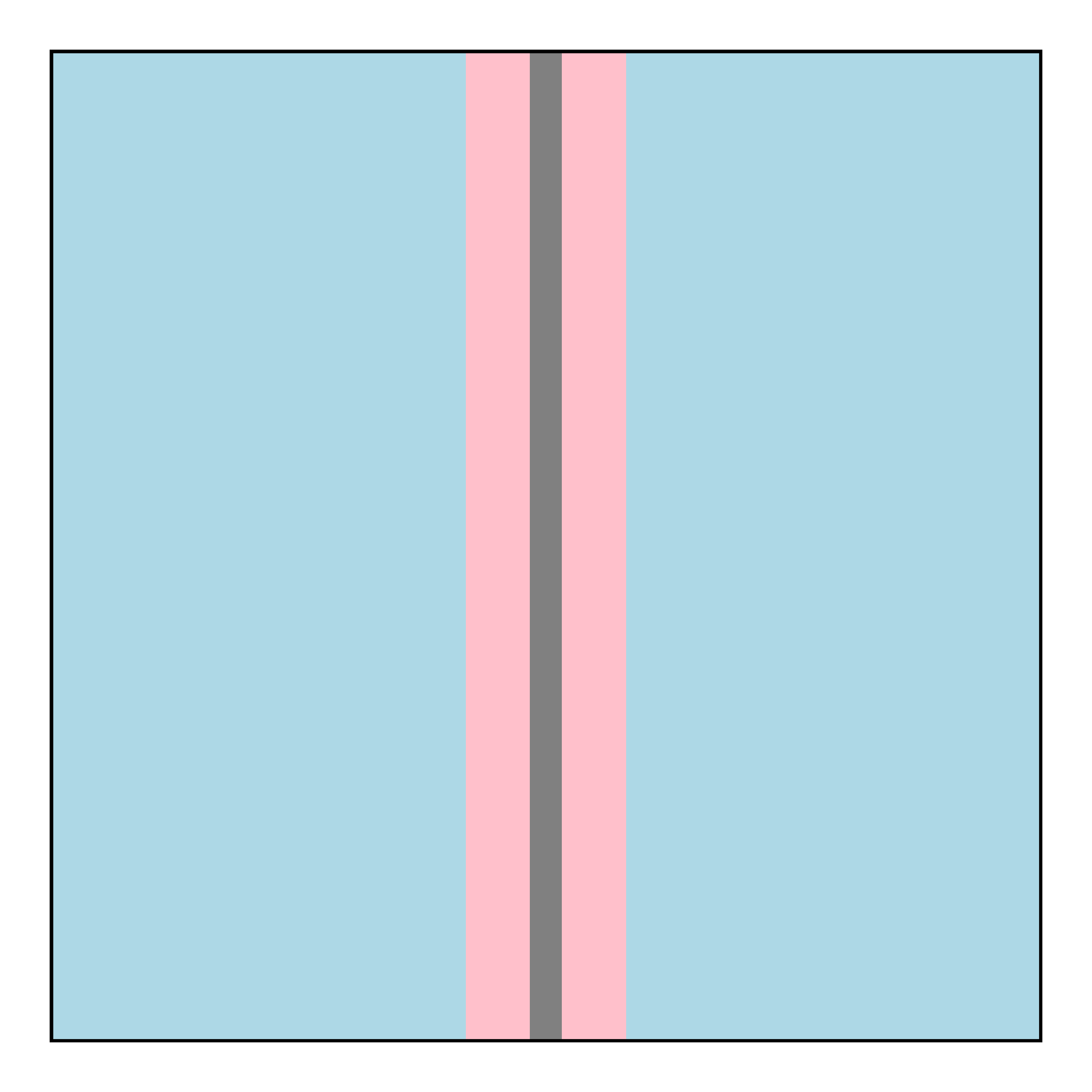}%
				\includegraphics[width=0.23\linewidth,angle=90]{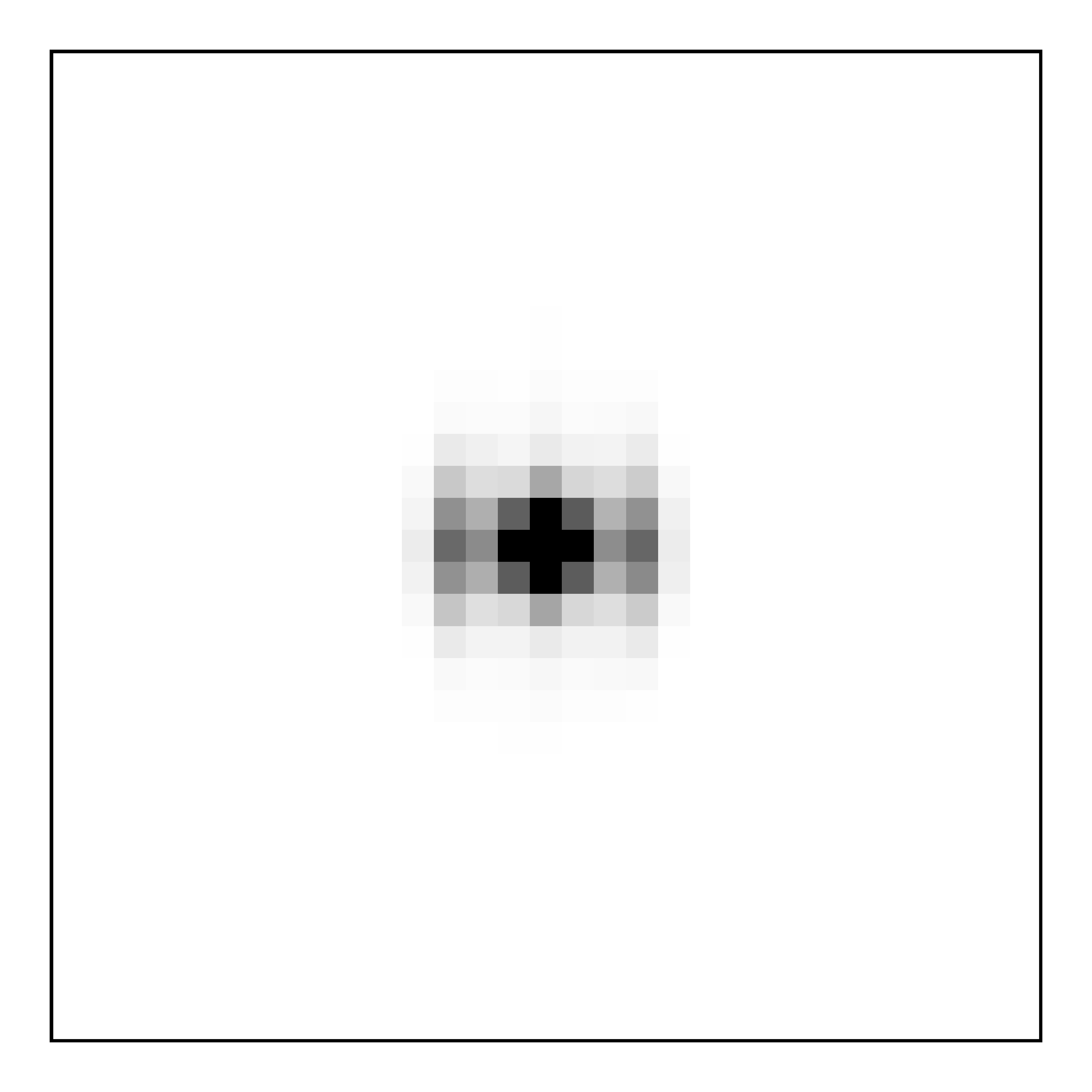}%
				\includegraphics[width=0.23\linewidth,angle=90]{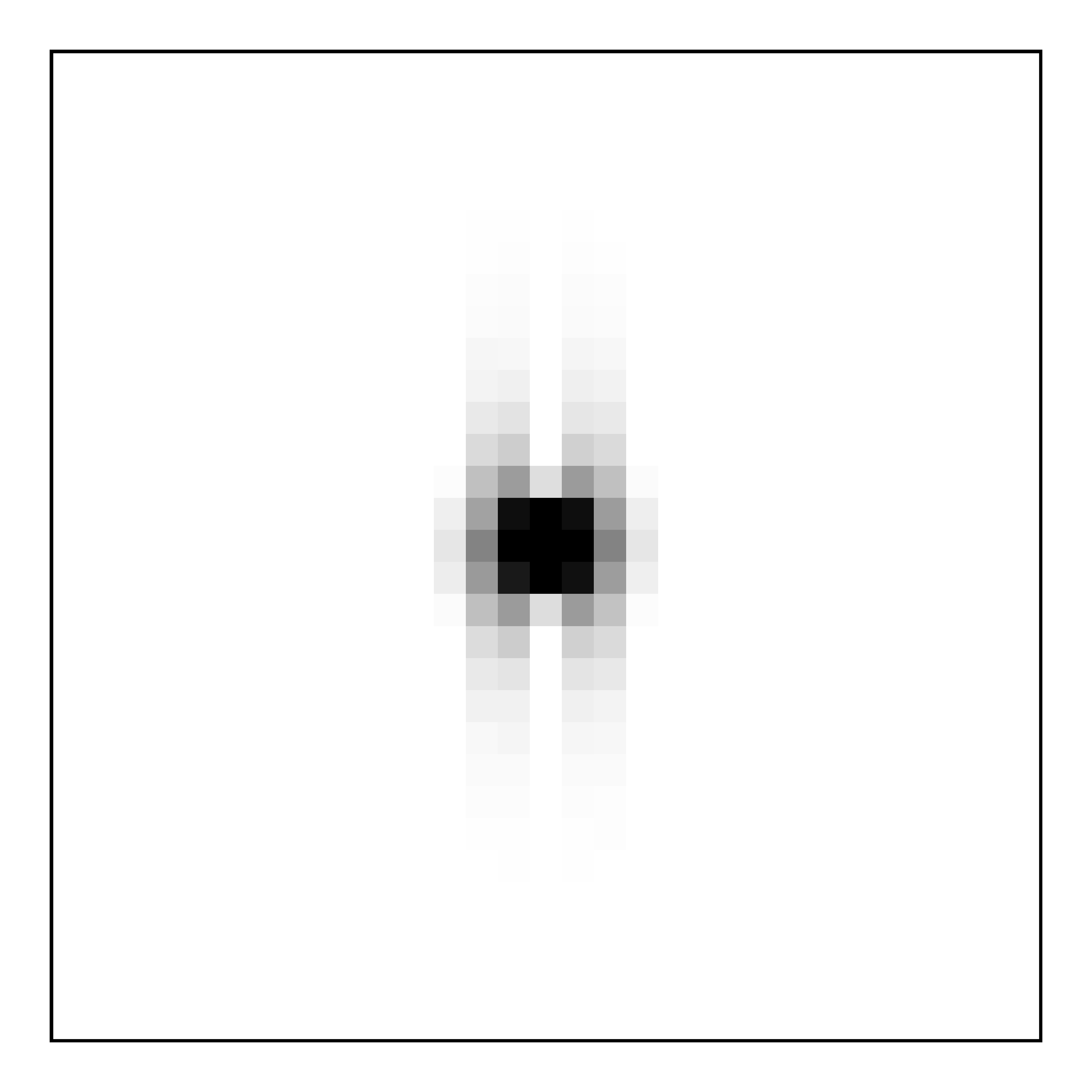}%
				\includegraphics[width=0.23\linewidth,angle=90]{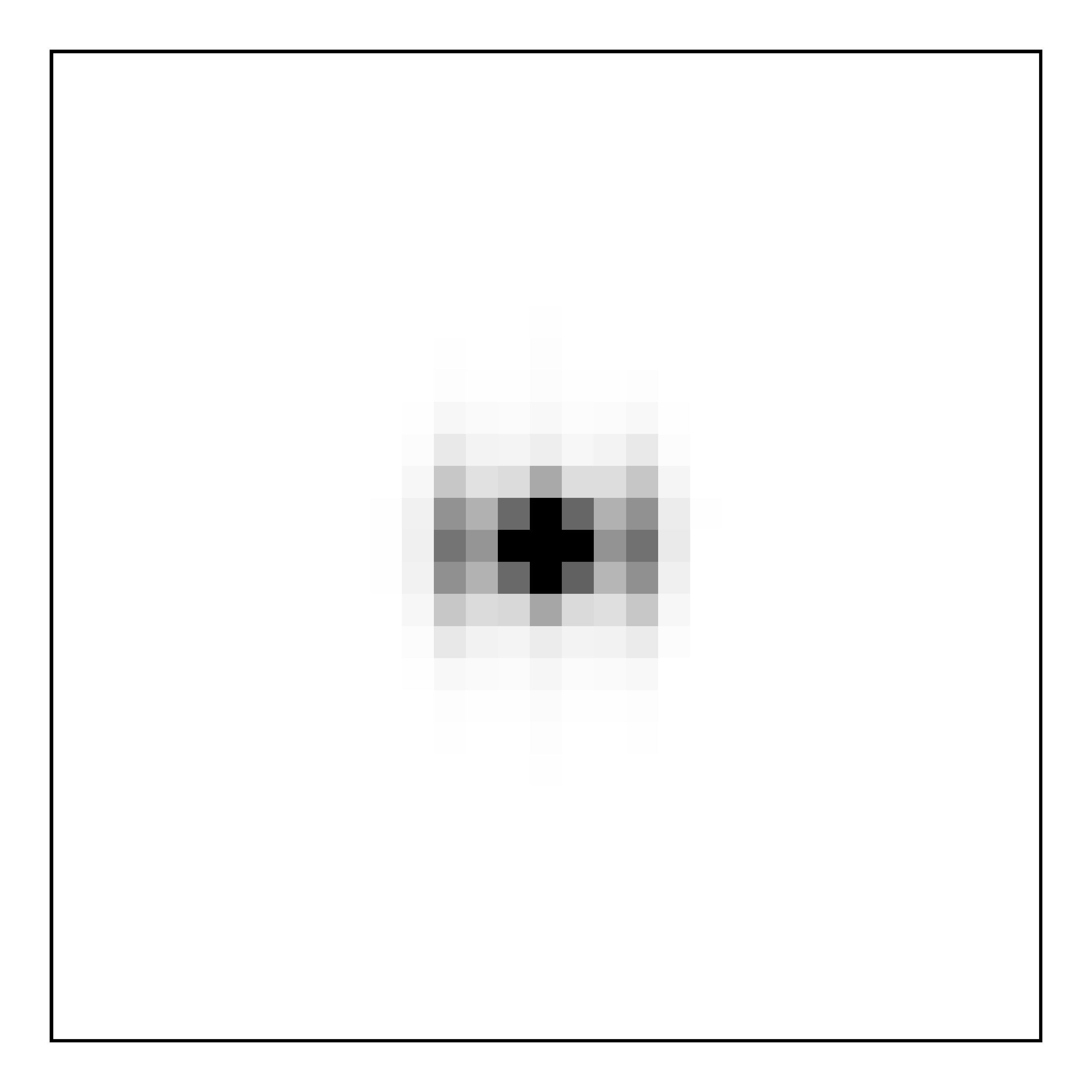}%
			\end{minipage}\\[0.5mm]
			
			\begin{minipage}{0.03\linewidth}
				\centering
				\rotatebox{90}{\tiny Phantom 5}
			\end{minipage}
			\begin{minipage}{0.95\linewidth}
				\centering
				\includegraphics[width=0.23\linewidth,angle=90]{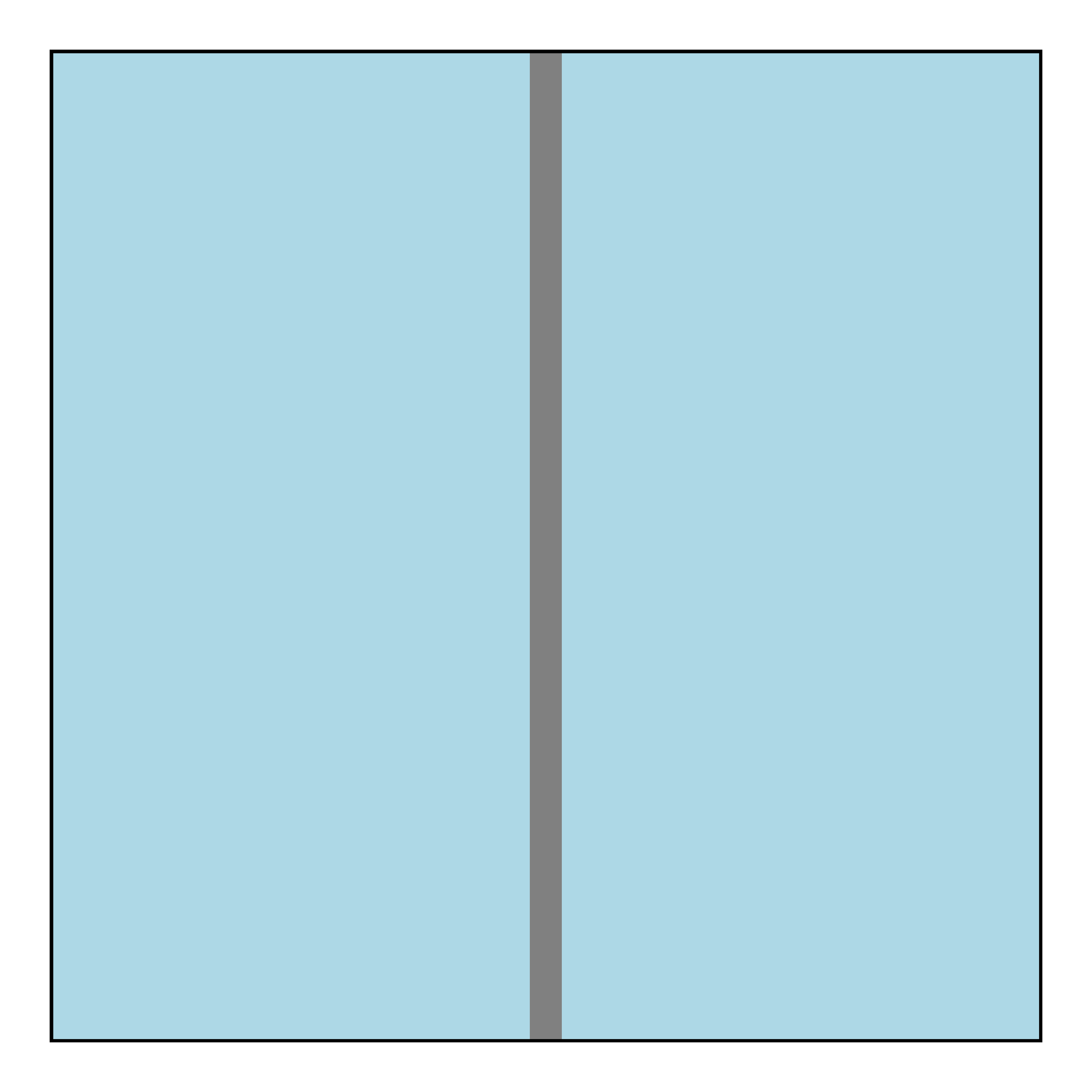}%
				\includegraphics[width=0.23\linewidth,angle=90]{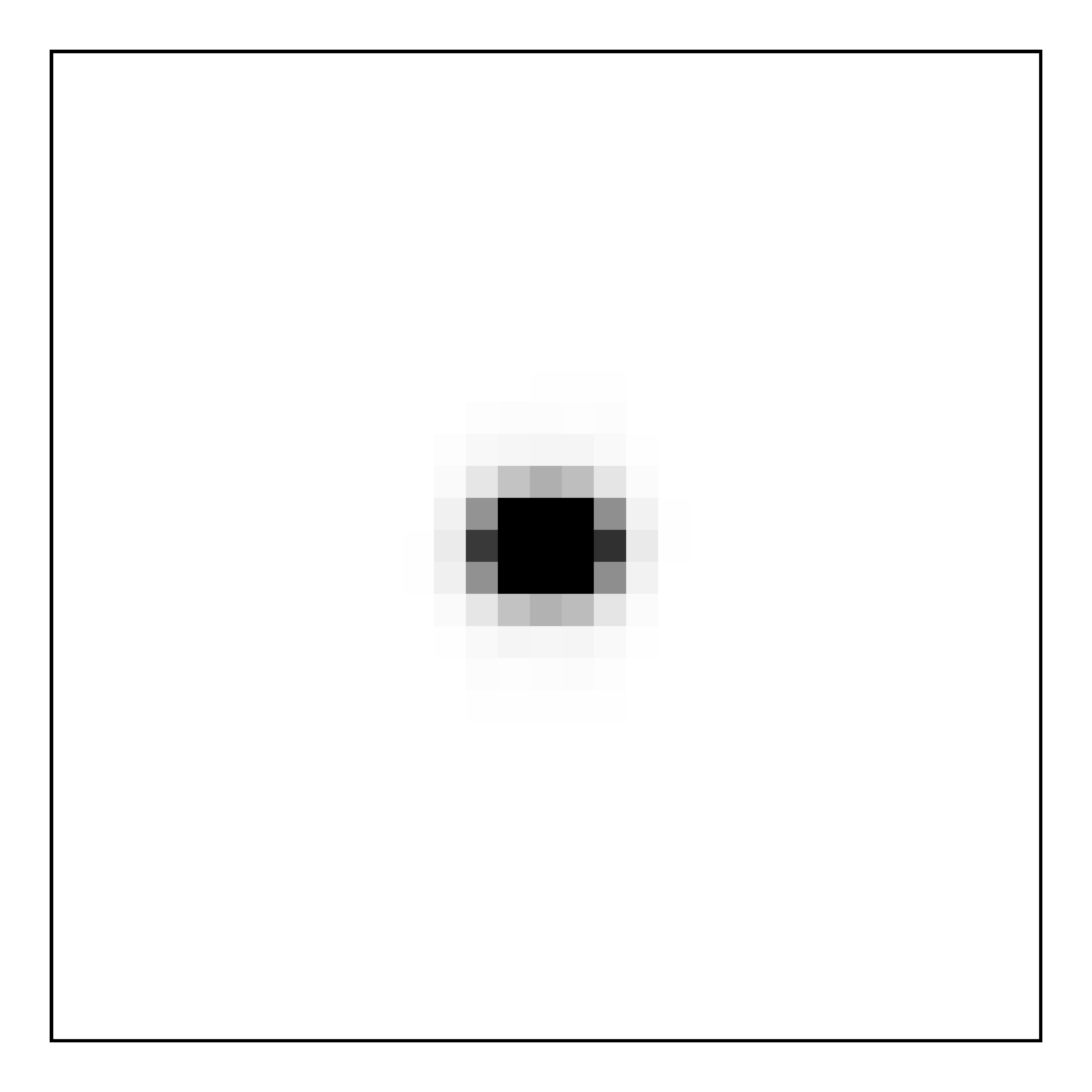}%
				\includegraphics[width=0.23\linewidth,angle=90]{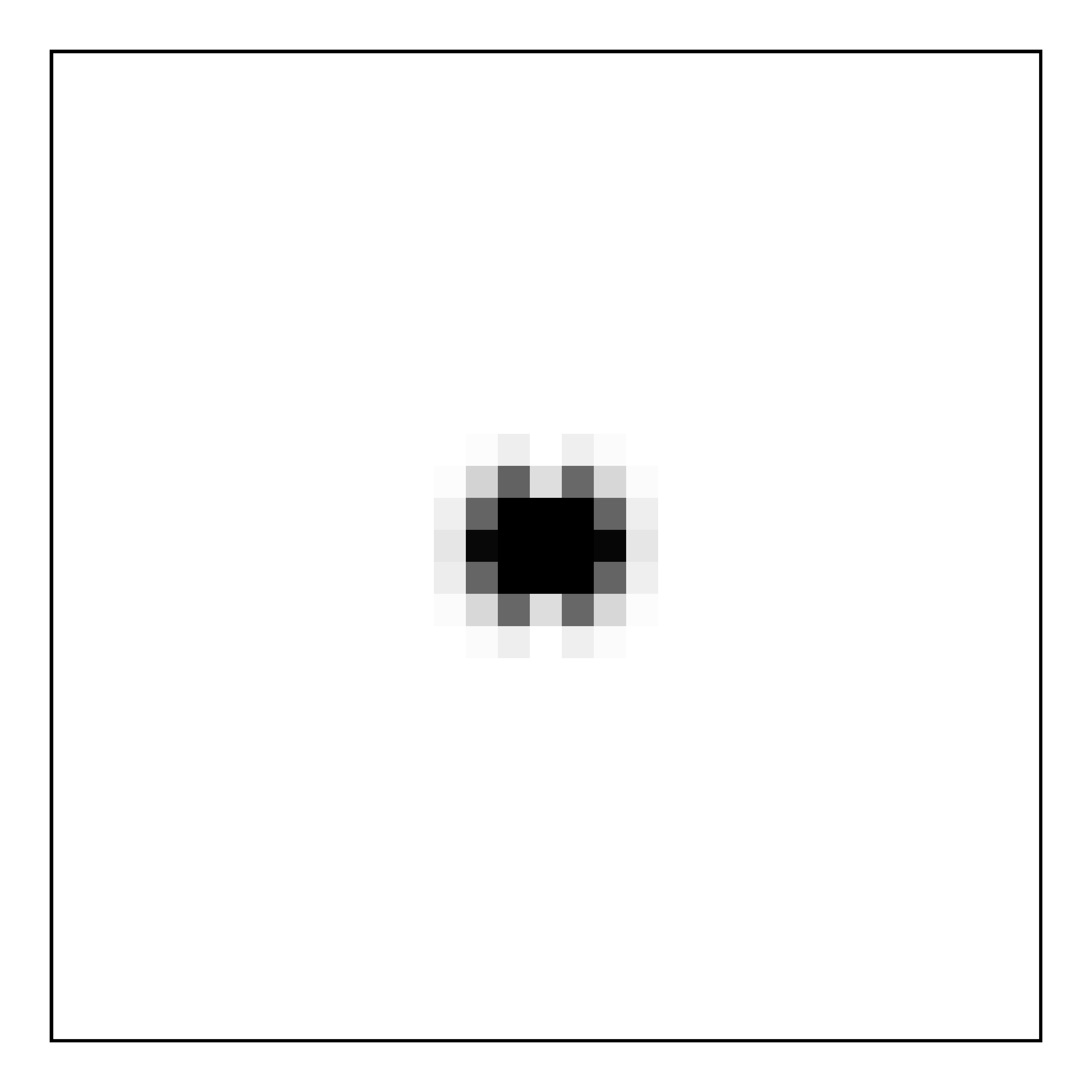}%
				\includegraphics[width=0.23\linewidth,angle=90]{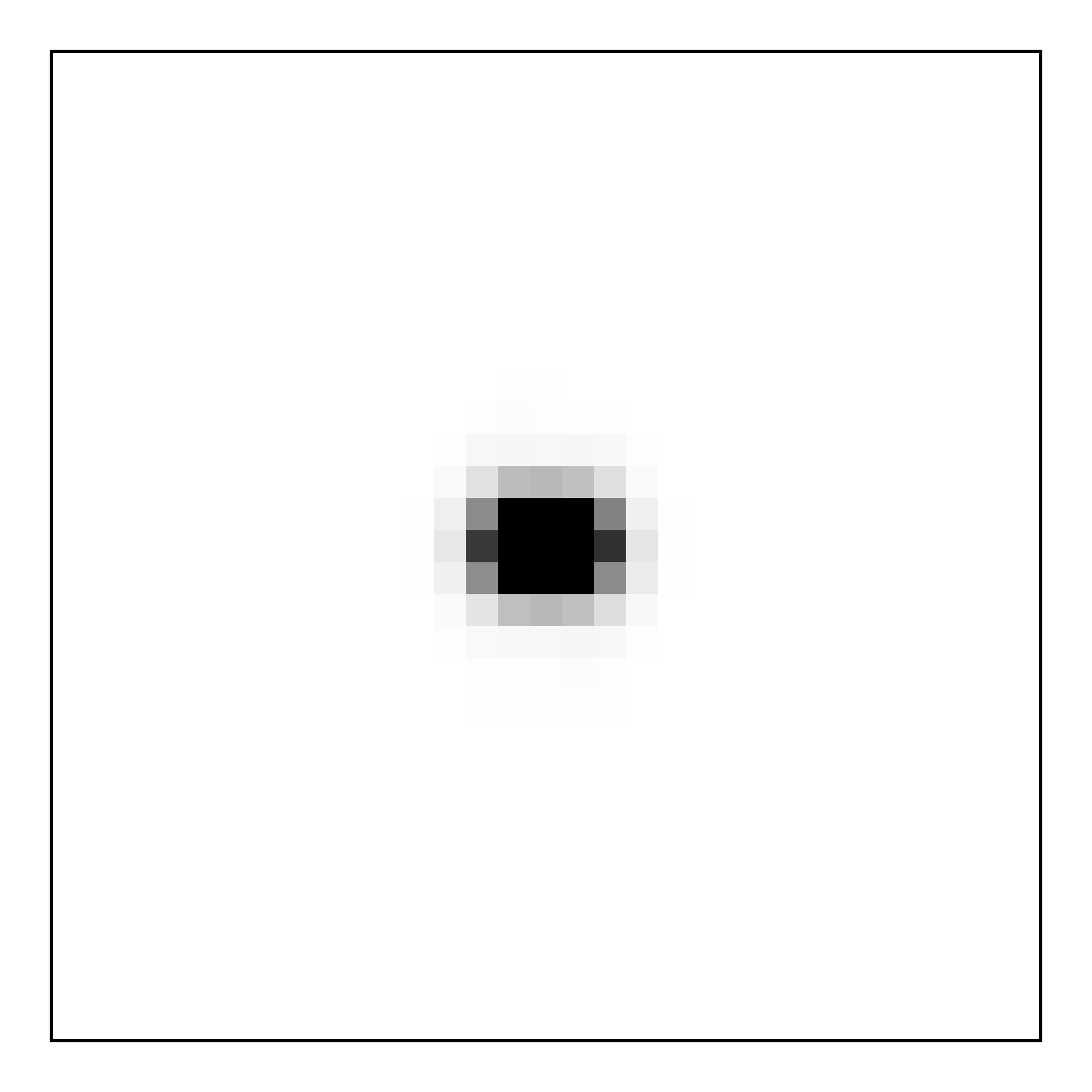}%
			\end{minipage}
		\end{minipage}%
	}
	\hfill
	\mynewroundedbox{
		\begin{minipage}[t]{0.3\textwidth}
			\centering
			\textbf{Axis 3 (sagittal)}\\[1mm]
			
			\begin{minipage}{0.03\linewidth}
				\centering
				~
			\end{minipage}
			\begin{minipage}{0.95\linewidth}
				\centering
				\begin{minipage}{0.23\linewidth}\centering\tiny Phantom\end{minipage}%
				\begin{minipage}{0.23\linewidth}\centering\tiny MC (reference)\end{minipage}%
				\begin{minipage}{0.23\linewidth}\centering\tiny SVTD\end{minipage}%
				\begin{minipage}{0.23\linewidth}\centering\tiny DDConv\end{minipage}%
			\end{minipage}\\[1mm]
			
			\begin{minipage}{0.03\linewidth}
				\centering
				\rotatebox{90}{\tiny Phantom 1}
			\end{minipage}
			\begin{minipage}{0.95\linewidth}
				\centering
				\includegraphics[width=0.23\linewidth,angle=90]{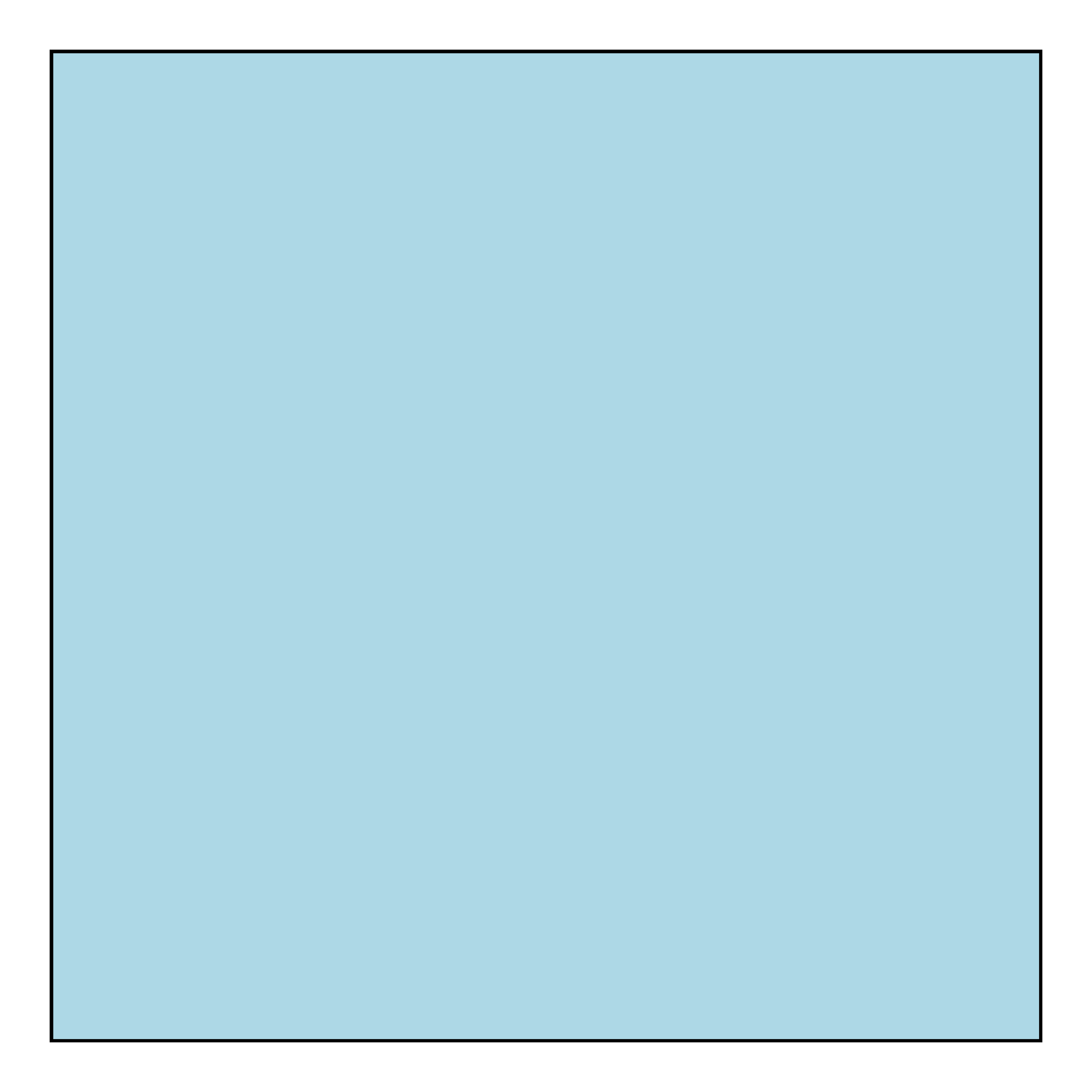}%
				\includegraphics[width=0.23\linewidth,angle=90]{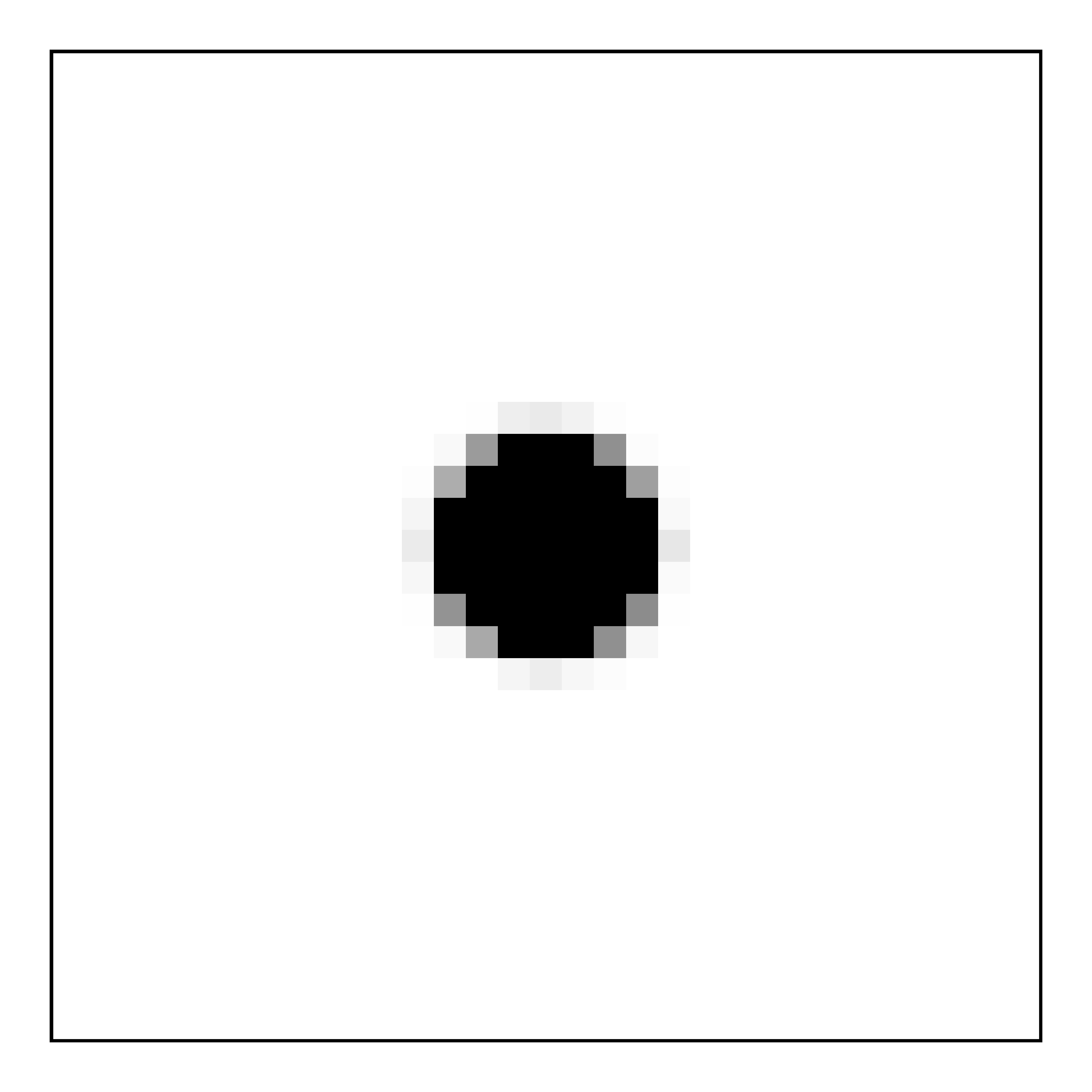}%
				\includegraphics[width=0.23\linewidth,angle=90]{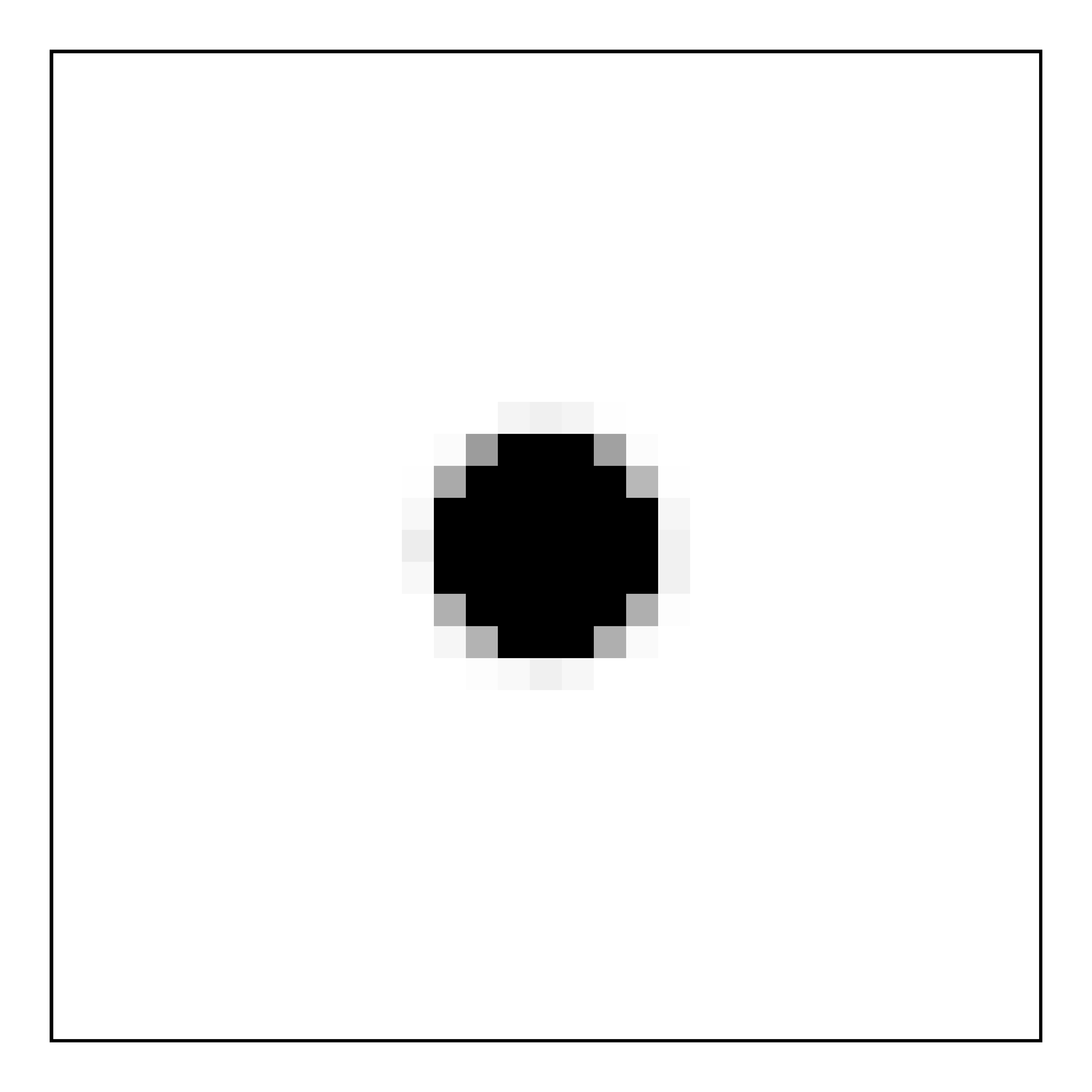}%
				\includegraphics[width=0.23\linewidth,angle=90]{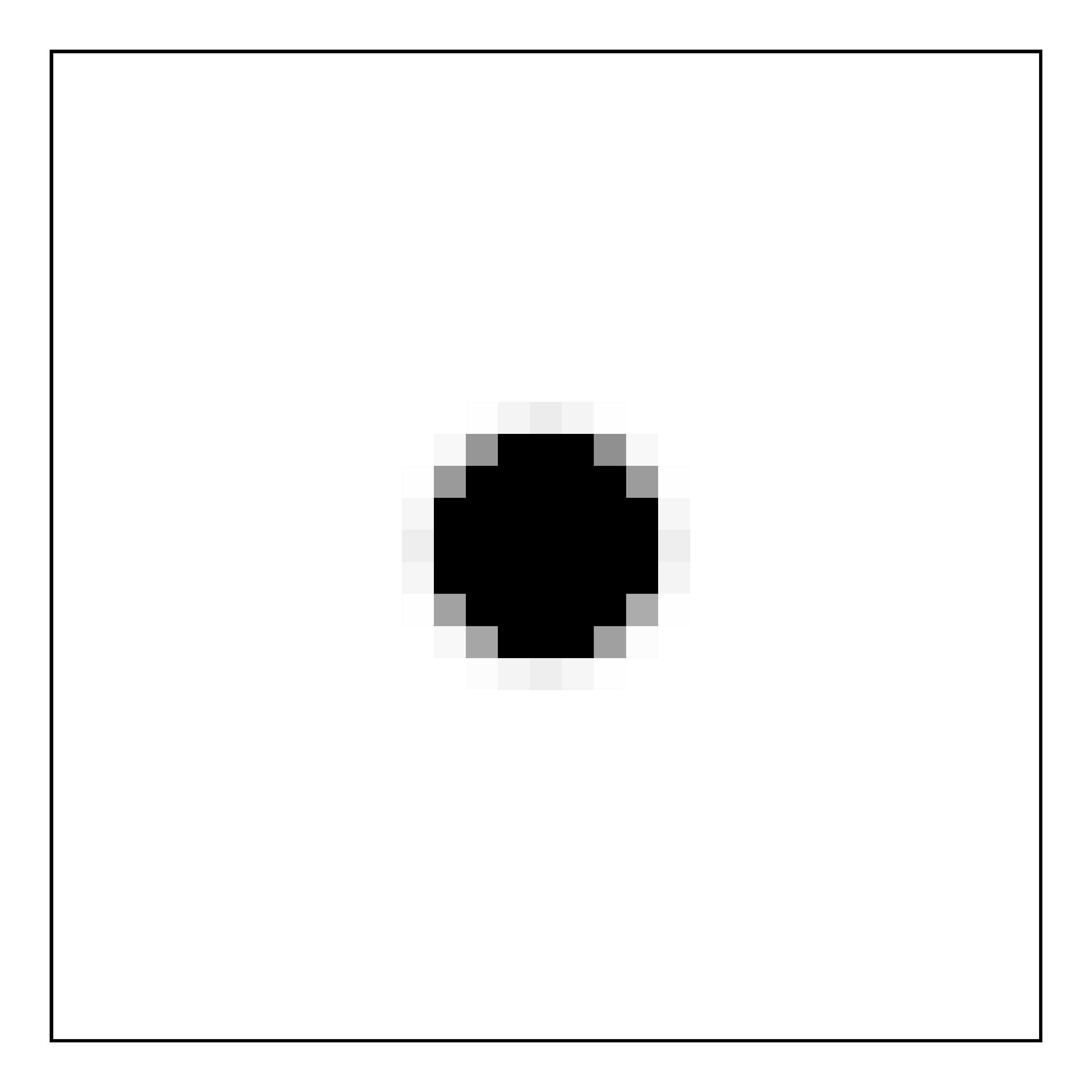}%
			\end{minipage}\\[0.5mm]
			
			\begin{minipage}{0.03\linewidth}
				\centering
				\rotatebox{90}{\tiny Phantom 2}
			\end{minipage}
			\begin{minipage}{0.95\linewidth}
				\centering
				\includegraphics[width=0.23\linewidth,angle=90]{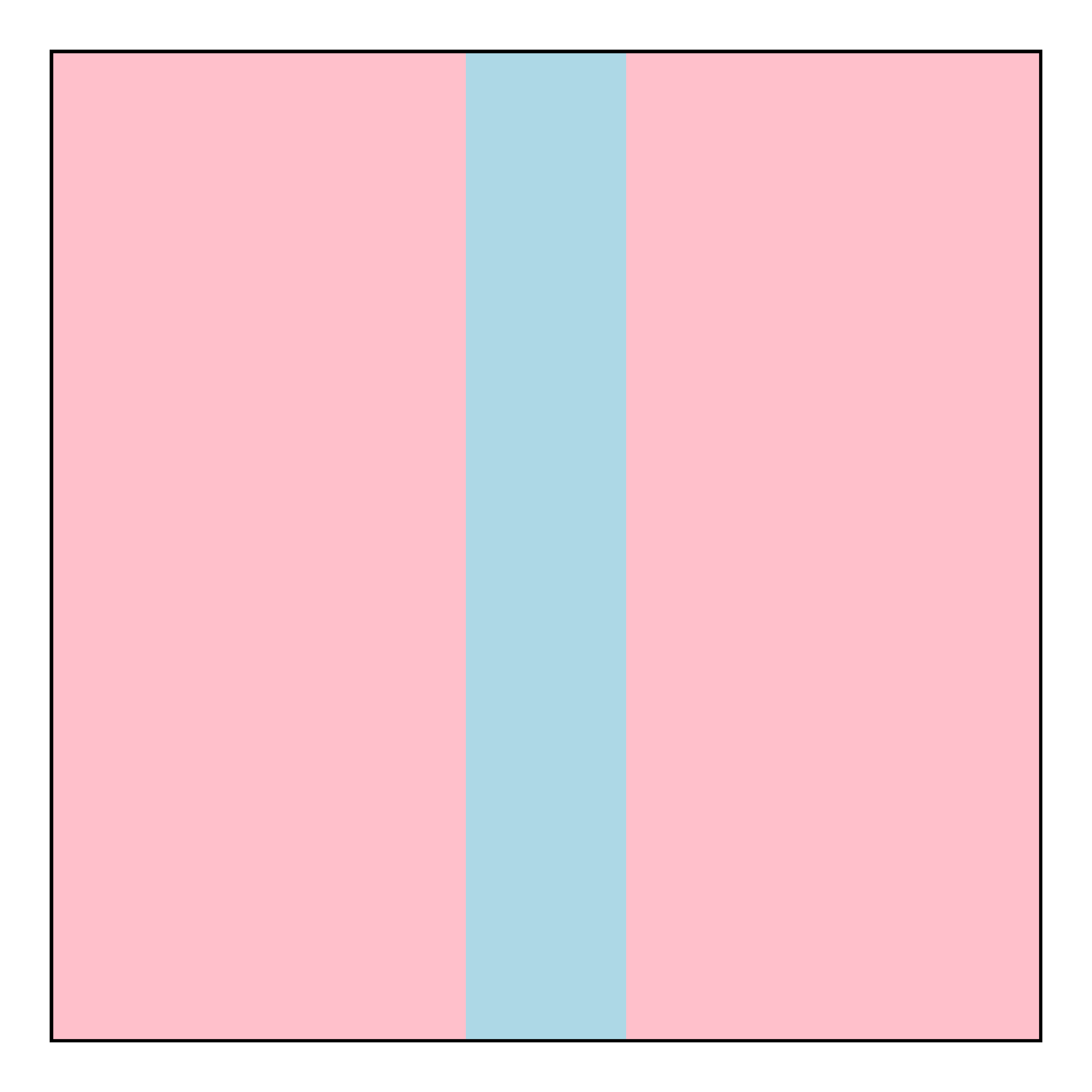}%
				\includegraphics[width=0.23\linewidth,angle=90]{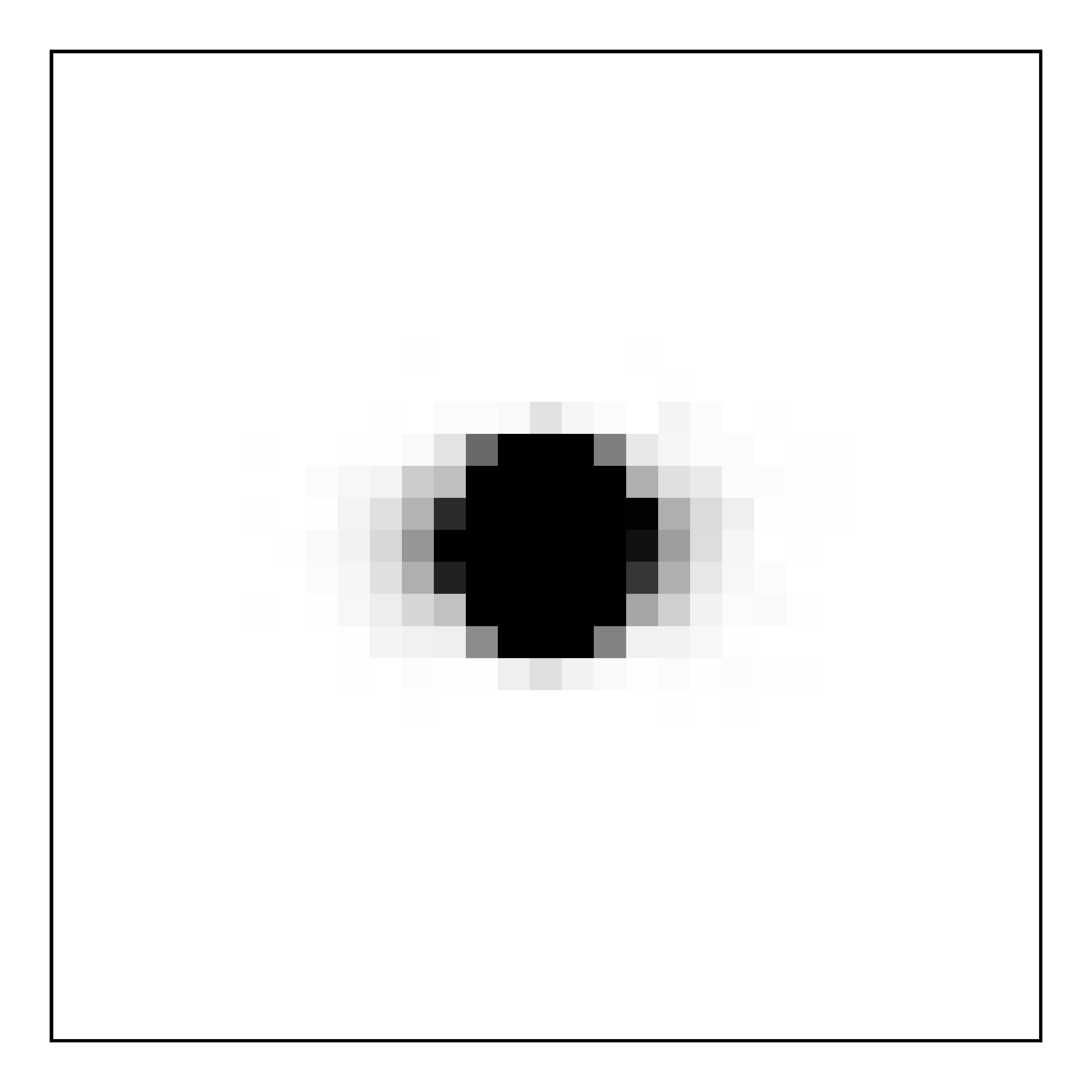}%
				\includegraphics[width=0.23\linewidth,angle=90]{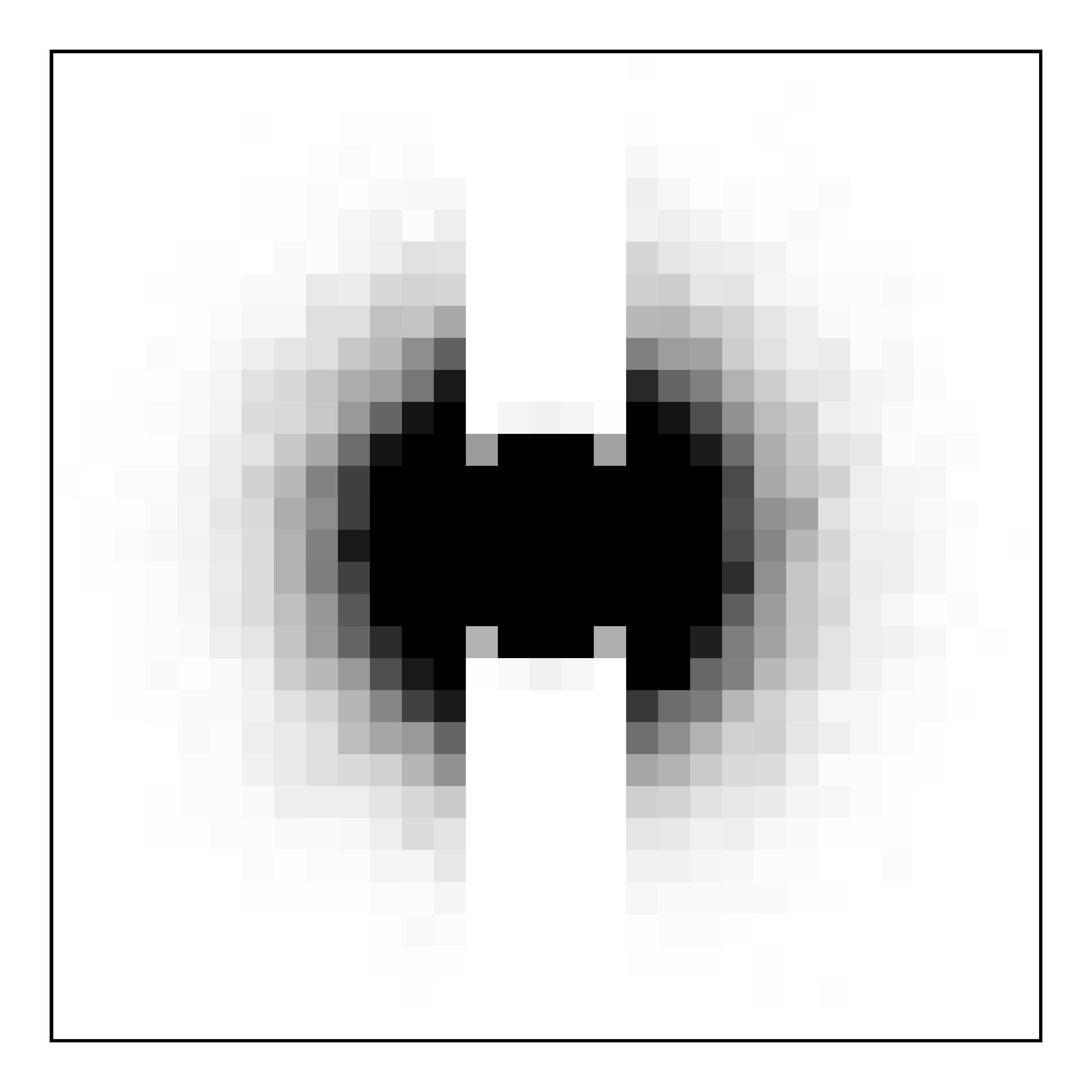}%
				\includegraphics[width=0.23\linewidth,angle=90]{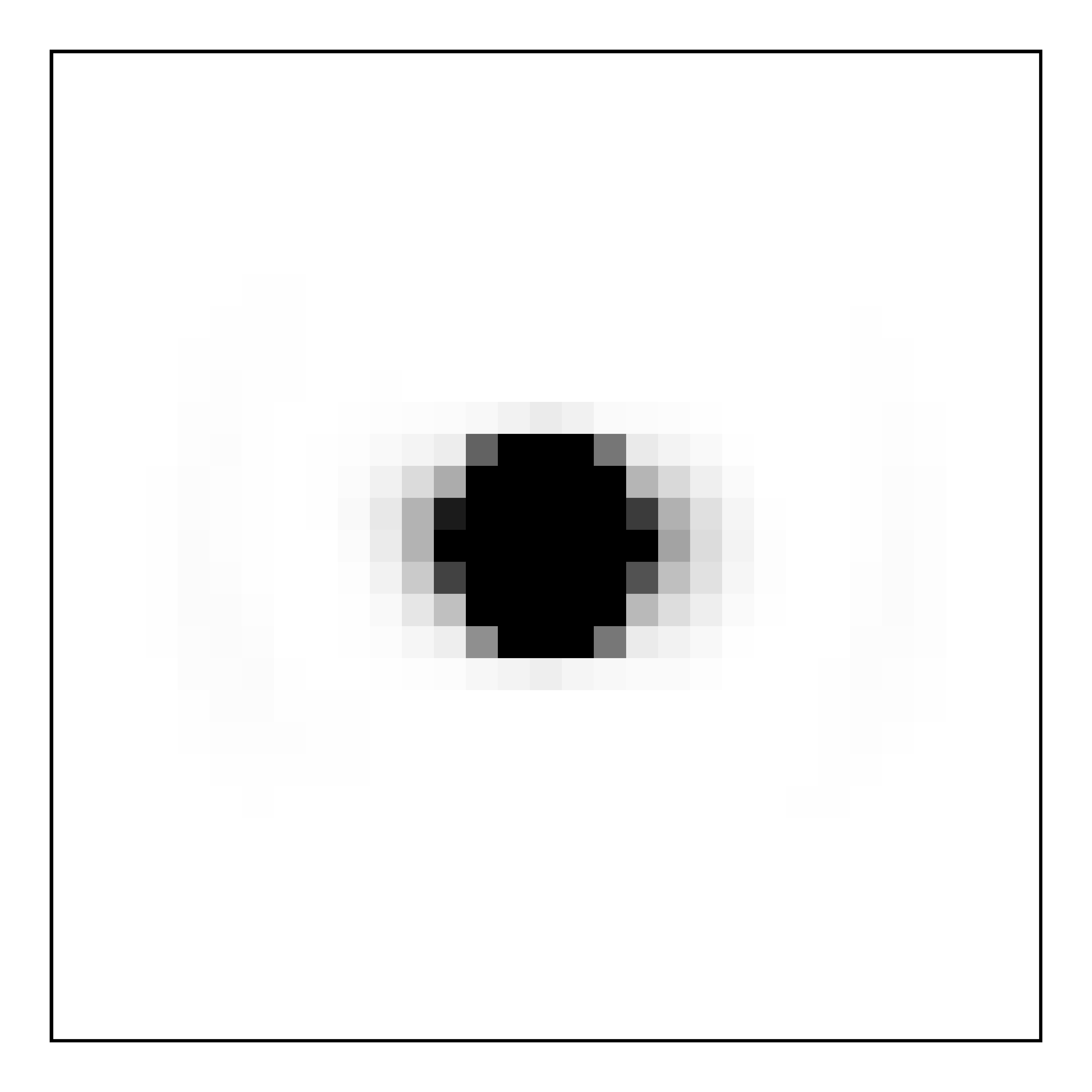}%
			\end{minipage}\\[0.5mm]
			
			\begin{minipage}{0.03\linewidth}
				\centering
				\rotatebox{90}{\tiny Phantom 3}
			\end{minipage}
			\begin{minipage}{0.95\linewidth}
				\centering
				\includegraphics[width=0.23\linewidth,angle=90]{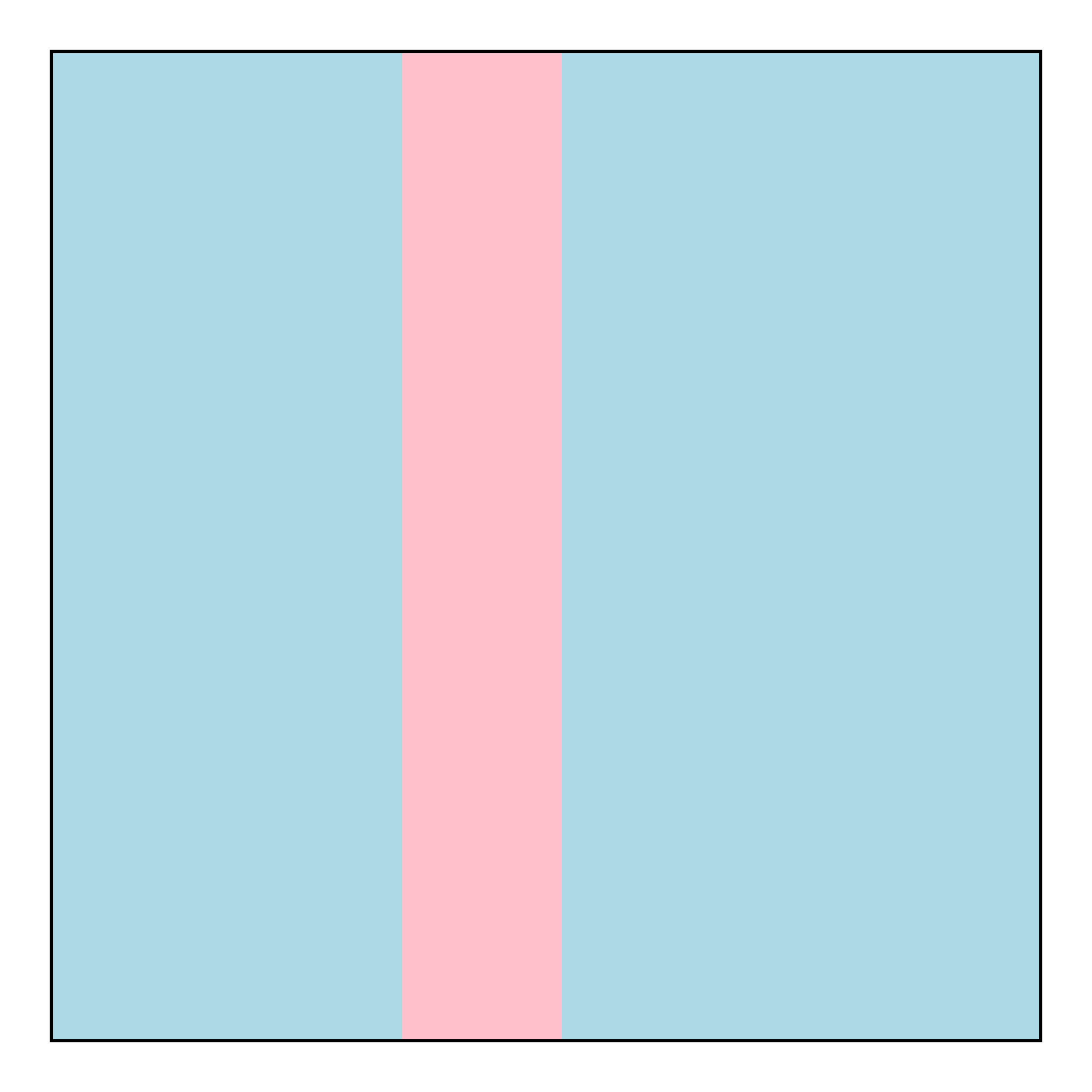}%
				\includegraphics[width=0.23\linewidth,angle=90]{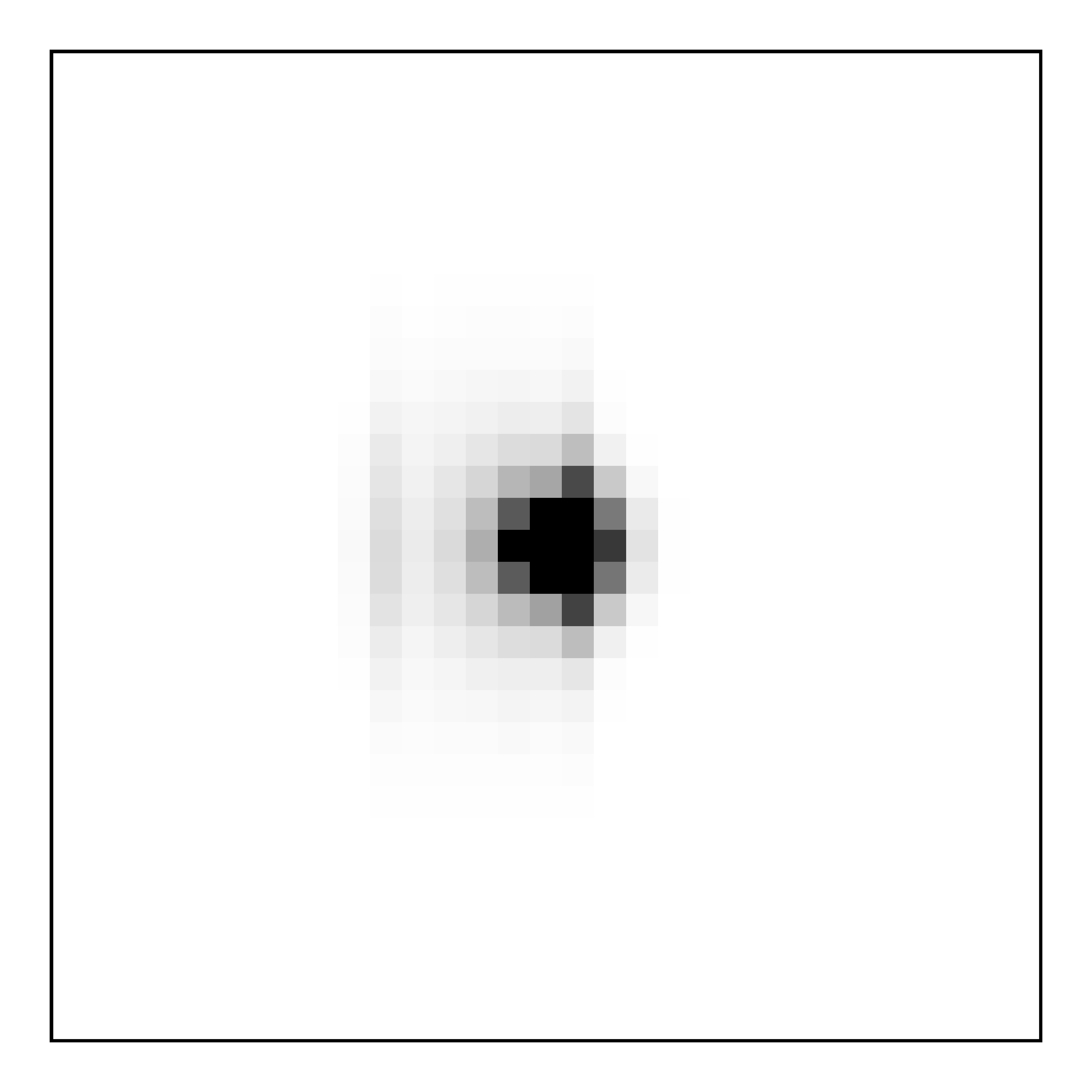}%
				\includegraphics[width=0.23\linewidth,angle=90]{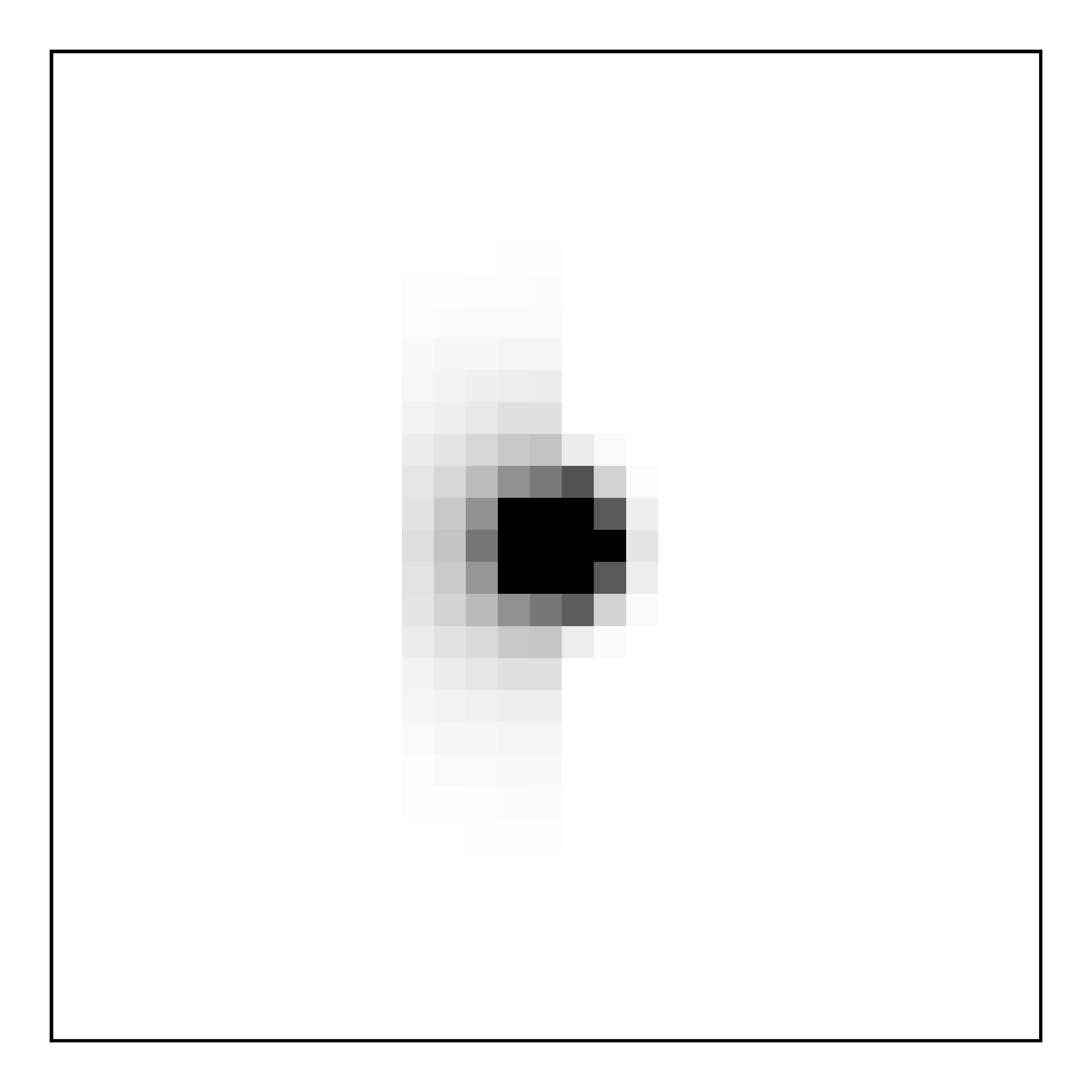}%
				\includegraphics[width=0.23\linewidth,angle=90]{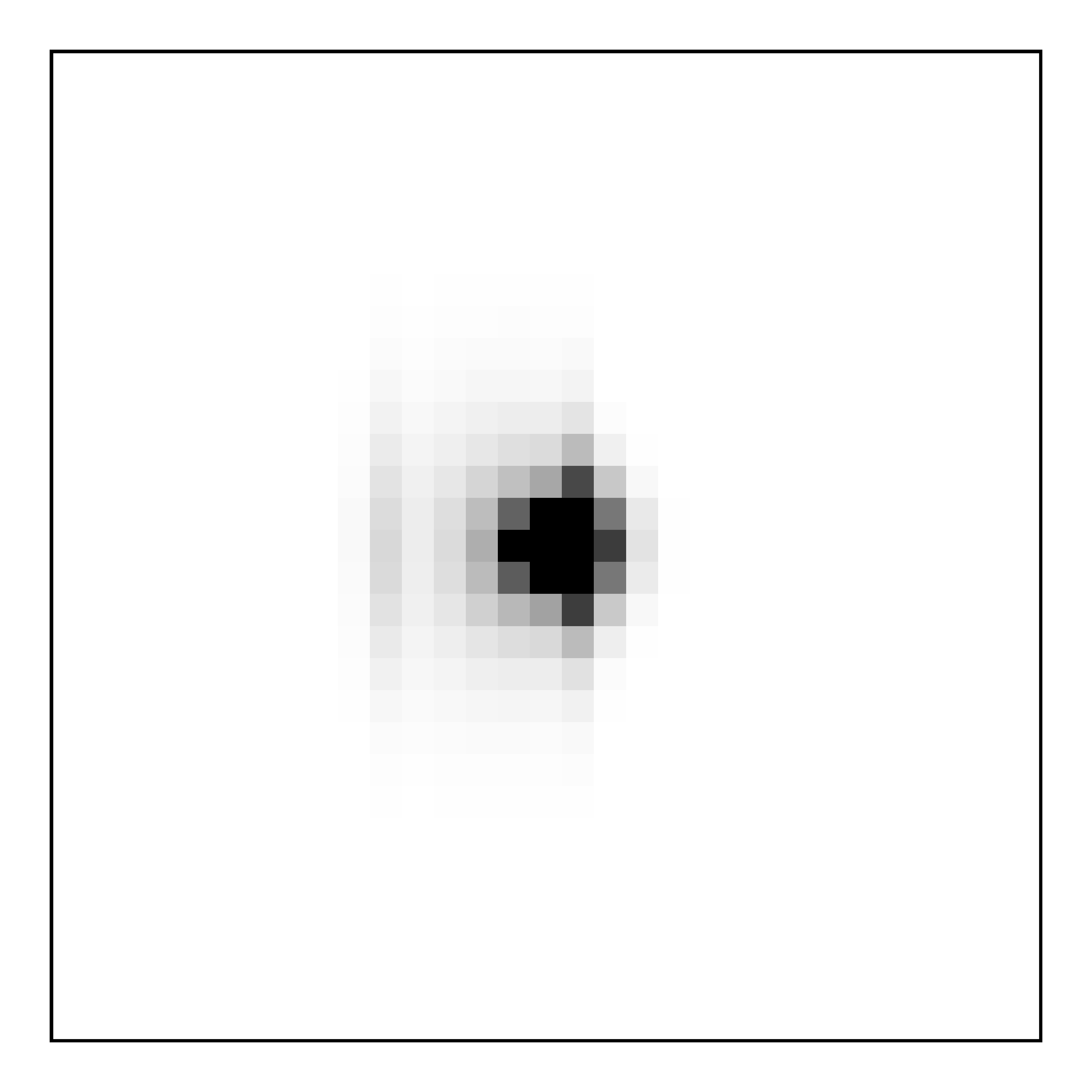}%
			\end{minipage}\\[0.5mm]
			
			\begin{minipage}{0.03\linewidth}
				\centering
				\rotatebox{90}{\tiny Phantom 4}
			\end{minipage}
			\begin{minipage}{0.95\linewidth}
				\centering
				\includegraphics[width=0.23\linewidth,angle=90]{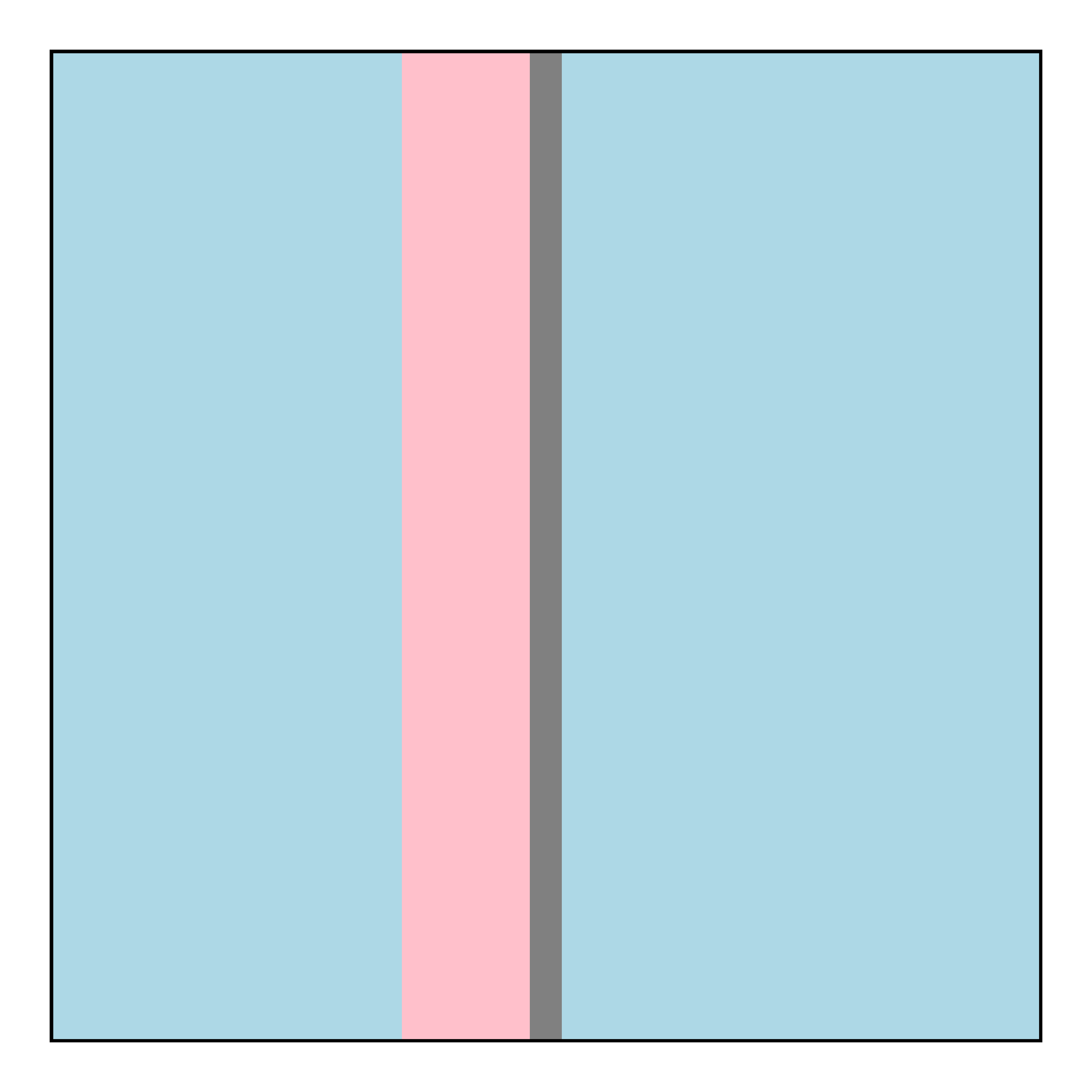}%
				\includegraphics[width=0.23\linewidth,angle=90]{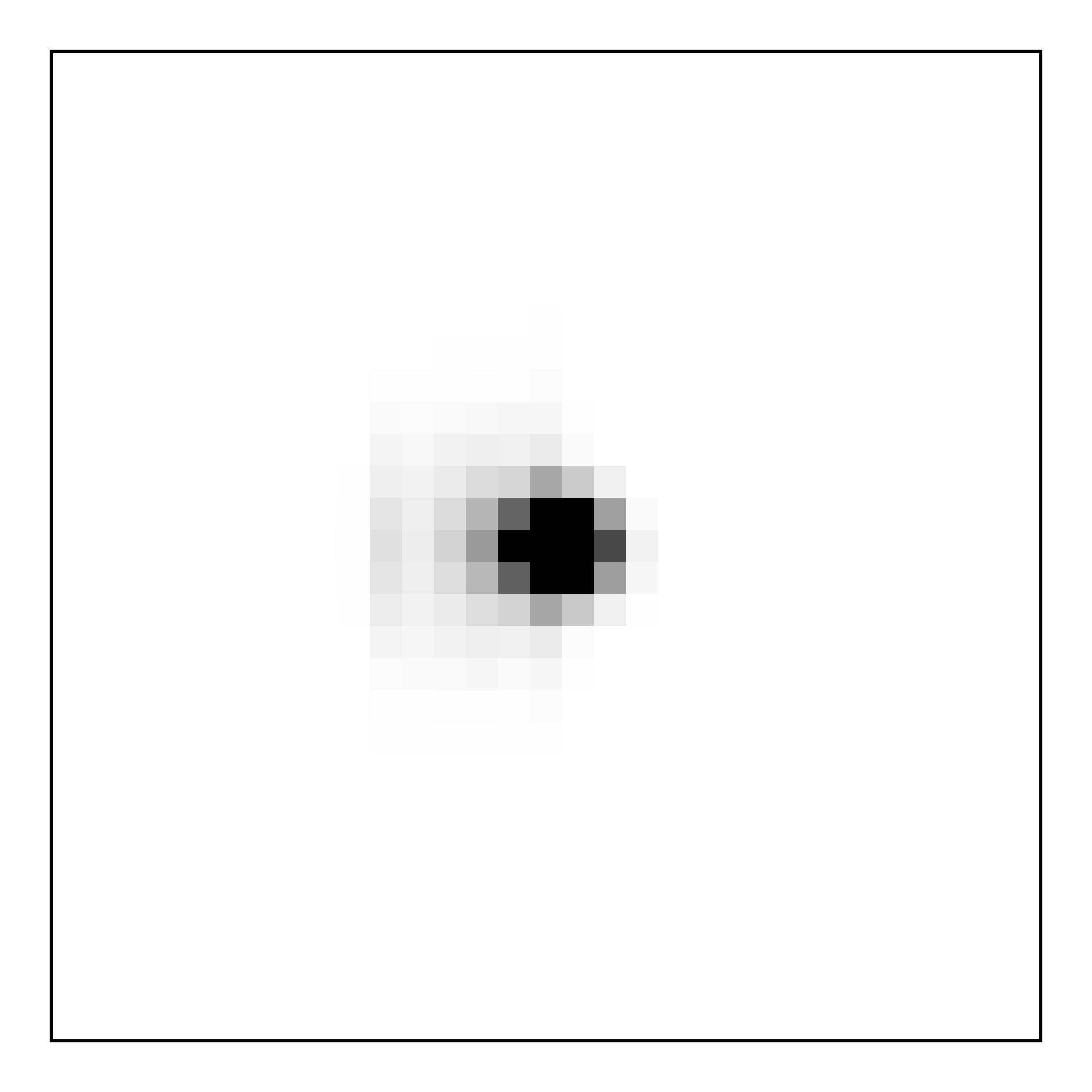}%
				\includegraphics[width=0.23\linewidth,angle=90]{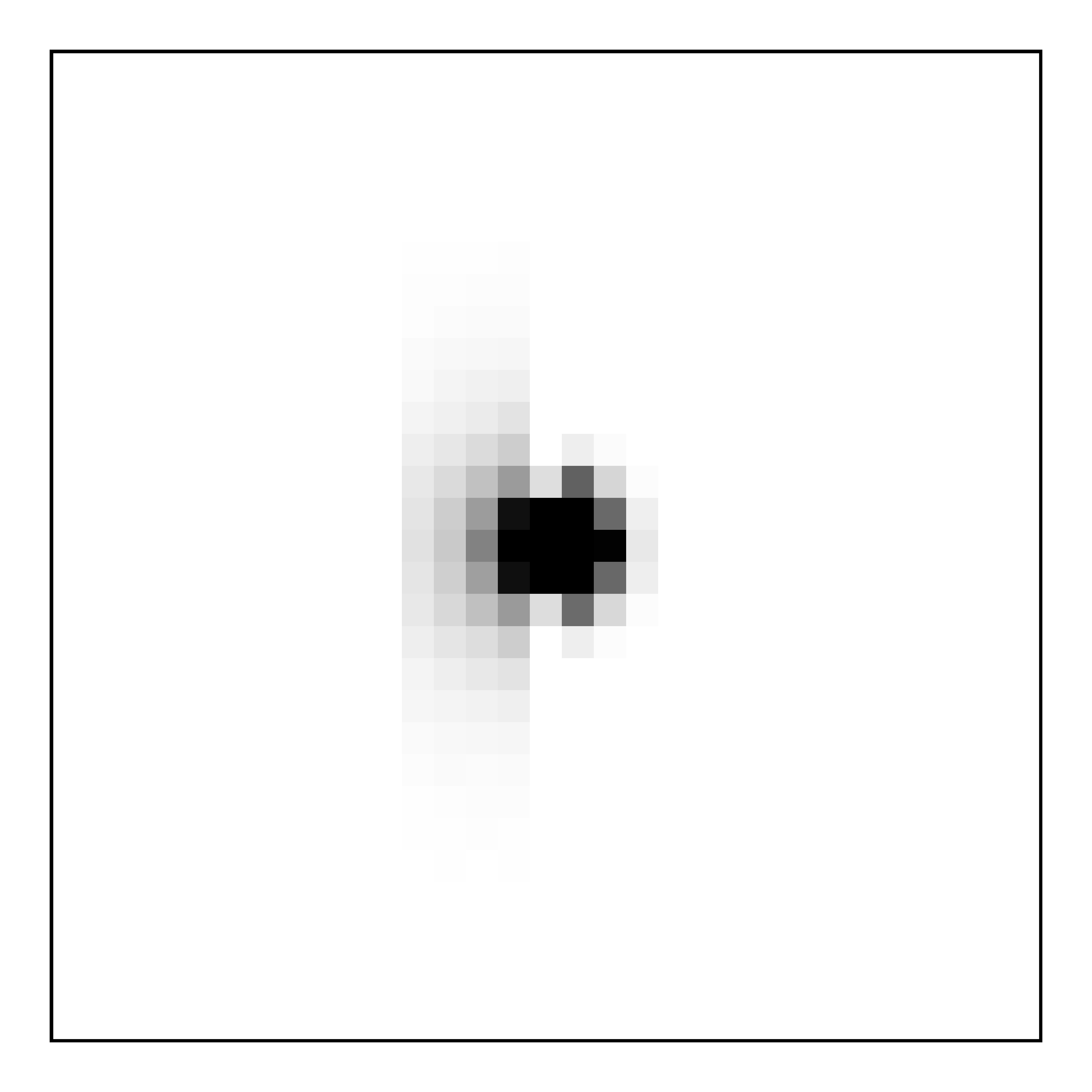}%
				\includegraphics[width=0.23\linewidth,angle=90]{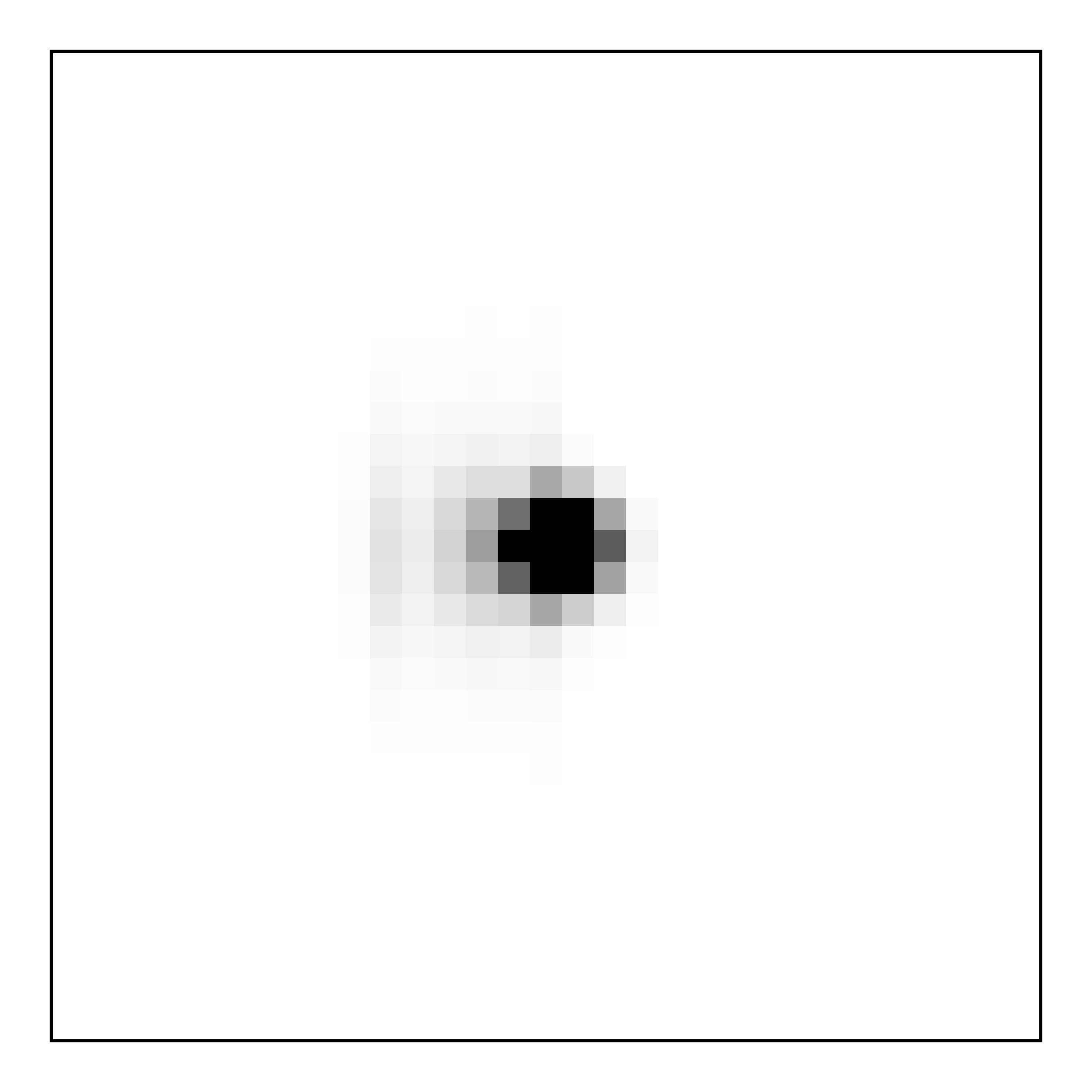}%
			\end{minipage}\\[0.5mm]
			
			\begin{minipage}{0.03\linewidth}
				\centering
				\rotatebox{90}{\tiny Phantom 5}
			\end{minipage}
			\begin{minipage}{0.95\linewidth}
				\centering
				\includegraphics[width=0.23\linewidth,angle=90]{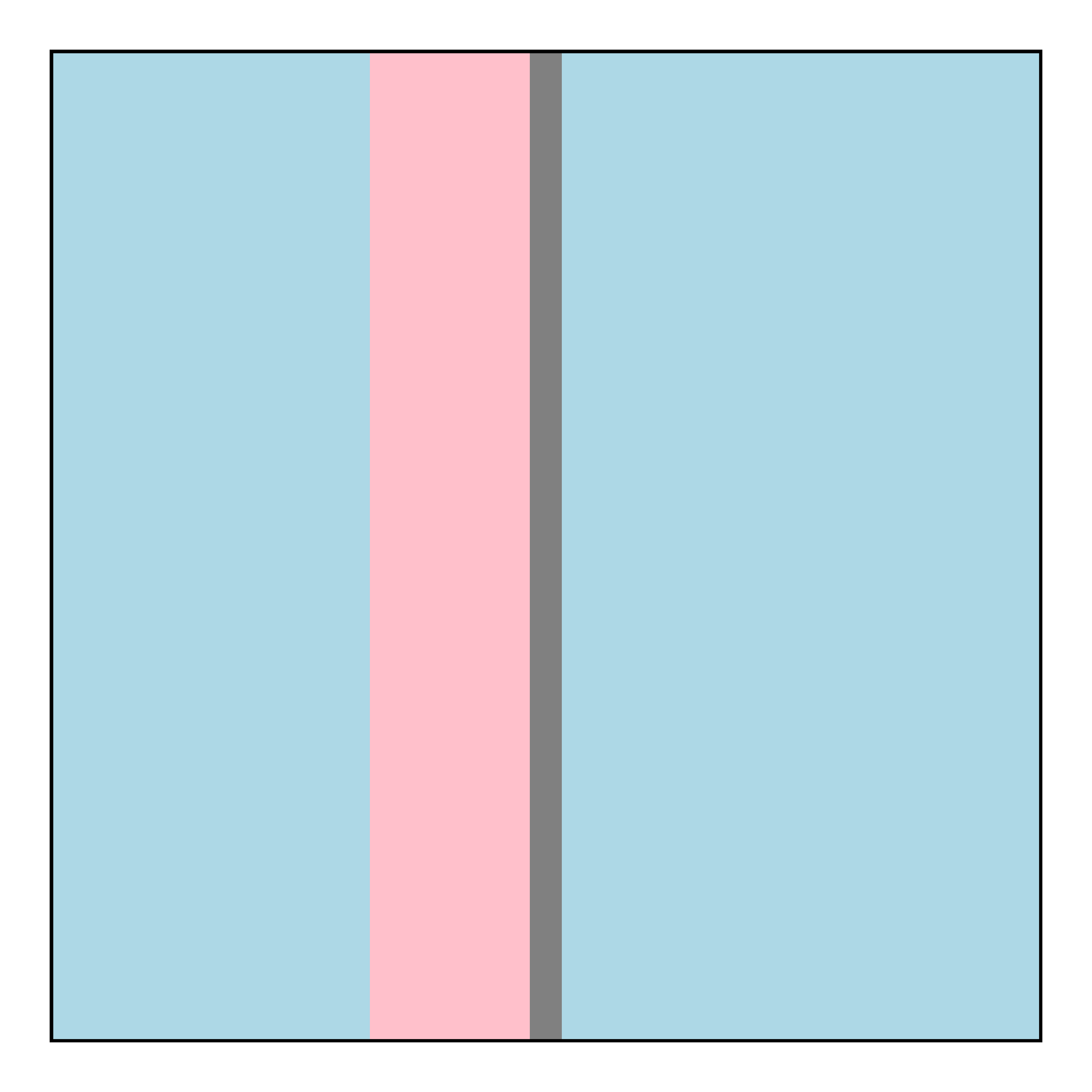}%
				\includegraphics[width=0.23\linewidth,angle=90]{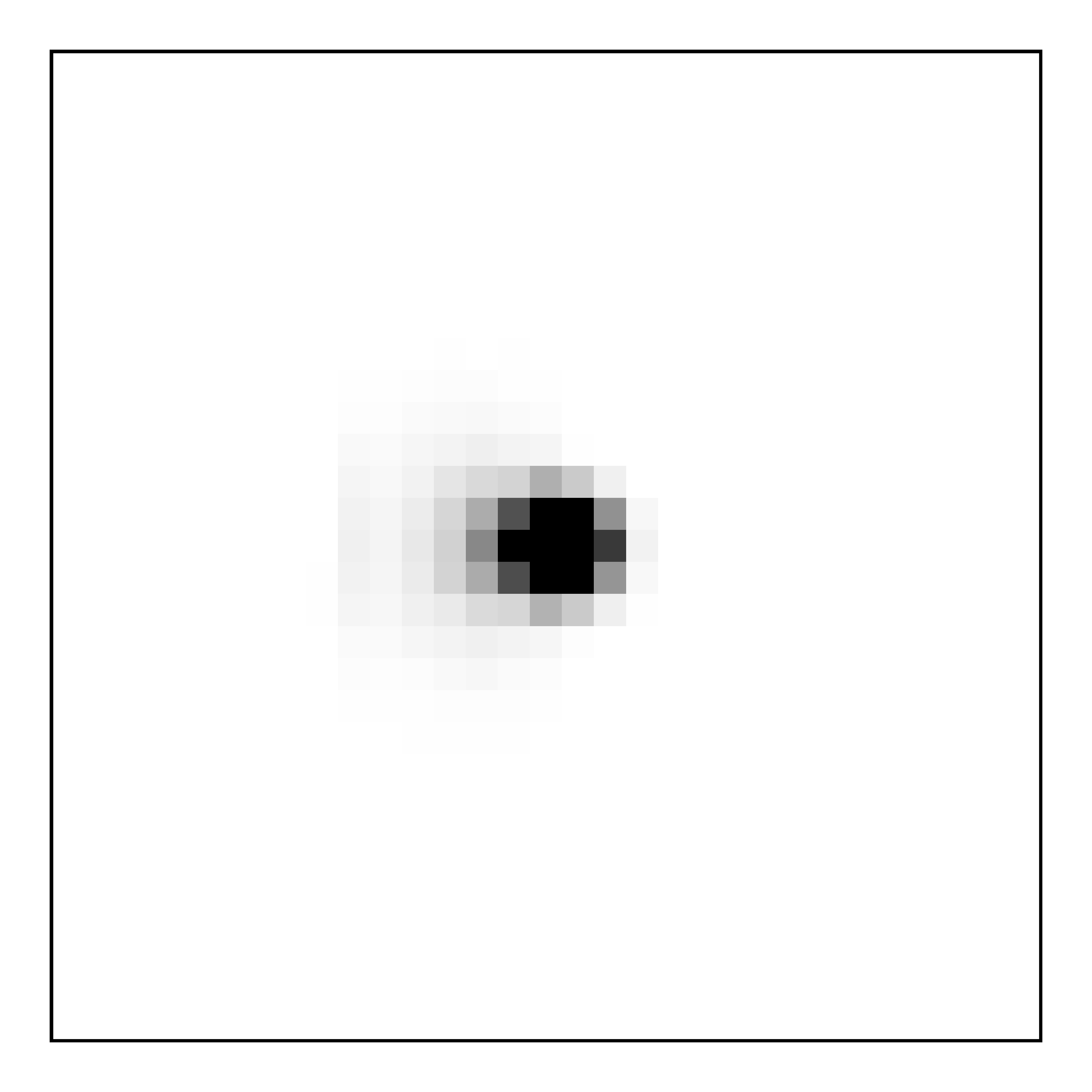}%
				\includegraphics[width=0.23\linewidth,angle=90]{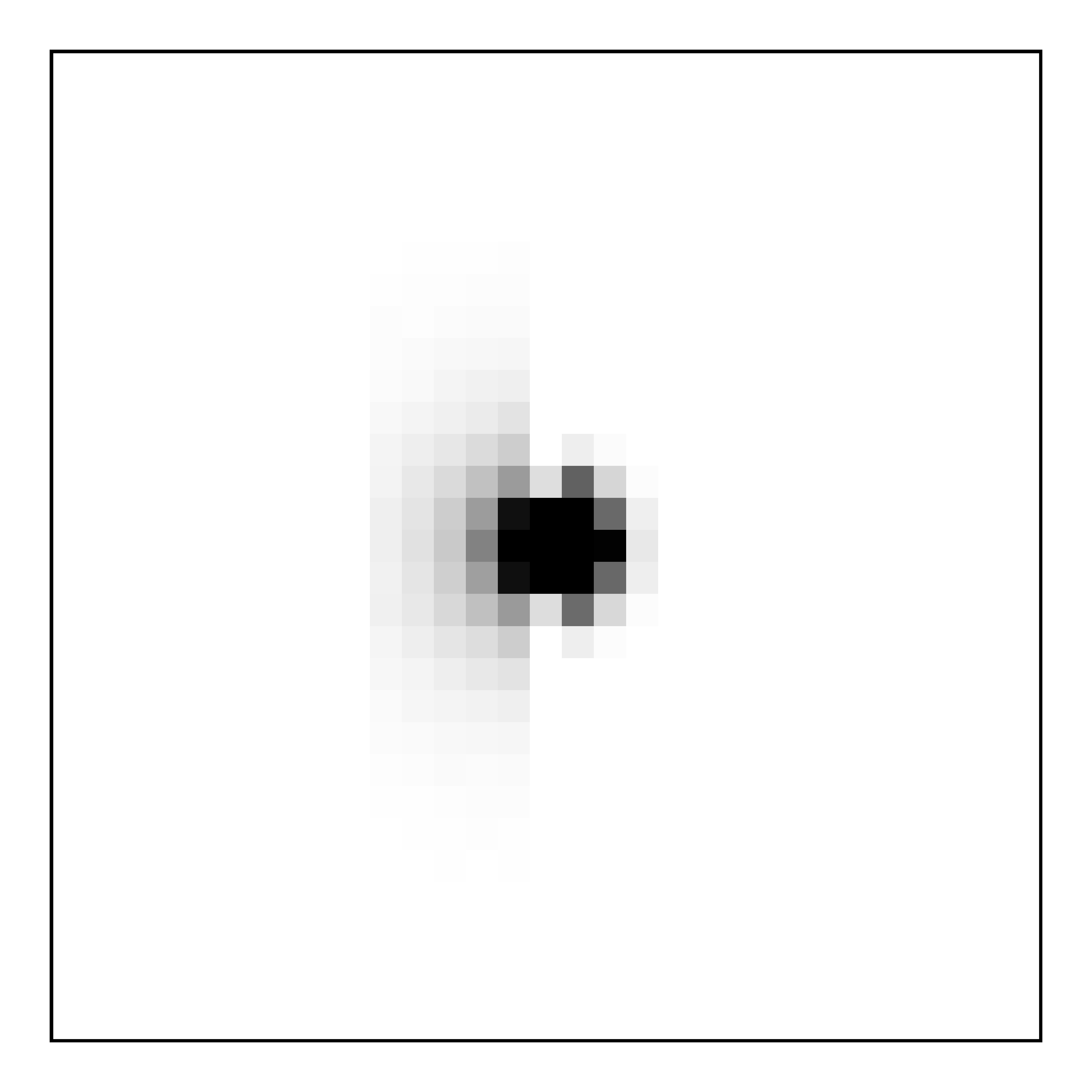}%
				\includegraphics[width=0.23\linewidth,angle=90]{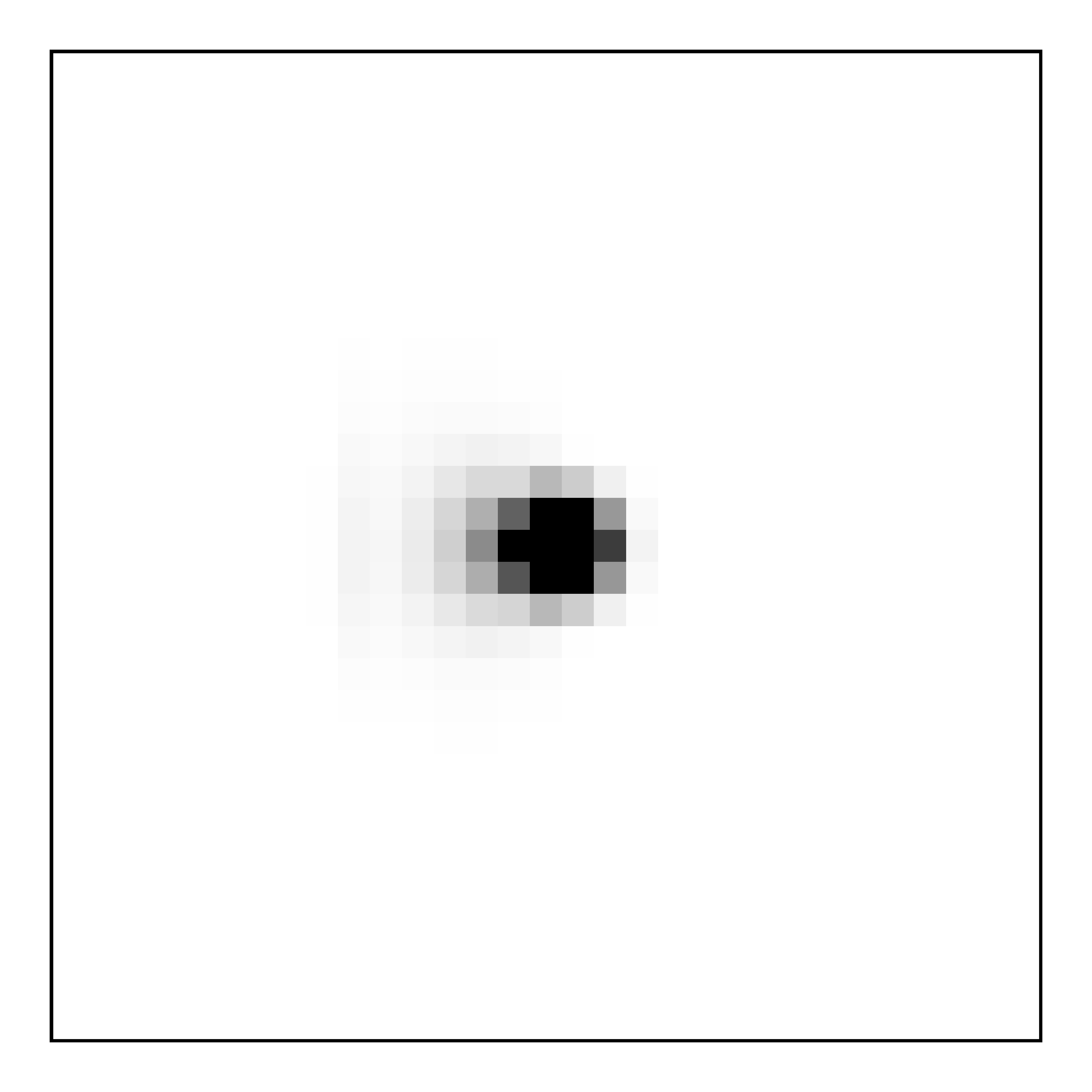}%
			\end{minipage}
		\end{minipage}%
	}

	\caption{
		Experiment 1---Overview of PR distributions across different viewing axes with the digital phantoms from Figure~\ref{fig:phantoms} (pink for lung, light blue for water) with \ac{MC} simulations (reference), SVTD and DDConv.
	}
	\label{fig:overview_axes}
\end{figure*}

\subsubsection{XCAT Phantom}\label{sec:xcat_pr}\label{sec:exp2}

We proceeded with a similar experiment, this time with an \ac{XCAT}-generated \ac{Ga68} activity distribution (Figure~\ref{subfig:xcat_act}) with the corresponding \ac{XCAT}-generated material image (Figure~\ref{subfig:xcat_mat}). The activity distribution contains four hot lesions: two in the lung (Lesion 1 and Lesion 2), one at the interface between the lung and soft tissues (Lesion 3), and one at the interface between the lung and the liver (Lesion 4). The radii of Lesions 1 through 4 are 8~mm, 5~mm, 2~mm, and 10~mm, respectively. While \ac{DDConv} is expected to model \ac{PR} accurately, \ac{SVTD} is expected to be inaccurate for Lesion 3 and Lesion 4 which are located in heterogeneous regions.

\newcommand{\spyOne}[1]{%
  \spy [vividorange] on ($(#1.south east)+(-3.15,1.08)$) in node [left] at ($(#1.north west)+(0.8,-0.3)$);
}

\newcommand{\spyTwo}[1]{%
  \spy [vividorange] on ($(#1.south east)+(-2.77,0.75)$) in node [left] at ($(#1.north west)+(0.8,-1.5)$);
  \draw[green, line width=0.5pt] (-0.6,-0.1) -- (-0.6,-0.6);
}

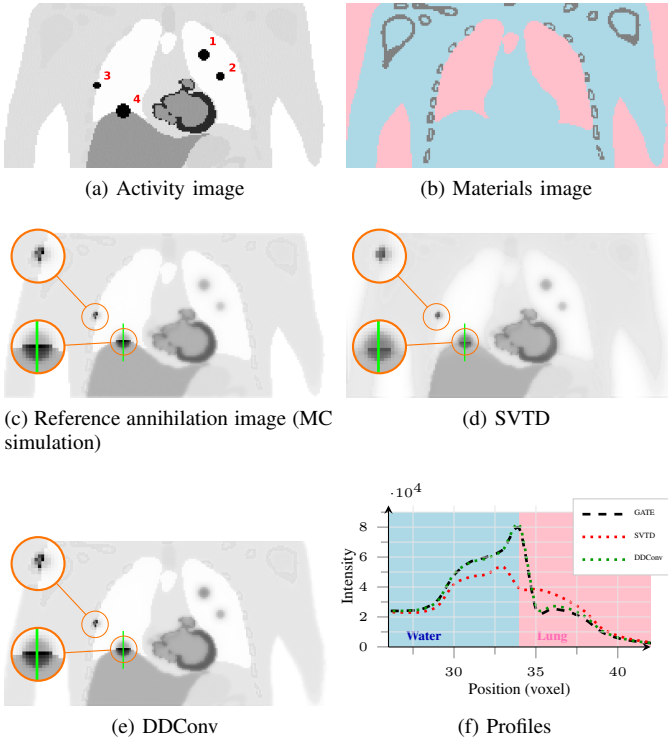
\begin{figure}
	\subfloat[Activity image \label{subfig:xcat_act}]{%
		\includegraphics[width=0.49\linewidth]{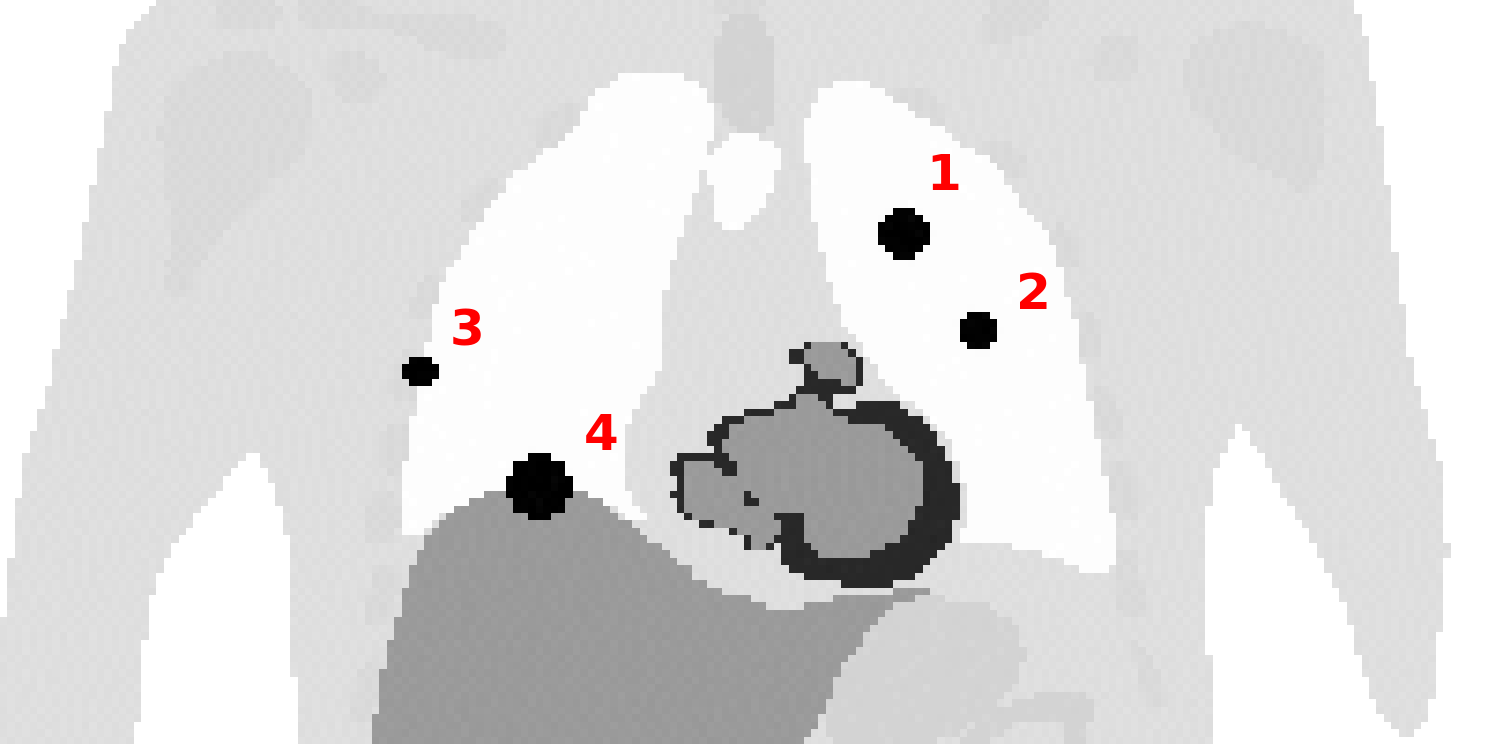}%
	}\hfill
	\subfloat[Materials image \label{subfig:xcat_mat}]{%
		\includegraphics[width=0.49\linewidth]{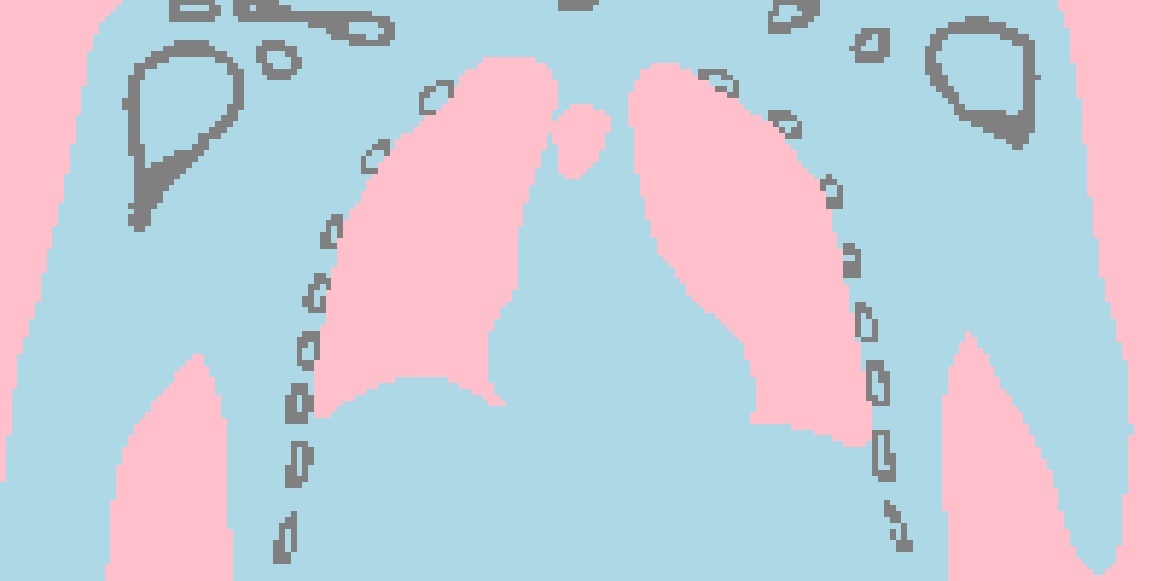}%
	}
	
	\subfloat[Reference annihilation image (\ac{MC} simulation) \label{subfig:xcat_mc}]{%
		\begin{tikzpicture}[spy using outlines={circle, magnification=2, size=0.7cm, connect spies}]
			\node[inner sep=0pt] (img) {\includegraphics[width=0.49\linewidth]{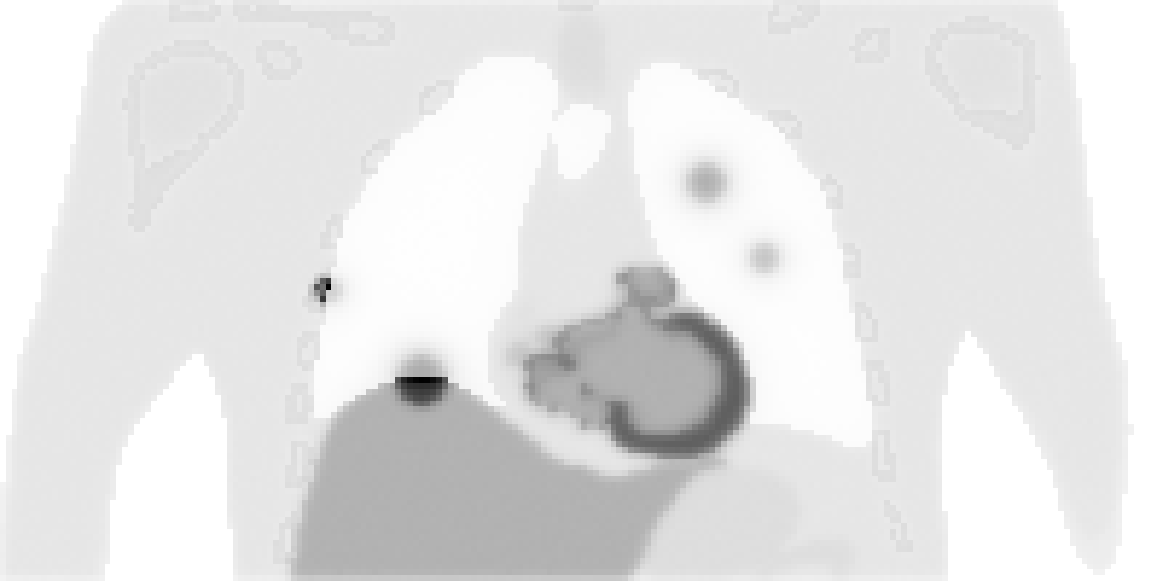}};
			\spyOne{img}
			\spyTwo{img}
		\end{tikzpicture}%
	}\hfill
	\subfloat[\Ac{SVTD} \label{subfig:xcat_svtd}]{%
		\begin{tikzpicture}[spy using outlines={circle, magnification=2, size=0.7cm, connect spies}]
			\node[inner sep=0pt] (img) {\includegraphics[width=0.49\linewidth]{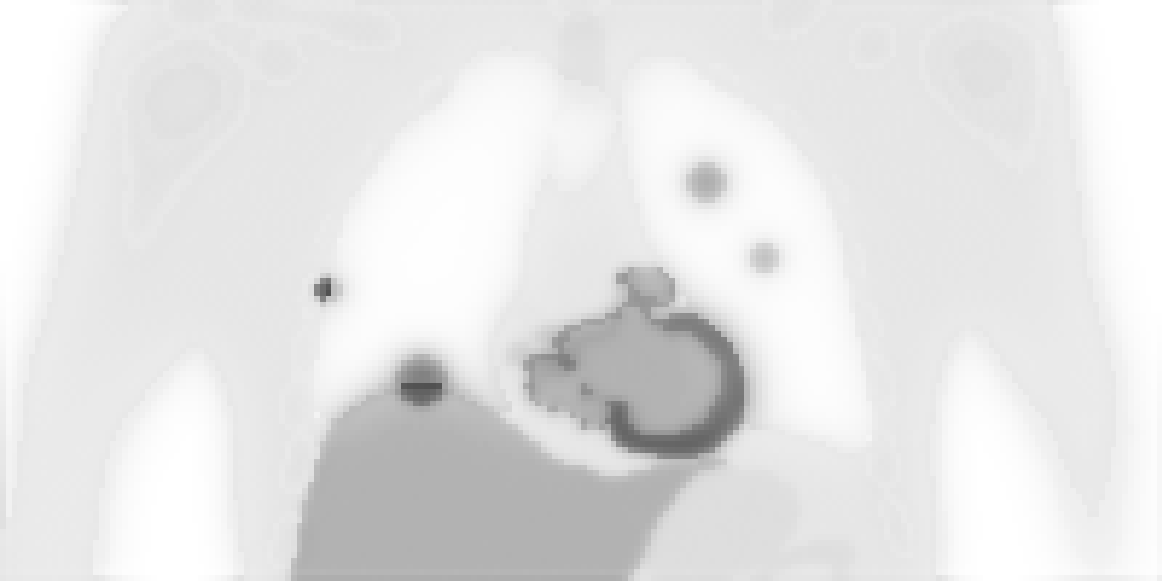}};
			\spyOne{img}
			\spyTwo{img}
		\end{tikzpicture}%
	}
	
	\subfloat[\Ac{DDConv} \label{subfig:xcat_ddconv}]{%
		\begin{tikzpicture}[spy using outlines={circle, magnification=2, size=0.7cm, connect spies}]
			\node[inner sep=0pt] (img) {\includegraphics[width=0.49\linewidth]{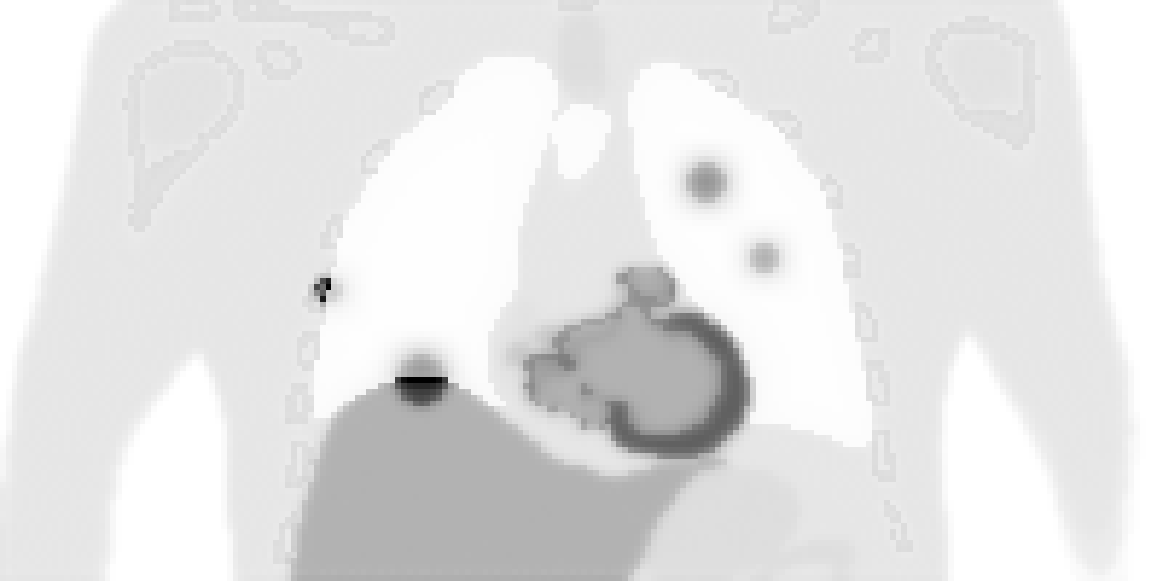}};
			\spyOne{img}
			\spyTwo{img}
		\end{tikzpicture}%
	}
	\subfloat[Profiles\label{subfig:xcat_profiles}]{%
	\begin{tikzpicture}
    	\begin{axis}[
    		width=0.57\linewidth,
    		height=0.39\linewidth,
    		axis x line=bottom,
    		axis y line=left,
    		xmin=26, xmax=42,
    		ymin=-50, ymax=90000,
    		enlarge y limits={upper, value=0.05},
    		clip marker paths=false,
    		grid=both,
    		grid style={line width=.1pt, draw=gray!20},
    		major grid style={line width=.2pt, draw=gray!20},
    		minor tick num=1,
    		tick label style={font=\fontsize{3}{3}\selectfont},
    		xlabel style={font=\fontsize{6}{6}\selectfont,yshift=5pt},
    		ylabel style={font=\fontsize{6}{6}\selectfont,yshift=-20pt},
    		xlabel={Position (voxel)},
    		ylabel={Intensity},
    		tick align=outside,
    		tick pos=left,
    		axis on top=true,
    		legend style={
    			at={(1.1,1.05)},
    			anchor=north east,
    			font=\fontsize{3}{3}\selectfont,
    			row sep=1pt,
    			cells={anchor=west},
    			draw=black!20,
    			fill=white!90
    			},
	    		legend cell align={left},
	    		every axis plot/.append style={line width=1pt},
	    		]
	    		\fill[lightblue] (axis cs:26,-50) rectangle (axis cs:33.999,90000); 
	    		\fill[softpink] (axis cs:34,-50) rectangle (axis cs:42,90000);      
	    		
	    		\node[anchor=north west, font=\tiny\bfseries, text=blue!70!black] at (axis cs:26.5,17000) {Water};
	            \node[anchor=north west, font=\tiny\bfseries, text=strongpink] at (axis cs:34.5,17000) {Lung};
	    			
	    		\addplot[black, dashed, smooth] 
	    			table [x index=0, y index=1, col sep=comma] {images/txt_files/Stops905.txt};
	    			
	    		\addplot[red,  dotted, smooth] 
	    			table [x index=0, y index=1, col sep=comma] {images/txt_files/SMNS905.txt};
	    			
	    		\addplot[green!60!black, dotted, smooth] 
	    			table [x index=0, y index=1, col sep=comma] {images/txt_files/MyPRC905_whole.txt};
	    		\legend{GATE, SVTD, DDConv}
    		\end{axis}
    	\end{tikzpicture}%
    }

	\caption{
		Experiment 1---\Ac{PR} blurring experiment with the \ac{XCAT} phantom: \protect\subref{subfig:xcat_act} activity phantom, \protect\subref{subfig:xcat_mat} material phantom, \protect\subref{subfig:xcat_mc} annihilation image (\ac{MC} simulation), \protect\subref{subfig:xcat_svtd} \ac{SVTD}-blurred activity, \protect\subref{subfig:xcat_ddconv} \ac{DDConv}-blurred activity and \protect\subref{subfig:xcat_profiles} profiles across the green line.
	}
\end{figure}

We observe that the blurring of Lesion 1 and Lesion 2 is accurately achieved by both \ac{SVTD} and \ac{DDConv}. However, \ac{SVTD} fails to blur Lesion 3 and Lesion 4 accurately due to its inability to model \ac{PR} in heterogeneous regions, whereas \ac{DDConv} remains precise.

Analysis of the line profile further highlights these differences. \Ac{SVTD} exhibits moderate broadening due to \ac{PR} but shows reduced intensity in heterogeneous regions, indicating an underestimation of localized activity, while \ac{DDConv} nearly coincides with the \ac{MC} reference.

\subsection{Experiment 2: Reconstruction from MC-simulated Data}\label{sec:recon_xcat}

Reconstruction was performed on \ac{MC}-simulated data from the same activity phantom as in Section~\ref{sec:xcat_pr} (same lesion numbering) with  120 \ac{EM} iterations on a 200\texttimes{}200\texttimes{}100 voxel grid (2\texttimes{}2\texttimes{}2 mm\textsuperscript{3}). Three strategies were compared: no \ac{PRC}, \ac{SVTD} and the  proposed \ac{DDConv} approach. 
Figure~\ref{fig:xcat_Reconstructions} shows the reconstructed images at different iterations.

For lesions entirely located in homogeneous lung tissue (Lesion 1 and Lesion 2), both \ac{SVTD} and \ac{DDConv} produced similar results. In contrast, Lesion 4---located in heterogeneous tissues---was accurately reconstructed with \ac{DDConv}, while \ac{SVTD} failed to capture the lung component and the interface between the lung and the liver. These observations are validated by line profiles through Lesion 4 (Figure~\ref{fig:activity_profiles}). The reconstruction performance varies between water and lung regions. In the water region, the no-PRC reconstruction method recovers activity close to the \ac{GT}, whereas the \ac{SVTD} method tends to overestimate activity. In the lung region, both no-PRC and \ac{SVTD} reconstructions exhibit loss of activity, failing to capture the true signal. In contrast, the \ac{DDConv} reconstruction method consistently approximates the true activity in both regions, offering a stable recovery and a smoother transition at the interface between water and the lung.

\definecolor{lightblue}{HTML}{ADD8E6}
\definecolor{softpink}{HTML}{FFC0CB}
\definecolor{strongpink}{HTML}{FF69B4}
\definecolor{vividorange}{HTML}{FF7300}

\newcommand{\xcatImgSpy}[2]{
\begin{tikzpicture}[spy using outlines={circle, magnification=2, size=0.5cm, connect spies, vividorange}]
  \node (img) {\includegraphics[width=0.2\linewidth]{#1}};
  \node[text=vividorange, font=\footnotesize\bfseries, anchor=north west]
       at (img.north west) {iter. #2};
  \spy [vividorange] on ($(img.south east)+(-2.1,0.65)$)
       in node [left] at ($(img.north west)+(0.8,-1.3)$);
\end{tikzpicture}%
}

\newcommand{\xcatImgSpyLast}[2]{
\begin{tikzpicture}[spy using outlines={circle, magnification=2, size=0.5cm, connect spies, vividorange}]
  \node (img) {\includegraphics[width=0.2\linewidth]{#1}};
  \node[text=vividorange, font=\footnotesize\bfseries, anchor=north west]
       at (img.north west) {iter. #2};
  \spy [vividorange] on ($(img.south east)+(-2.1,0.65)$)
       in node [left] at ($(img.north west)+(0.8,-1.3)$);
  \draw[green, line width=0.5pt] (-0.43,-0.1) -- (-0.43,-0.35);
\end{tikzpicture}%
}

\newcommand{\xcatImgSpyLastWithCbar}[3]{%
\begin{tikzpicture}[spy using outlines={circle, magnification=2, size=0.5cm, connect spies, vividorange}]
  \node (imgbox) {%
    \begin{overpic}[width=0.2\linewidth]{#1}
      \put(101.5,-100){\makebox[0pt][l]{%
        \includegraphics[width=0.035\textwidth]{#3}%
      }}
    \end{overpic}%
  };
  \node[text=vividorange, font=\footnotesize\bfseries, anchor=north west]
       at (imgbox.north west) {iter. #2};
  \spy [vividorange] on ($(imgbox.south east)+(-2.1,0.65)$)
       in node [left] at ($(imgbox.north west)+(0.8,-1.3)$);
  \draw[green, line width=0.5pt] (-0.43,-0.1) -- (-0.43,-0.35);
\end{tikzpicture}%
}

\begin{figure*}[htbp]
\centering
\begin{minipage}[t]{0.9\textwidth}

  \begin{minipage}{0.03\linewidth}
    \centering
    \rotatebox{90}{\tiny  no-PRC}
  \end{minipage}%
  \begin{minipage}{0.95\linewidth}
    \centering
    \xcatImgSpy{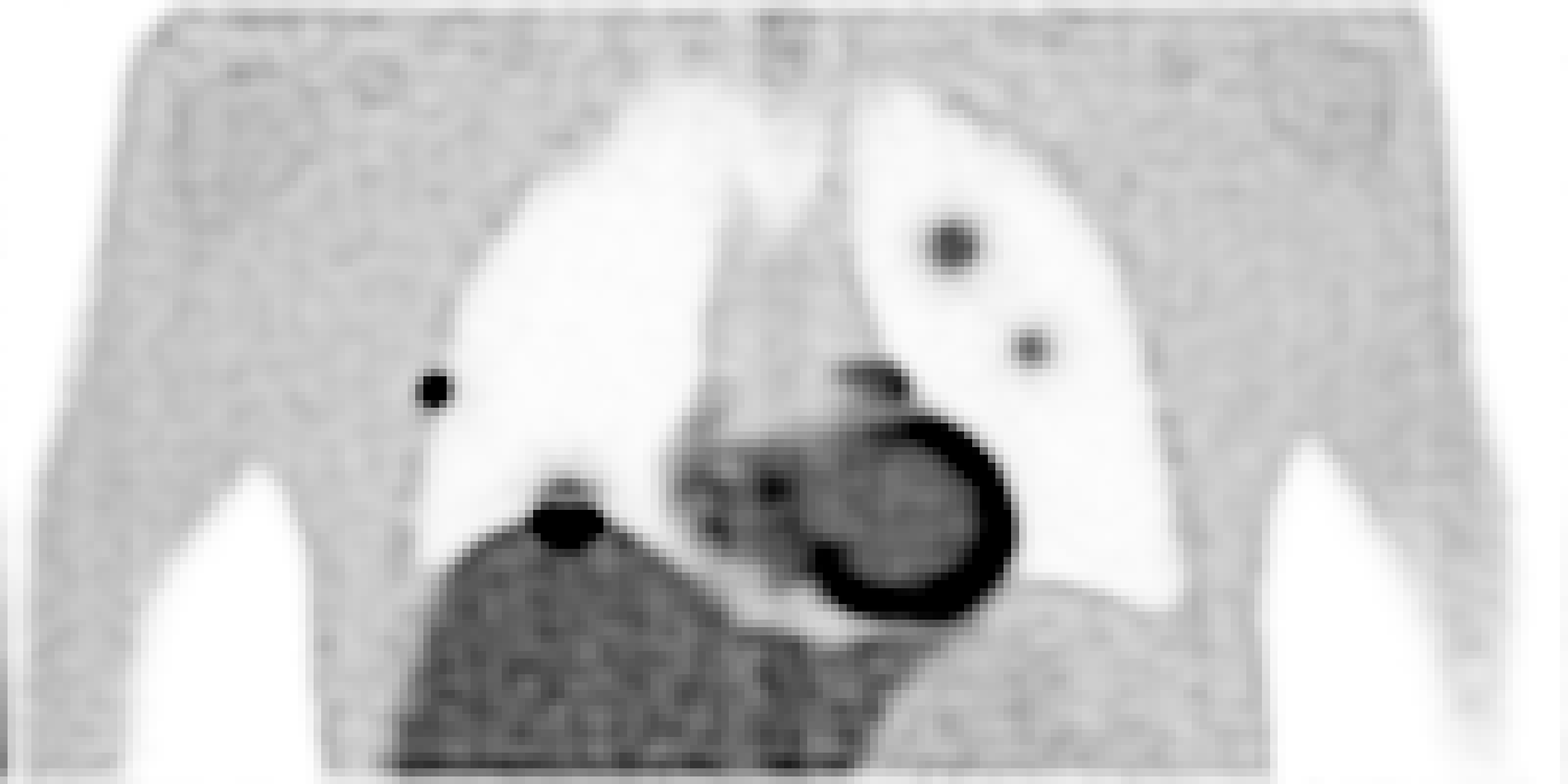}{20}%
    \xcatImgSpy{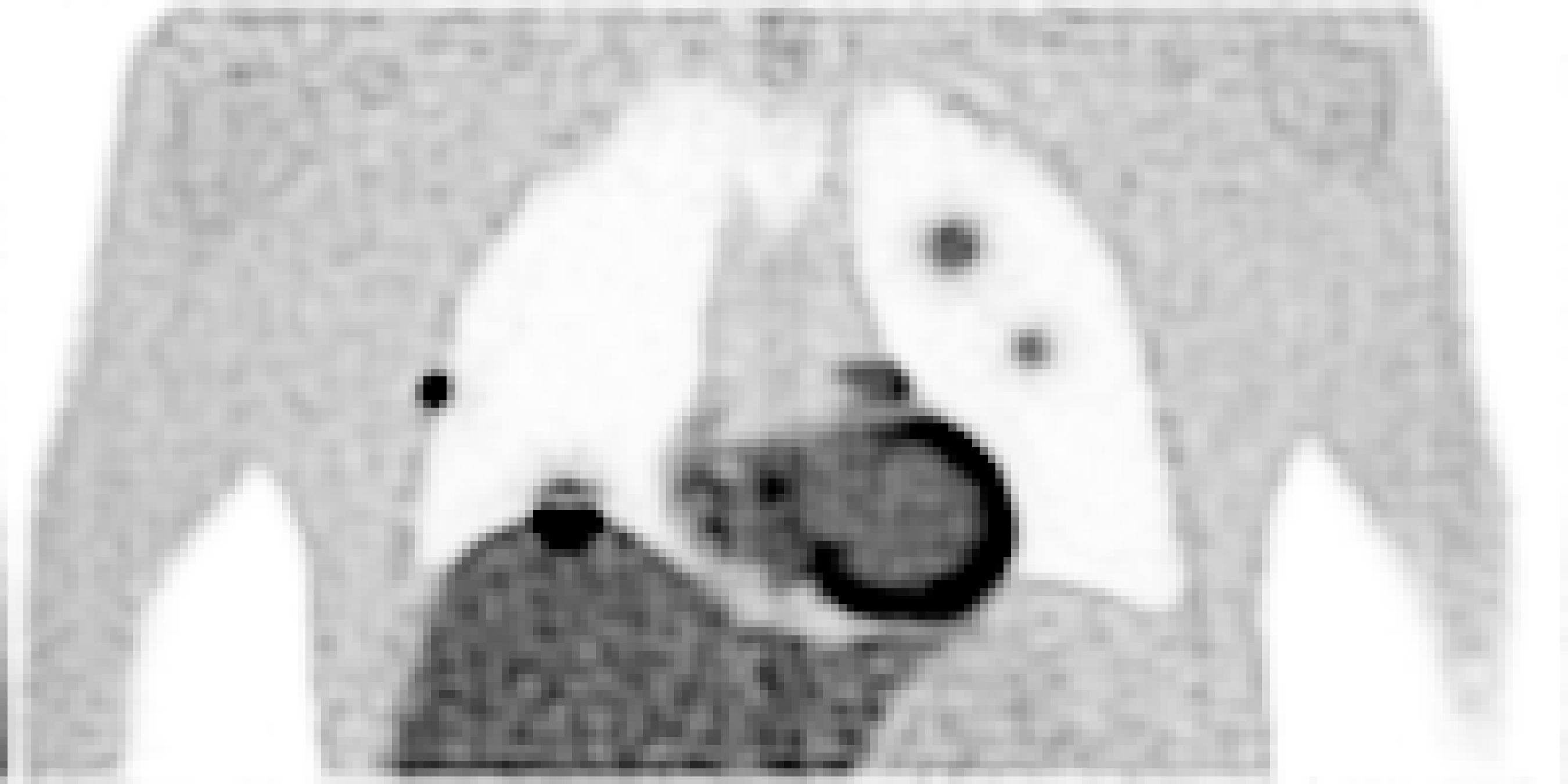}{40}%
    \xcatImgSpy{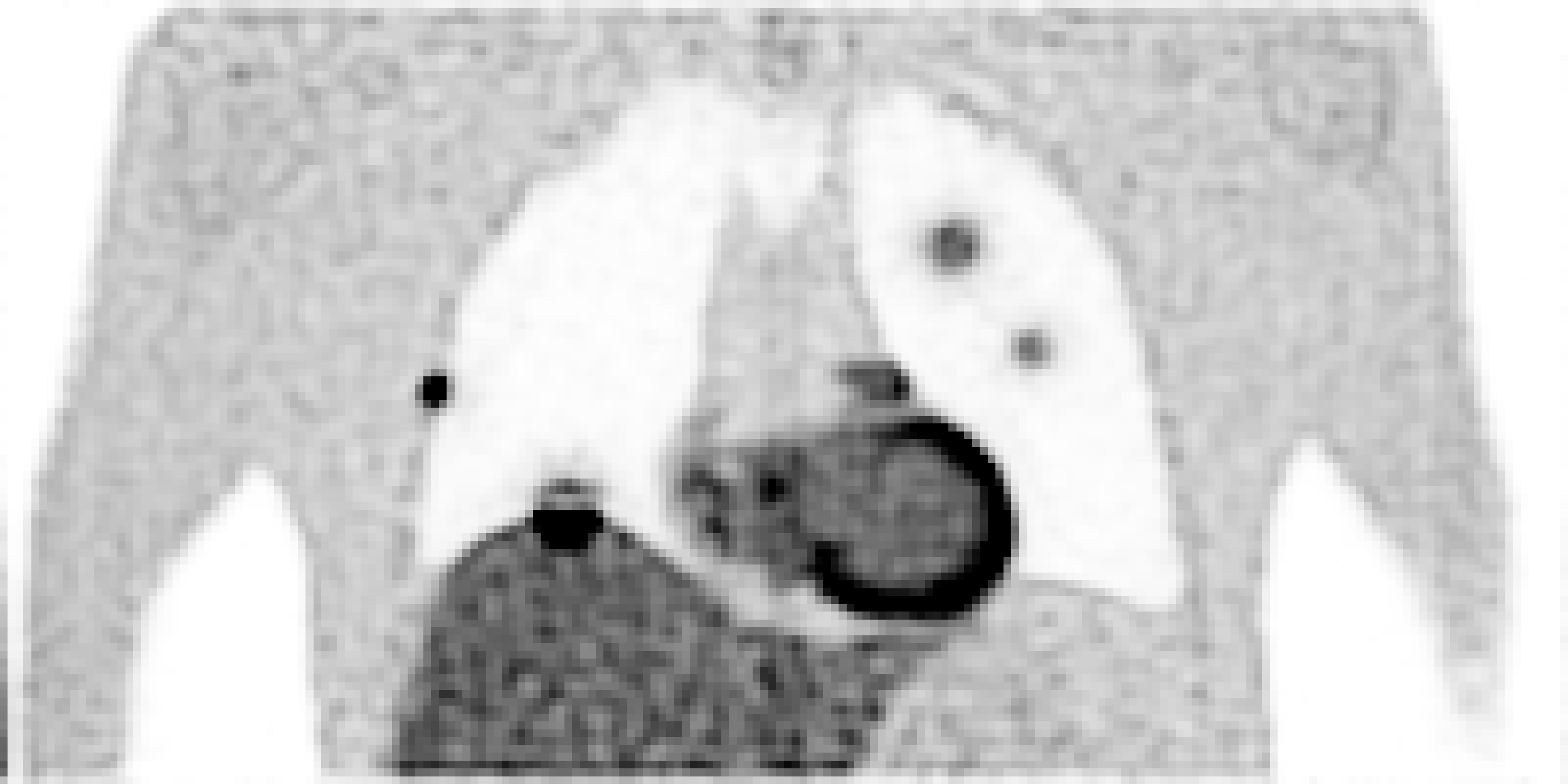}{60}%
    \xcatImgSpy{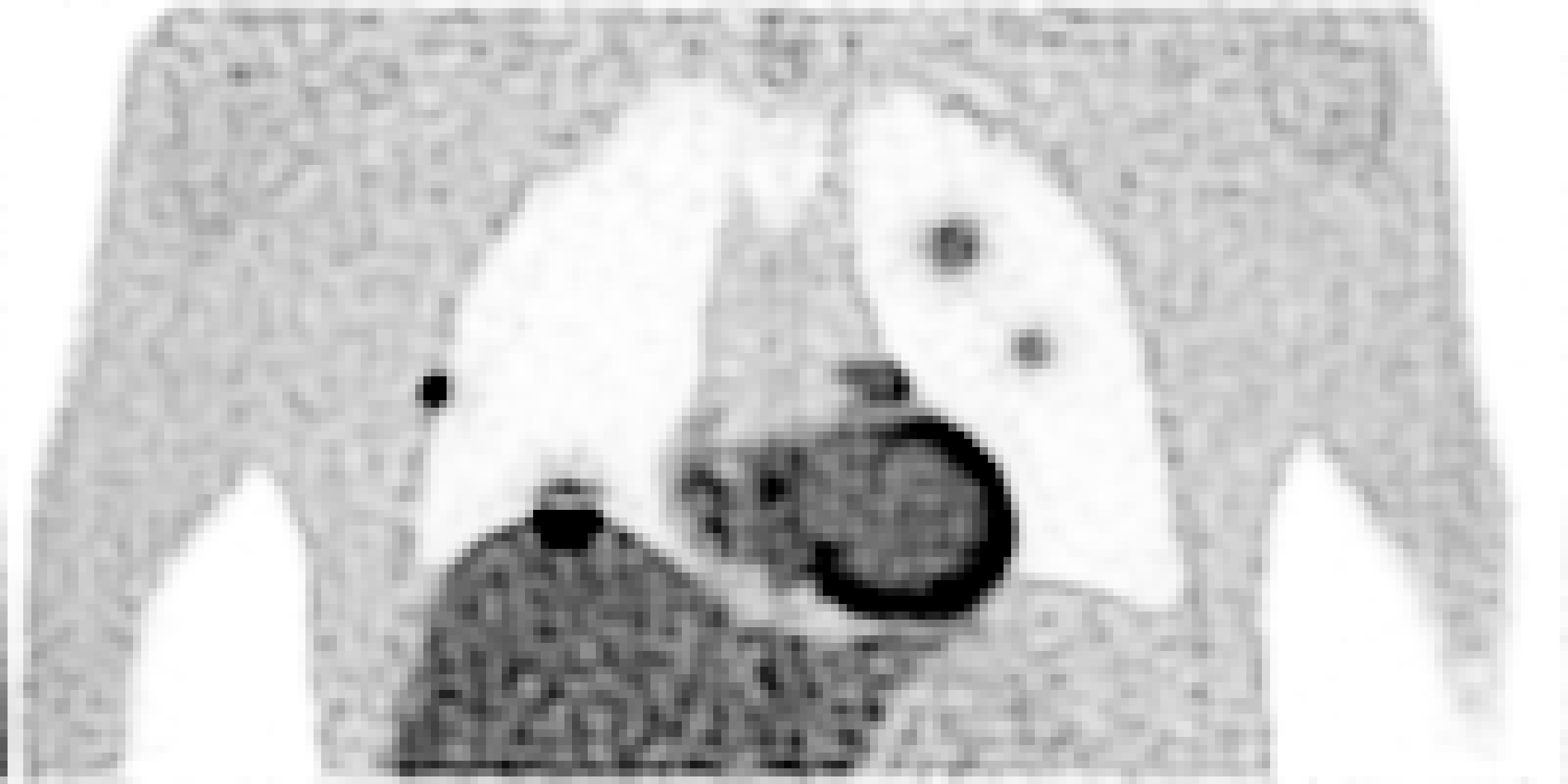}{80}%
    \xcatImgSpyLastWithCbar{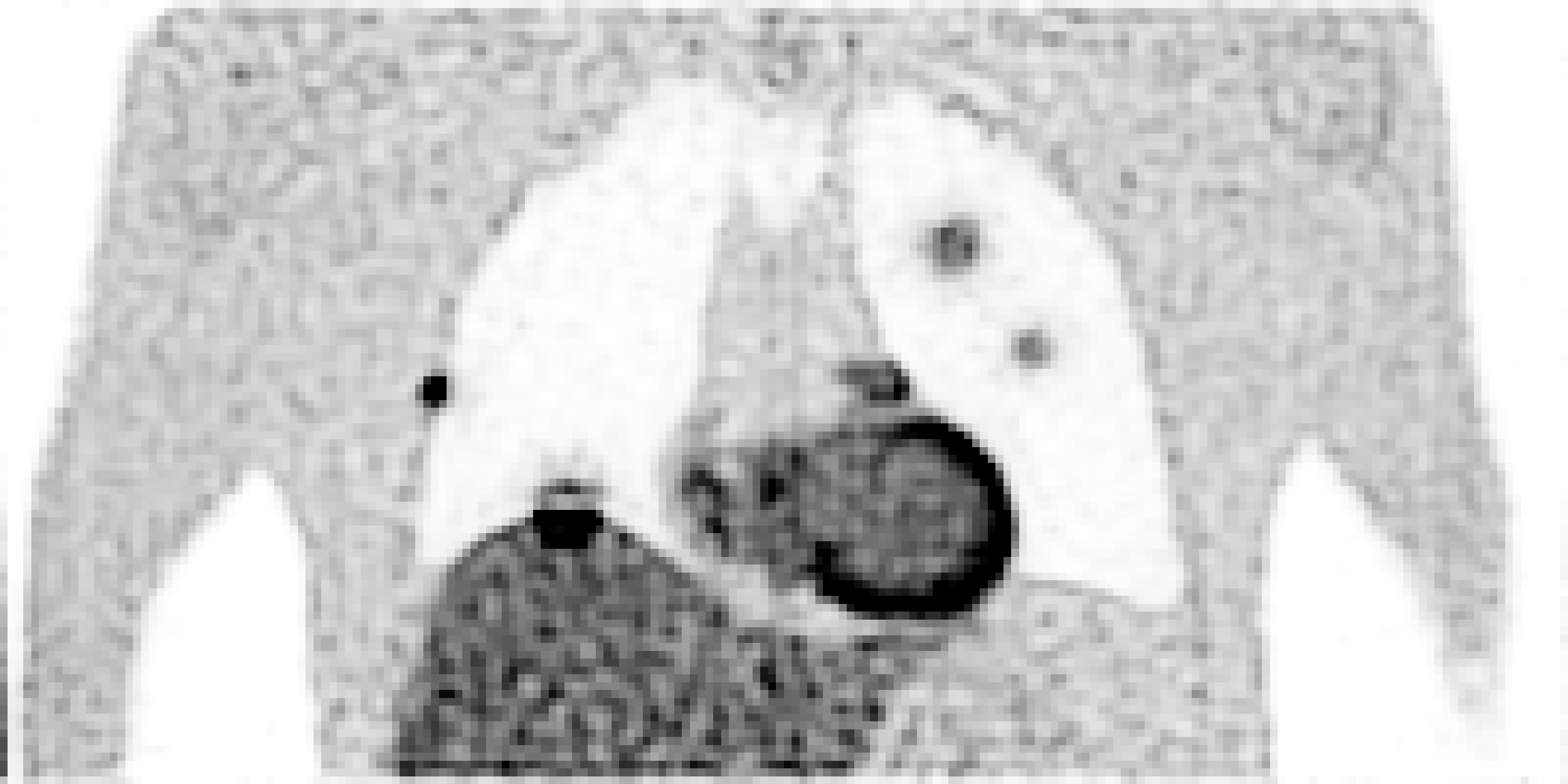}{120}{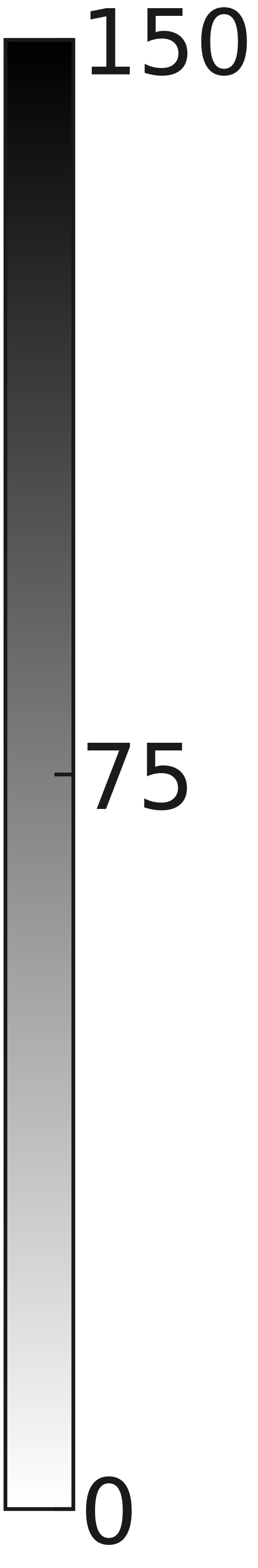}%
  \end{minipage}\\[3mm]

  \begin{minipage}{0.03\linewidth}
    \centering
    \rotatebox{90}{\tiny SVTD}
  \end{minipage}%
  \begin{minipage}{0.95\linewidth}
    \centering
    \xcatImgSpy{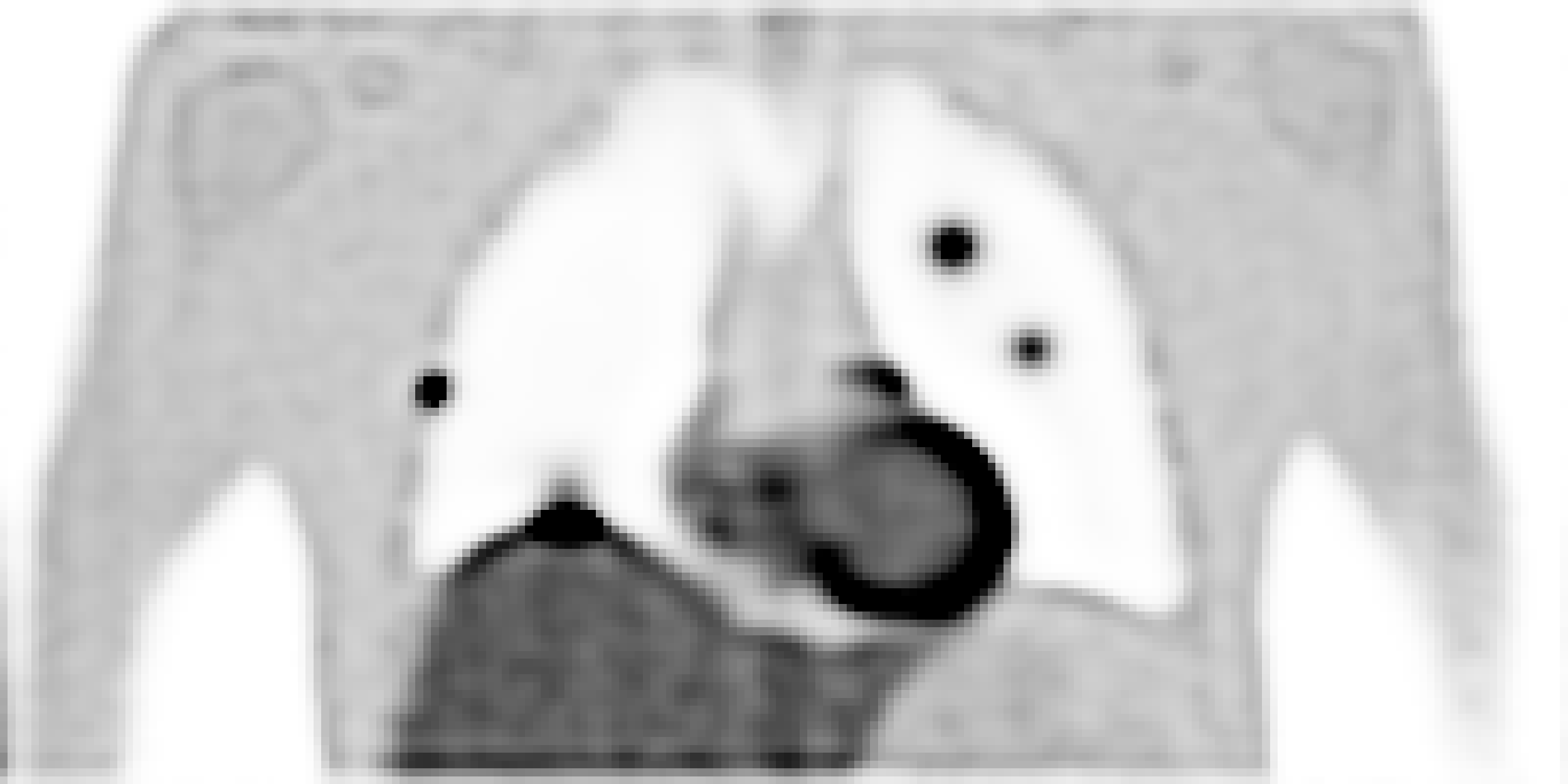}{20}%
    \xcatImgSpy{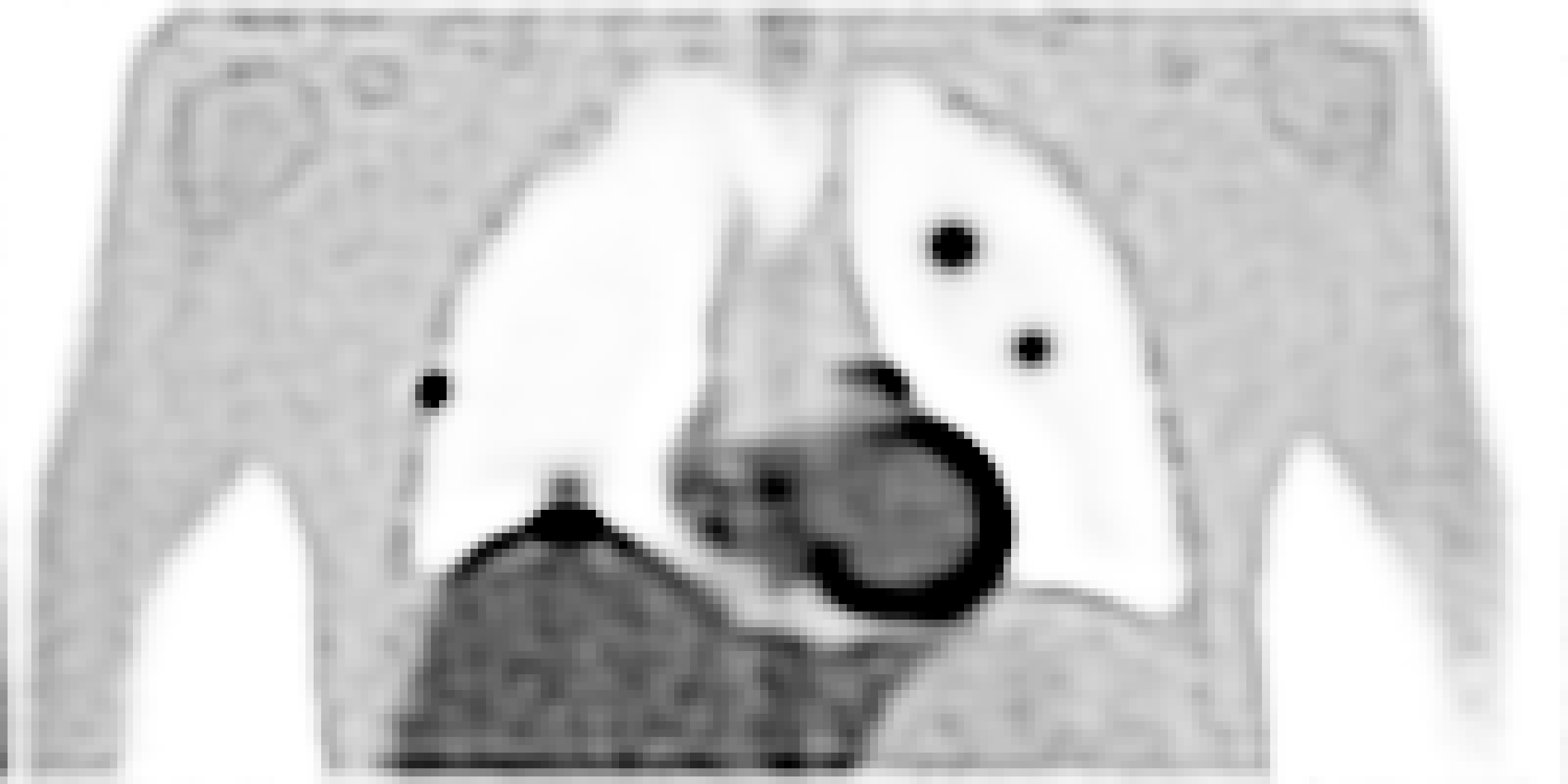}{40}%
    \xcatImgSpy{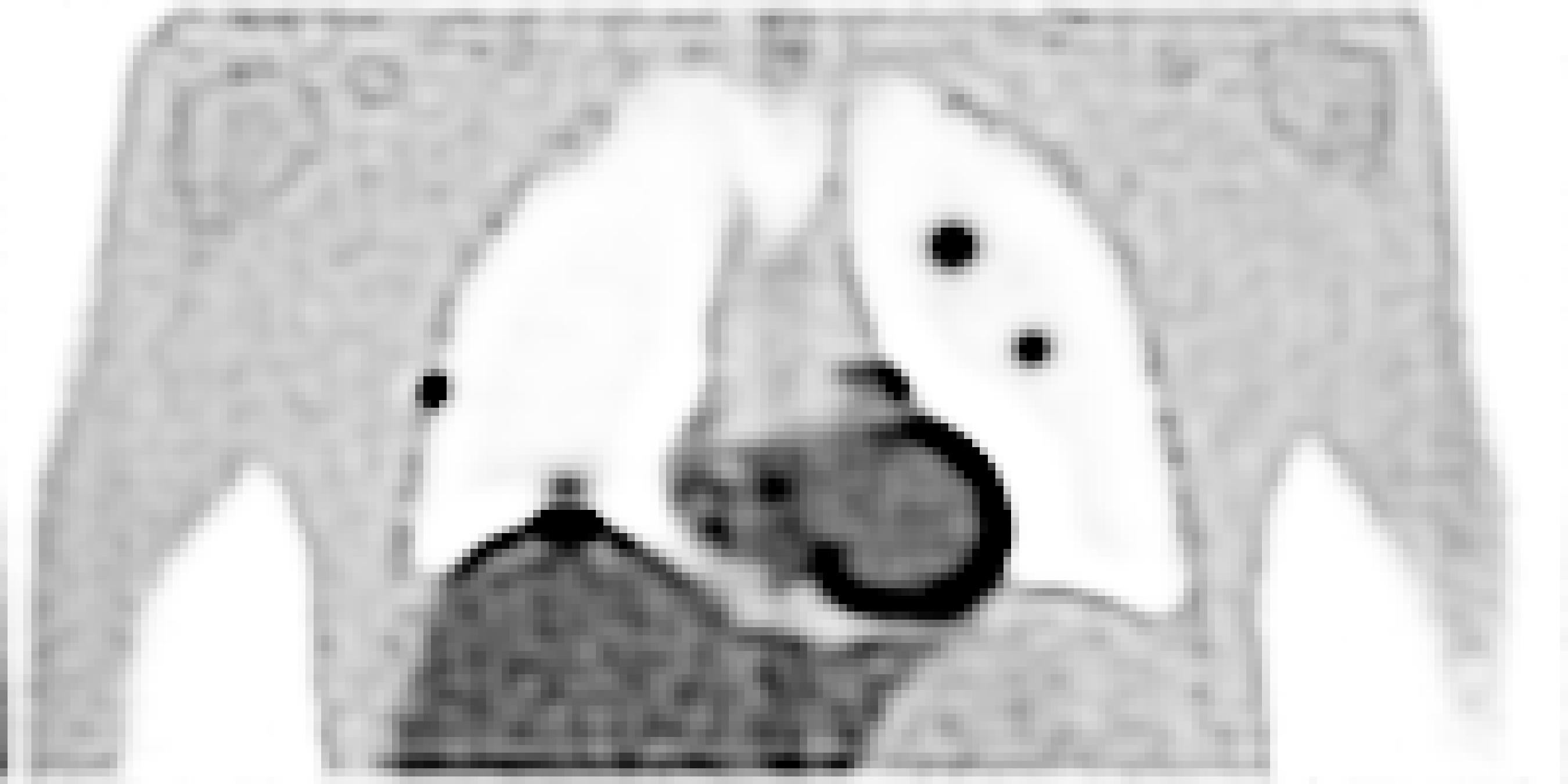}{60}%
    \xcatImgSpy{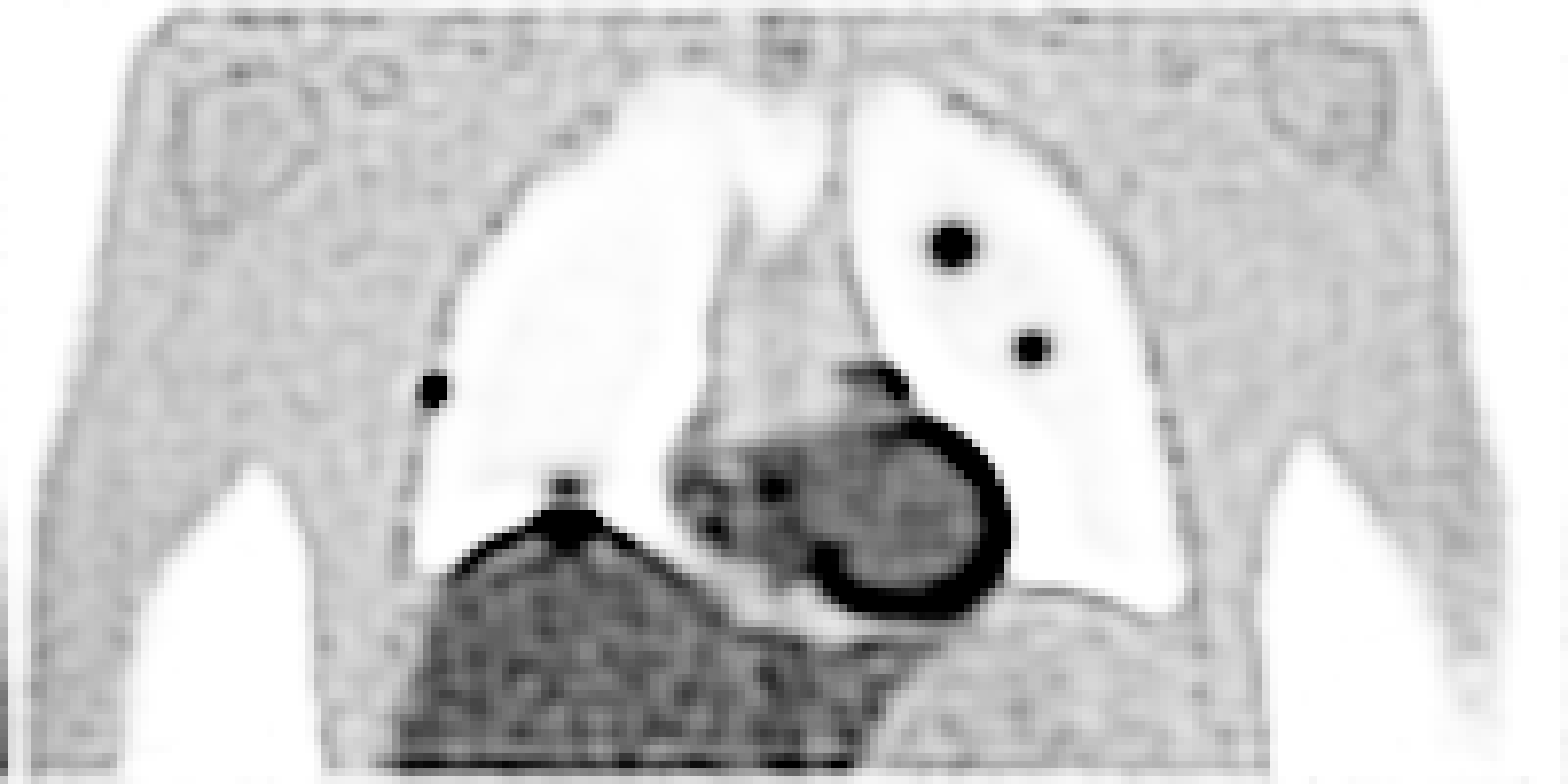}{80}%
    \xcatImgSpyLast{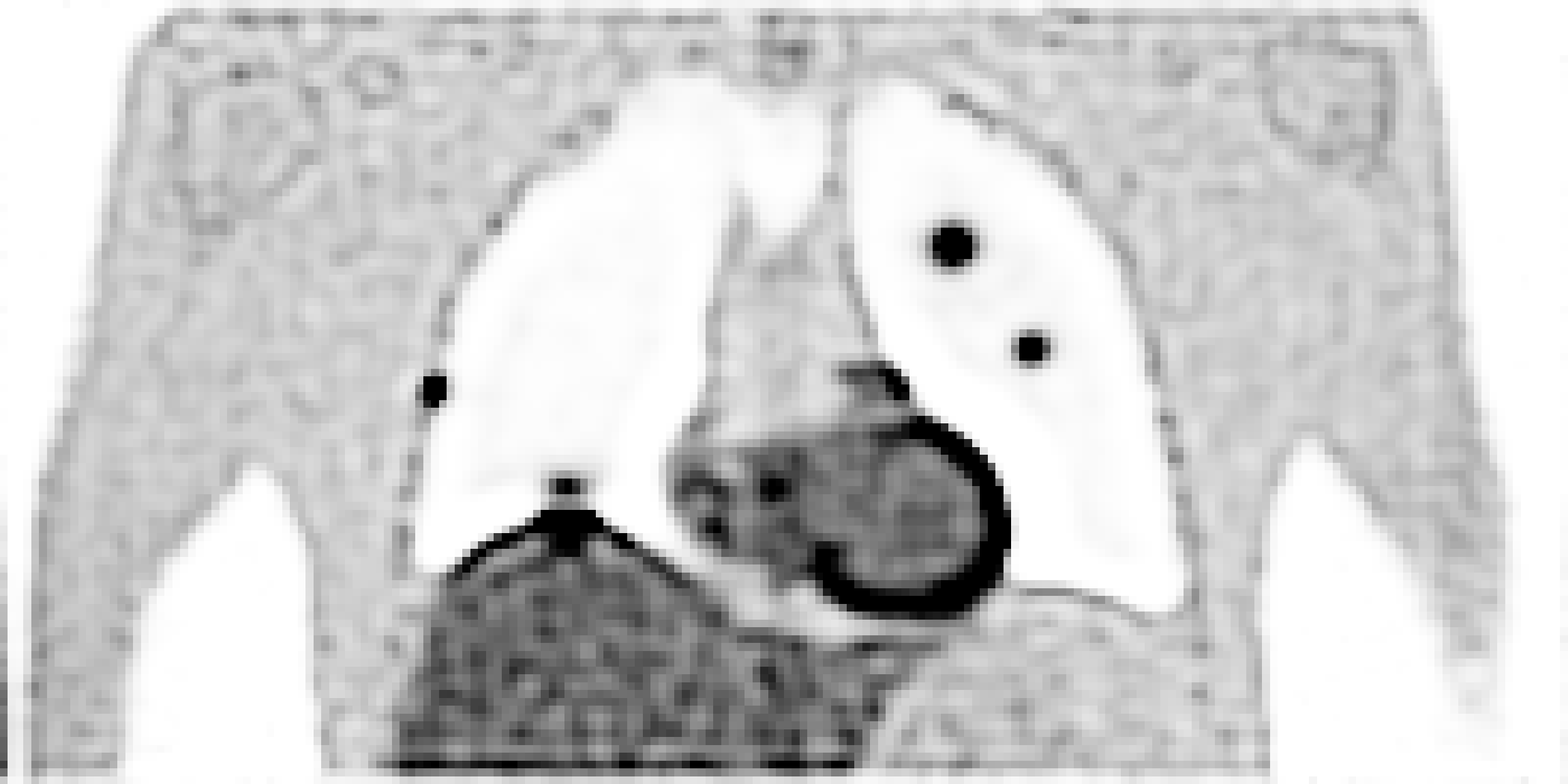}{120}%
  \end{minipage}\\[3mm]

  \begin{minipage}{0.03\linewidth}
    \centering
    \rotatebox{90}{\tiny DDConv}
  \end{minipage}%
  \begin{minipage}{0.95\linewidth}
    \centering
    \xcatImgSpy{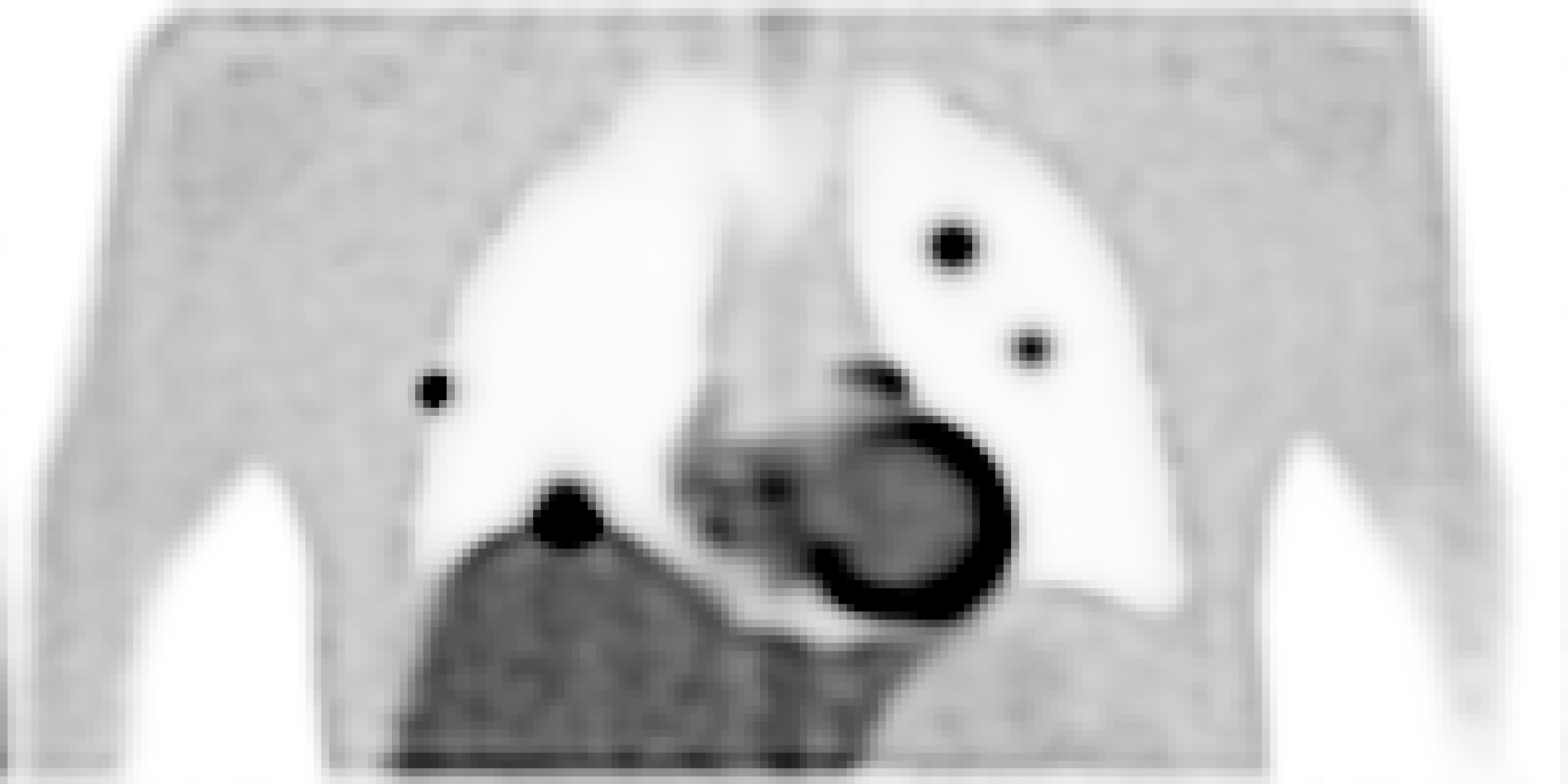}{20}%
    \xcatImgSpy{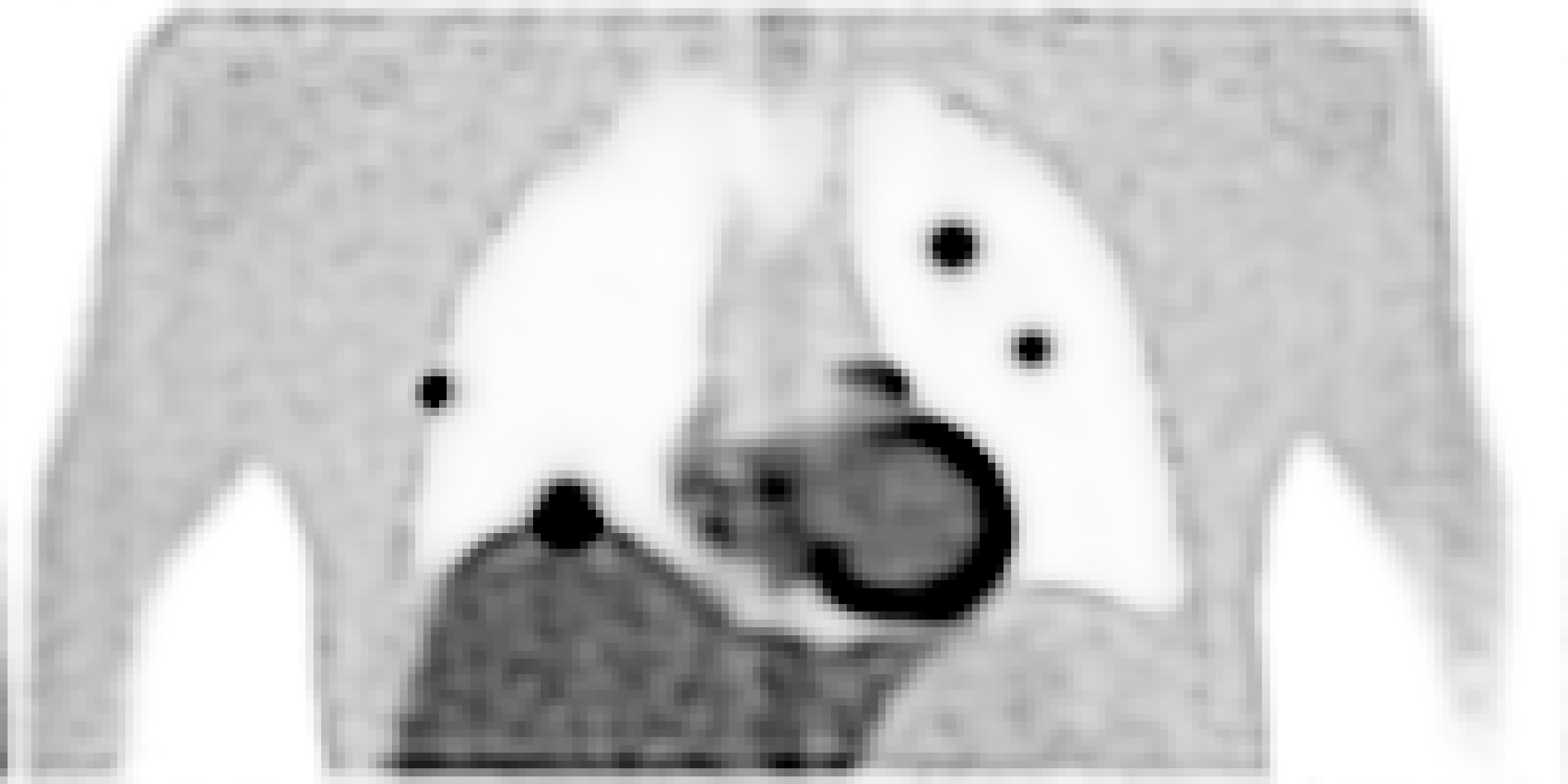}{40}%
    \xcatImgSpy{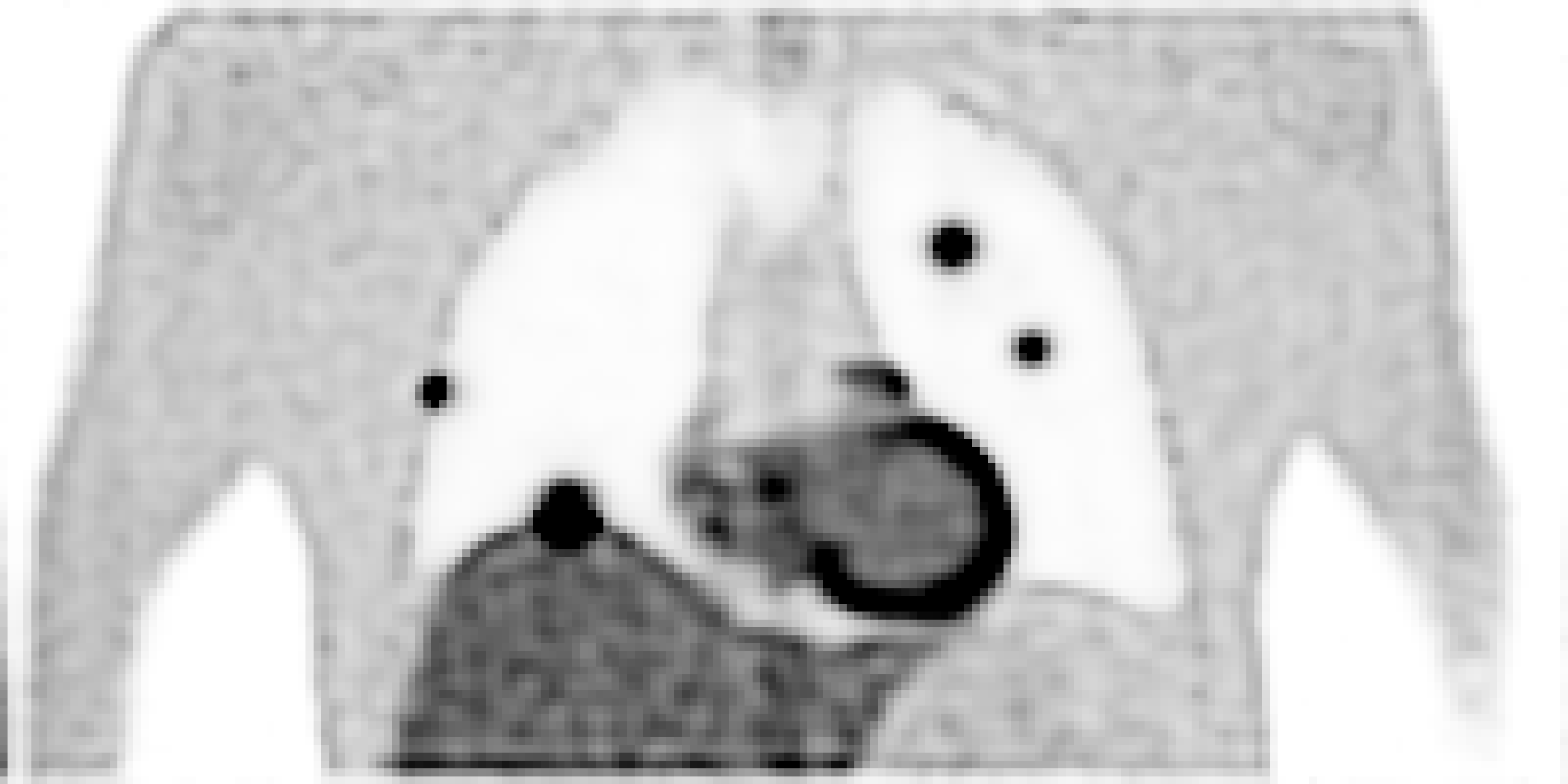}{60}%
    \xcatImgSpy{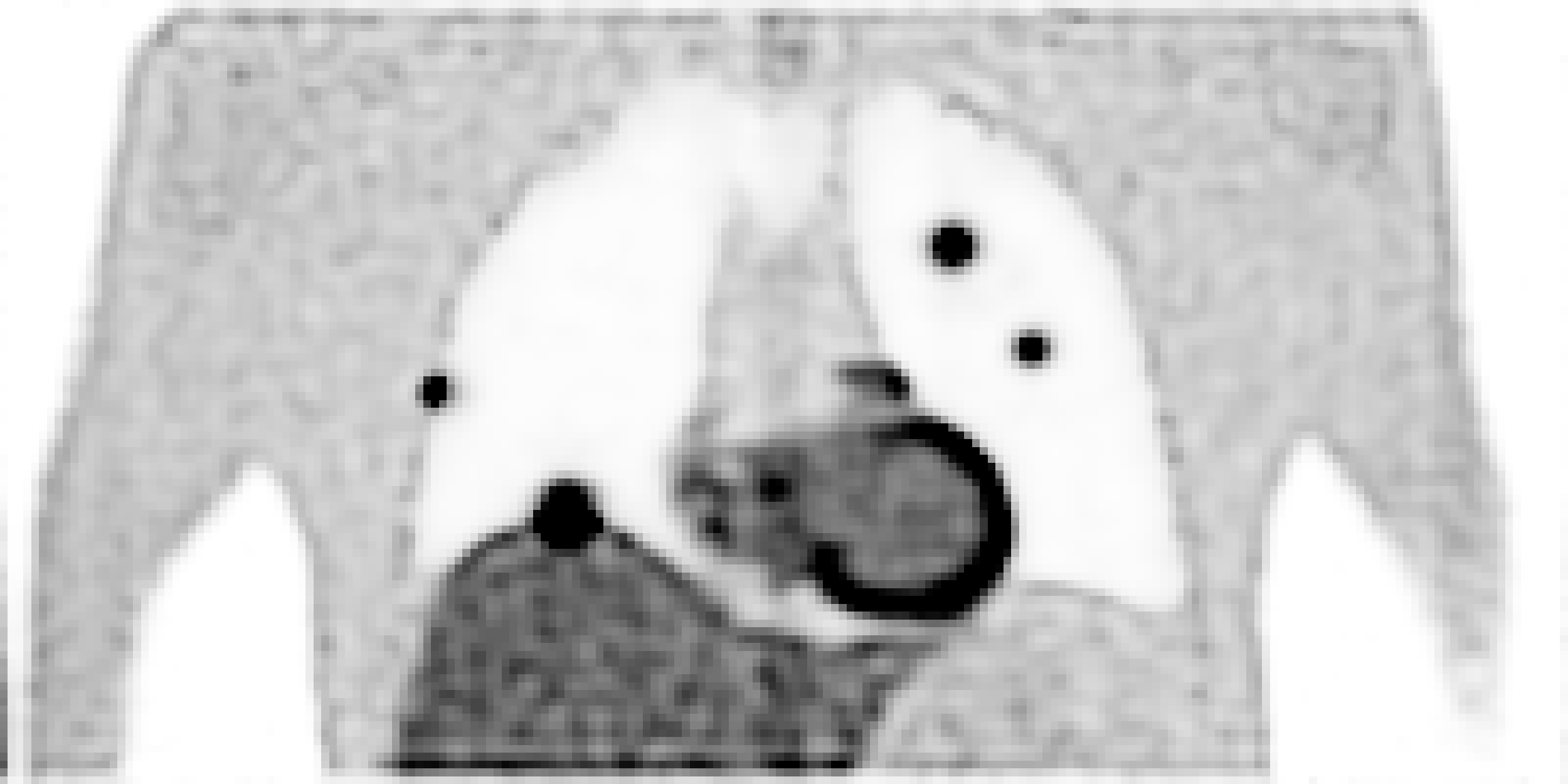}{80}%
    \xcatImgSpyLast{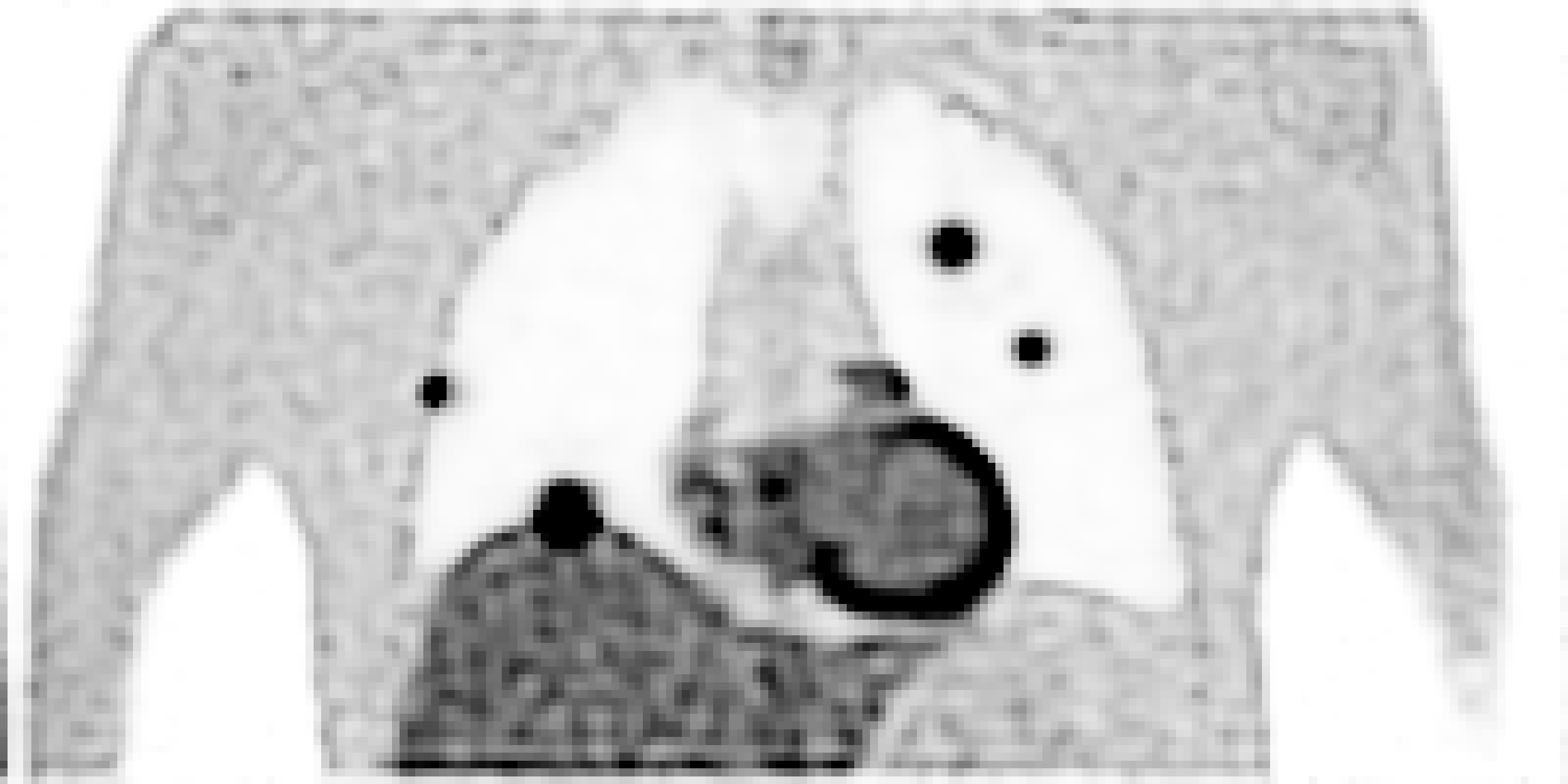}{120}%
  \end{minipage}

\end{minipage}

\caption{Experiment 2—Reconstructed images (\ac{MC}-simulated data) with no \ac{PRC}, \ac{SVTD}, and \ac{DDConv} at different iterations.}
\label{fig:xcat_Reconstructions}
\end{figure*}


\begin{figure}[htbp]
	\centering
	\begin{tikzpicture}
		\begin{axis}[
		    width=9cm,
		    height=6cm,
		    xlabel={Position (voxel)},
		    ylabel={Intensity},
		    xlabel style={font=\small},
		    ylabel style={font=\small,yshift=-7pt},
		    tick label style={font=\footnotesize},
		    xmin=26, xmax=42,
		    ymin=0, ymax=350,
		    grid=both,
		    grid style={line width=.1pt, draw=gray!20},
		    major grid style={line width=.2pt,draw=gray!20},
		    minor tick num=1,
		    legend style={
		        at={(0.98,0.98)},
		        anchor=north east,
		        font=\tiny
		    },
		    legend cell align={left},
		    every axis plot/.append style={line width=1pt},
		    axis on top=true,
		    tick align=outside,
		    tick pos=left,
			axis x line=bottom,
			axis y line=left,
			axis line style={draw=black},
		]
		
		\fill[lightblue] (axis cs:24,-50) rectangle (axis cs:33.999,350); 
		\fill[softpink] (axis cs:34,-50) rectangle (axis cs:46,350);      
		
		
		\node[anchor=north west, font=\small\bfseries, text=blue] at (axis cs:26.5,350) {Water};
		\node[anchor=north west, font=\small\bfseries, text=strongpink] at (axis cs:34,350) {Lung};
		
		\addplot[
		    color=black,
		    solid,
		    mark=*,
		    mark size=1.5pt,
		    mark options={solid, fill=black}
		] table[col sep=comma] {images/txt_files/activity_profile_x71.txt};
		
		\addplot[
		    color=blue!70!black,
		    dotted,
		    line width=1.2pt,
		    mark=triangle*,
		    mark size=1.8pt,
		    mark options={solid, fill=blue!70!black, rotate=180}
		] table[col sep=comma] {images/txt_files/nopr_profile_x71.txt};
		
		\addplot[
		    color=red!80!black,
		    dashed,
		    mark=square*,
		    mark size=1.5pt,
		    mark options={solid, fill=red!80!black}
		] table[col sep=comma] {images/txt_files/smns_profile_x71.txt};
		
		\addplot[
		    color=green!60!black,
		    dashdotted,
		    mark=diamond*,
		    mark size=2pt,
		    mark options={solid, fill=green!60!black}
		] table[col sep=comma] {images/txt_files/my_prc_profile_x71.txt};
		
		\legend{Activity (True), No PRC, \ac{SVTD}, \ac{DDConv}}
	\end{axis}
\end{tikzpicture}
\caption{Experiment 2---Line profiles of the reconstructed images  (\ac{MC}-simulated data, 120 EM iterations, cf. the green line in Figure~\ref{fig:xcat_Reconstructions}) through Lesion 4   with no \ac{PRC}, \ac{SVTD} and \ac{DDConv}, at the interface between the water (light blue) and lung (soft pink) regions.}
\label{fig:activity_profiles}
\end{figure}
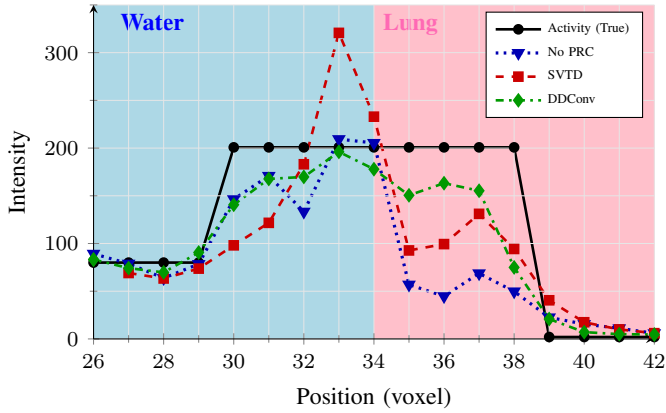

Figure~\ref{fig:metrics_with_axislabels} summarizes the evolution of the metrics described in Section~\ref{sec:setup} at each iteration for the four lesions. In the homogeneous lung regions (Lesions~1 and 2), the \ac{RC} and \ac{MAPE} curves of \ac{SVTD} and \ac{DDConv} almost overlap, both increasing smoothly before stabilizing after about ten iterations. \ac{RC} values remain around 0.5--0.6 and \ac{MAPE} around 45--50\%. The \ac{SUV}\textsubscript{max} error also converges to similar values for both methods (approximately 60--80\%), while no \ac{PRC} stays consistently lower for \ac{RC} and higher for \ac{MAPE}, confirming lower quantitative performance. These results show that \ac{SVTD} and \ac{DDConv} achieve comparable quantification when tissue properties are uniform.

At the tissue interfaces (Lesions~3 and 4), clearer differences appear. The \ac{SVTD} method produces over-enhancement, with \ac{RC} values exceeding 1.0 and \ac{SUV}\textsubscript{max} errors above 200\%. In contrast, \ac{DDConv} maintains \ac{RC} values close to 1.0 and limits SUV\textsubscript{max} errors below 120\%, avoiding over-correction at material boundaries. \ac{DDConv} also provides the lowest \ac{MAPE}, between 35\% and 45\%, while \ac{SVTD} and no \ac{PRC} yield higher voxelwise errors. Overall, \ac{DDConv} demonstrates the most stable and reliable quantification across all regions, maintaining accurate recovery in homogeneous tissues and preventing overestimation at heterogeneous interfaces.

\pgfplotsset{set layers=axis on top}

\definecolor{ddgreen}{RGB}{0,158,115}
\definecolor{svorange}{RGB}{230,159,0}
\definecolor{nored}{HTML}{d52900}

\newlength\axish
\setlength{\axish}{3.1cm}

\pgfplotsset{
	myaxis/.style={
		width=0.125\textwidth,
		height=0.7\axish,
		xlabel style={font=\scriptsize, at={(0.5,0)}, yshift=1.2ex},
		y label style={at={(0.25,0.5)}, anchor=center, font=\scriptsize\bfseries, align=center},
		tick label style={font=\scriptsize},
		unbounded coords=discard,
		scale only axis=true,
		enlargelimits=false
	}
}

\newcommand{\epsRC}{0.03}
\newcommand{\epsSUV}{5.0}
\newcommand{\epsMAPE}{4.0}

\newcommand{\dRC}{0.04}
\newcommand{\dSUV}{4.0}
\newcommand{\dMAPE}{3.5}

\makeatletter
\newcommand{\metricEps}[1]{%
	\ifx#1RC\epsRC\else
	\ifx#1SUVmaxErr\epsSUV\else
	\ifx#1MAPE\epsMAPE\else 0.0\fi
	\fi
	\fi}
\newcommand{\metricDel}[1]{%
	\ifx#1RC\dRC\else
	\ifx#1SUVmaxErr\dSUV\else
	\ifx#1MAPE\dMAPE\else 0.0\fi
	\fi
	\fi}
\makeatother

\newcommand{\dashOn}{3pt}
\newcommand{\dashOff}{1.2pt}
\newcommand{\markszSmall}{1.2pt}
\newcommand{\markszStar}{1.6pt}
\newcommand{\markszRed}{1.6pt}
\newcommand{\preactionWidth}{1.4pt}

\newcommand{\plotNoPRC}[2]{%
	\addplot[
	nored, line width=1.2pt,
	dash pattern=on \dashOn\space off \dashOff,
	line cap=butt, line join=round,
	mark=diamond*, mark size=\markszRed,
	mark options={draw=nored, fill=white, line width=0.7pt},
	mark repeat=2
	] table [x=iteration, y=#1_NoPRC, col sep=comma] {#2};}

\newcommand{\plotDDConvOffset}[2]{%
	\addplot[
	ddgreen, line width=1.1pt,
	dash pattern=on 0.8pt off 1.6pt,
	line cap=round, line join=round,
	mark=*, mark size=\markszSmall,
	mark options={draw=ddgreen, fill=ddgreen, line width=0.6pt},
	mark repeat=2,
	preaction={draw=white, line width=1.6pt, opacity=0.8}
	] table [
	x=iteration,
	y expr={(\thisrow{#1_MyPRC}) +
		(abs(\thisrow{#1_SMNS}-\thisrow{#1_MyPRC}) < \metricEps{#1} ? -\metricDel{#1} : 0)},
	col sep=comma] {#2};}

\newcommand{\plotSVTDOffset}[2]{%
	\addplot[
	svorange, line width=1.05pt,
	dash pattern=on \dashOn\space off \dashOff,
	dash phase=1.5pt,
	line cap=round, line join=round,
	mark=star*, mark size=\markszStar,
	mark options={draw=white, fill=svorange, line width=0.5pt},
	mark repeat=2,
	preaction={draw=white, line width=\preactionWidth}
	] table [
	x=iteration,
	y expr={(\thisrow{#1_SMNS}) +
		(abs(\thisrow{#1_SMNS}-\thisrow{#1_MyPRC}) < \metricEps{#1} ?  \metricDel{#1} : 0)},
	col sep=comma] {#2};}

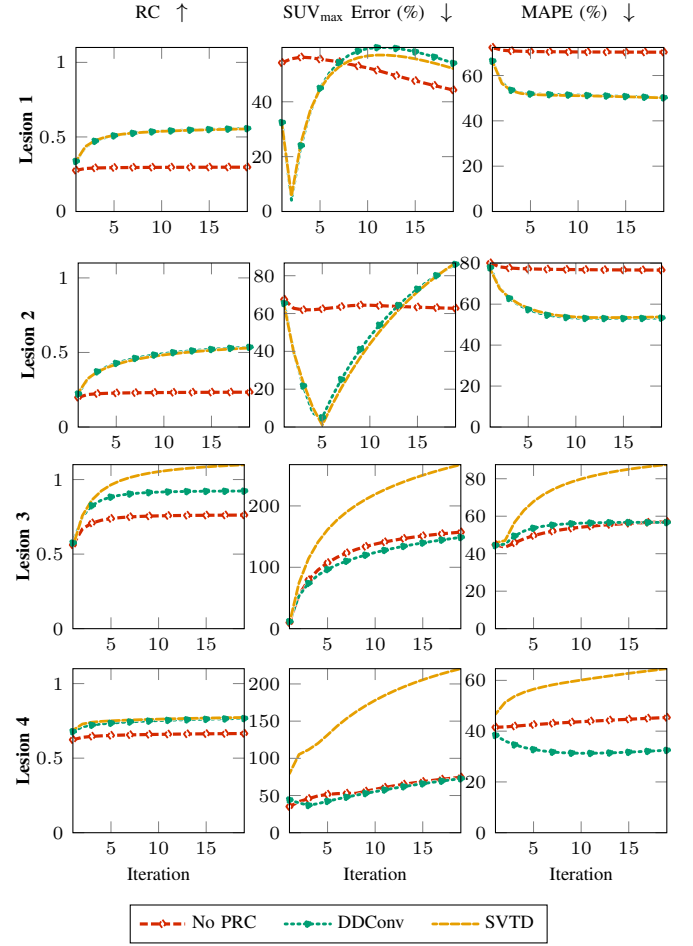
\begin{figure}[htbp]
	\centering
	
	\newcommand{\sphfile}{images/KPIs_txt/sphere1.txt}
	\begin{tikzpicture}
		\begin{axis}[myaxis, ymin=0, ymax=1.1,
			ylabel={Lesion 1},
			title={\scriptsize RC \, {\footnotesize$\uparrow$}},
			title style={at={(0.5,1)}, anchor=south, font=\scriptsize},
			legend to name=globallegend, legend columns=3,
			legend style={/tikz/every even column/.style={column sep=0.8em}}]
			\plotNoPRC{RC}{\sphfile}
			\plotDDConvOffset{RC}{\sphfile}
			\plotSVTDOffset{RC}{\sphfile}
			\addlegendentry{\scriptsize No \ac{PRC}}
			\addlegendentry{\scriptsize \ac{DDConv}}
			\addlegendentry{\scriptsize \ac{SVTD}}
		\end{axis}
	\end{tikzpicture}\hspace{-0.3em}%
	\begin{tikzpicture}
		\begin{axis}[myaxis, ymin=0,
			title={\scriptsize SUV$_{\text{max}}$ Error (\%) \, {\footnotesize$\downarrow$}},
			title style={at={(0.5,1)}, anchor=south, font=\scriptsize}]
			\plotNoPRC{SUVmaxErr}{\sphfile}
			\plotDDConvOffset{SUVmaxErr}{\sphfile}
			\plotSVTDOffset{SUVmaxErr}{\sphfile}
		\end{axis}
	\end{tikzpicture}\hspace{-0.3em}%
	\begin{tikzpicture}
		\begin{axis}[myaxis, ymin=0,
			title={\scriptsize MAPE (\%) \, {\footnotesize$\downarrow$}},
			title style={at={(0.5,1)}, anchor=south, font=\scriptsize}]
			\plotNoPRC{MAPE}{\sphfile}
			\plotDDConvOffset{MAPE}{\sphfile}
			\plotSVTDOffset{MAPE}{\sphfile}
		\end{axis}
	\end{tikzpicture}
	
	\vspace{0.25em}
	
	\renewcommand{\sphfile}{images/KPIs_txt/sphere2.txt}
	\begin{tikzpicture}
		\begin{axis}[myaxis, ymin=0, ymax=1.1,
			ylabel={Lesion 2}]
			\plotNoPRC{RC}{\sphfile}\plotDDConvOffset{RC}{\sphfile}\plotSVTDOffset{RC}{\sphfile}
		\end{axis}
	\end{tikzpicture}\hspace{-0.3em}%
	\begin{tikzpicture}
		\begin{axis}[myaxis, ymin=0]
			\plotNoPRC{SUVmaxErr}{\sphfile}\plotDDConvOffset{SUVmaxErr}{\sphfile}\plotSVTDOffset{SUVmaxErr}{\sphfile}
		\end{axis}
	\end{tikzpicture}\hspace{-0.3em}%
	\begin{tikzpicture}
		\begin{axis}[myaxis, ymin=0]
			\plotNoPRC{MAPE}{\sphfile}\plotDDConvOffset{MAPE}{\sphfile}\plotSVTDOffset{MAPE}{\sphfile}
		\end{axis}
	\end{tikzpicture}
	
	\vspace{0.25em}
	
	\renewcommand{\sphfile}{images/KPIs_txt/sphere3.txt}
	\begin{tikzpicture}
		\begin{axis}[myaxis, ymin=0, ymax=1.1,
			ylabel={Lesion 3}]
			\plotNoPRC{RC}{\sphfile}\plotDDConvOffset{RC}{\sphfile}\plotSVTDOffset{RC}{\sphfile}
		\end{axis}
	\end{tikzpicture}\hspace{-0.3em}%
	\begin{tikzpicture}
		\begin{axis}[myaxis, ymin=0]
			\plotNoPRC{SUVmaxErr}{\sphfile}\plotDDConvOffset{SUVmaxErr}{\sphfile}\plotSVTDOffset{SUVmaxErr}{\sphfile}
		\end{axis}
	\end{tikzpicture}\hspace{-0.3em}%
	\begin{tikzpicture}
		\begin{axis}[myaxis, ymin=0]
			\plotNoPRC{MAPE}{\sphfile}\plotDDConvOffset{MAPE}{\sphfile}\plotSVTDOffset{MAPE}{\sphfile}
		\end{axis}
	\end{tikzpicture}
	
	\vspace{0.25em}
	
	\renewcommand{\sphfile}{images/KPIs_txt/sphere4.txt}
	\begin{tikzpicture}
		\begin{axis}[myaxis, ymin=0, ymax=1.1,
			ylabel={Lesion 4},
			xlabel={\scriptsize Iteration}]
			\plotNoPRC{RC}{\sphfile}\plotDDConvOffset{RC}{\sphfile}\plotSVTDOffset{RC}{\sphfile}
		\end{axis}
	\end{tikzpicture}\hspace{-0.3em}%
	\begin{tikzpicture}
		\begin{axis}[myaxis, ymin=0,
			xlabel={\scriptsize Iteration}]
			\plotNoPRC{SUVmaxErr}{\sphfile}\plotDDConvOffset{SUVmaxErr}{\sphfile}\plotSVTDOffset{SUVmaxErr}{\sphfile}
		\end{axis}
	\end{tikzpicture}\hspace{-0.3em}%
	\begin{tikzpicture}
		\begin{axis}[myaxis, ymin=0,
			xlabel={\scriptsize Iteration}]
			\plotNoPRC{MAPE}{\sphfile}\plotDDConvOffset{MAPE}{\sphfile}\plotSVTDOffset{MAPE}{\sphfile}
		\end{axis}
	\end{tikzpicture}
	
	\par\vspace{2mm}
	\pgfplotslegendfromname{globallegend}
	
	\caption{Evolution of three metrics (columns) for each lesion (rows) during reconstruction. 
		Axis labels indicate lesion number (rows) and iteration (bottom). 
		}
	\label{fig:metrics_with_axislabels}
\end{figure}

\subsection{Experiment 3: Reconstruction from Real Phantom Data}\label{sec:exp3}

\begin{figure}
	\subfloat[CT image with profile line \label{subfig:ct_image1}]{%
	\includegraphics[width=0.48\linewidth]{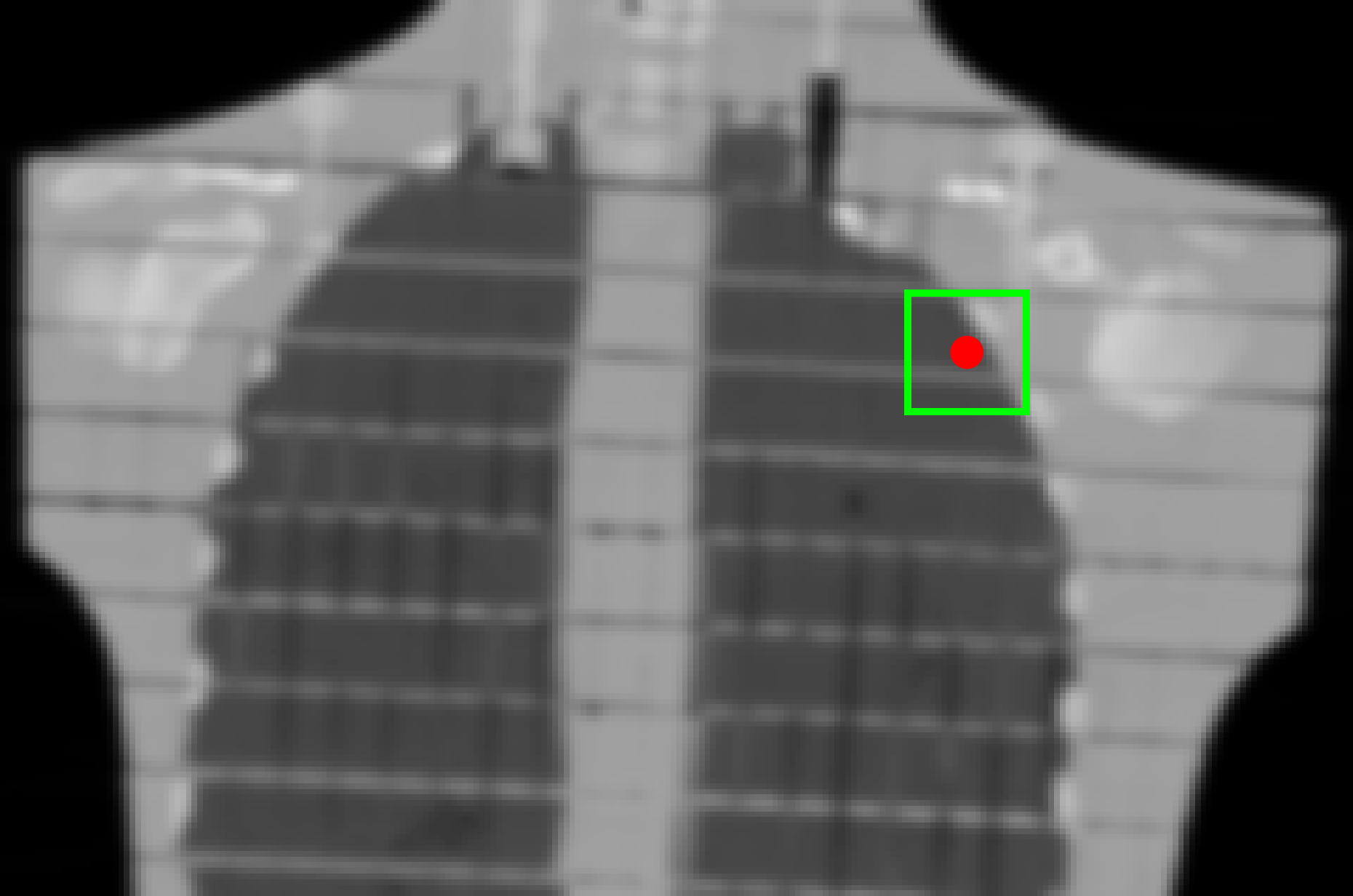}%
	}\hfill
	\subfloat[Line profiles across tumor region \label{subfig:profiles1}]{%
	\begin{tikzpicture}
		\begin{axis}[
		    width=0.6\linewidth,
		    height=0.45\linewidth,
		    axis x line=bottom,
		    axis y line=left,
		    xmin=0, xmax=31,
		    ymin=0, ymax=42295032,
		    clip marker paths=false,
		    grid=both,
		    grid style={line width=.1pt, draw=gray!20},
			major grid style={line width=.2pt,draw=gray!20},
		    minor tick num=1,
		    tick label style={font=\fontsize{5}{5}\selectfont},
		    xticklabel style={yshift=-1pt},
		    xlabel={Position (voxel)},
		    ylabel={Intensity},
		    xlabel style={font=\fontsize{6}{6}\selectfont,yshift=3pt},
		    ylabel style={font=\fontsize{6}{6}\selectfont,yshift=-20pt,},
		    tick align=outside,
		    tick pos=left,
		    axis on top=true,
		    legend style={
		        at={(1,1)},
		        anchor=north east,
		        font=\fontsize{5}{5}\selectfont,
		        row sep=0.5pt,
		        cells={anchor=west},
				inner sep=1pt,
		        outer sep=0pt
		    },
		    legend cell align={left},
		    every axis plot/.append style={line width=1pt},
		]
		\fill[softpink] (axis cs:0,0) rectangle (axis cs:15,42295032); 
		\fill[lightblue] (axis cs:15,0) rectangle (axis cs:30,42295032);      
		\node[anchor=north west, font=\scriptsize\bfseries, text=strongpink] at (axis cs:1,22295032) {Lung};
		\node[anchor=north west, font=\scriptsize\bfseries, text=blue] at (axis cs:20,22295032) {Water};
		\addplot[black, solid, smooth] table [x expr=\coordindex, y index=0, col sep=comma] 
		    {images/RealPhantom/tumor1/window_tumor1_noprc.txt};
		\addplot[red, dashed, smooth] table [x expr=\coordindex, y index=0, col sep=comma] 
		    {images/RealPhantom/tumor1/window_tumor1_svtd.txt};
		\addplot[green!60!black, dotted, smooth] table [x expr=\coordindex, y index=0, col sep=comma] 
		    {images/RealPhantom/tumor1/window_tumor1_ddconv.txt};
		\legend{No \ac{PRC}, \ac{SVTD}, \ac{DDConv}}
		\end{axis}
	\end{tikzpicture}%
	}
	
	\subfloat[No \ac{PRC} \label{subfig:noprc1}]{%
		\includegraphics[width=0.31\linewidth]{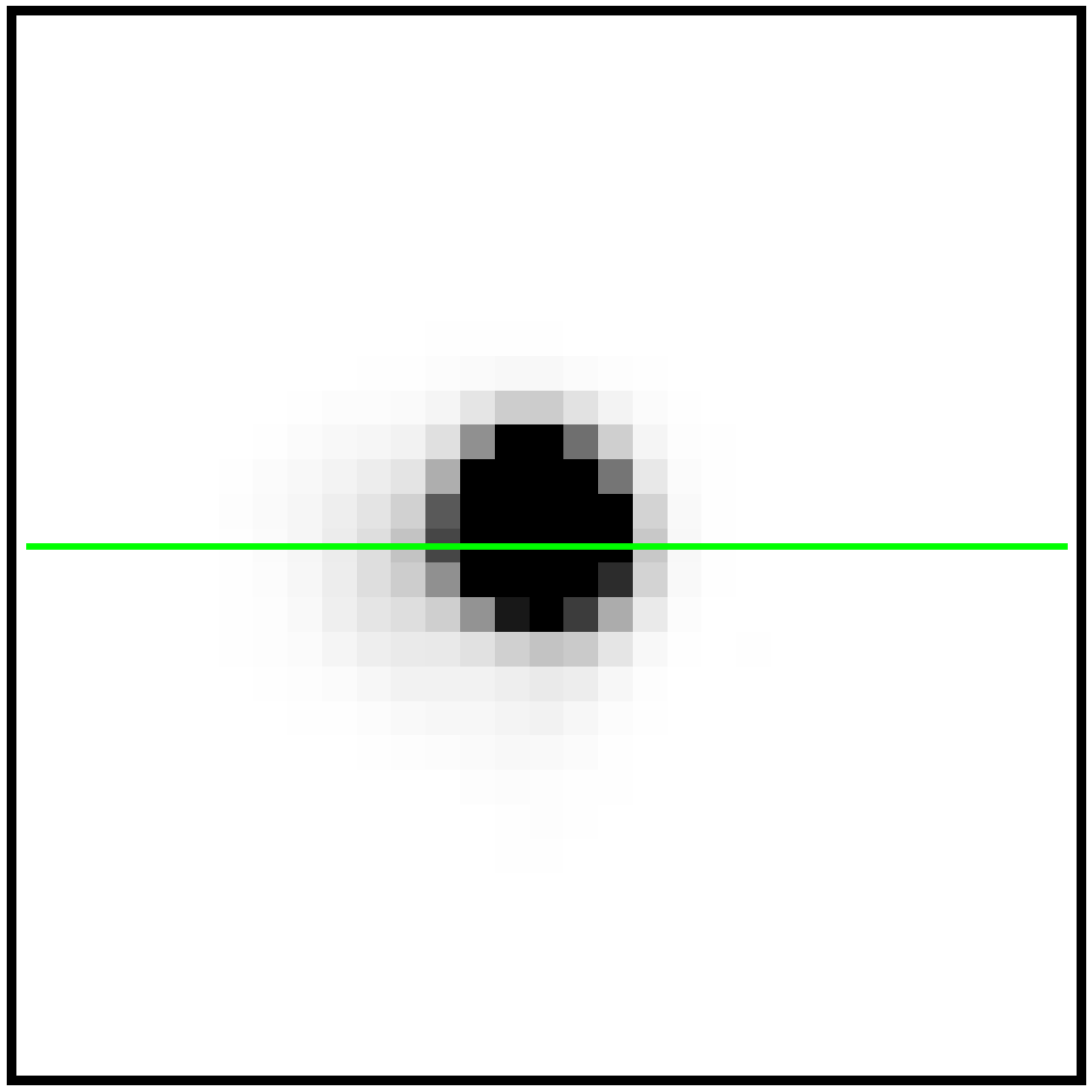}%
	}\hfill
	\subfloat[\Ac{SVTD} \label{subfig:svtd1}]{%
		\includegraphics[width=0.31\linewidth]{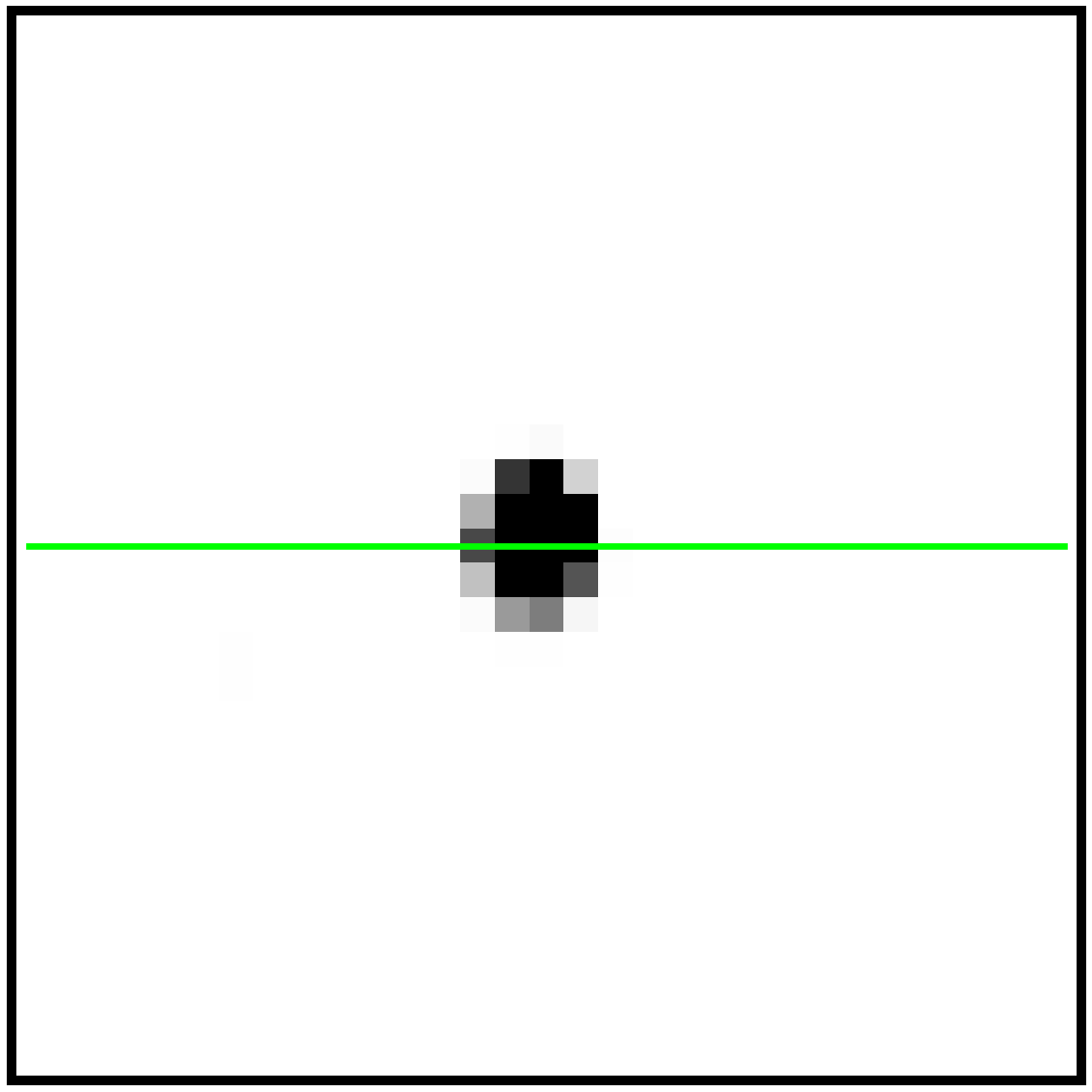}%
	}\hfill
	\subfloat[\ac{DDConv} \label{subfig:ddconv1}]{%
		\includegraphics[width=0.31\linewidth]{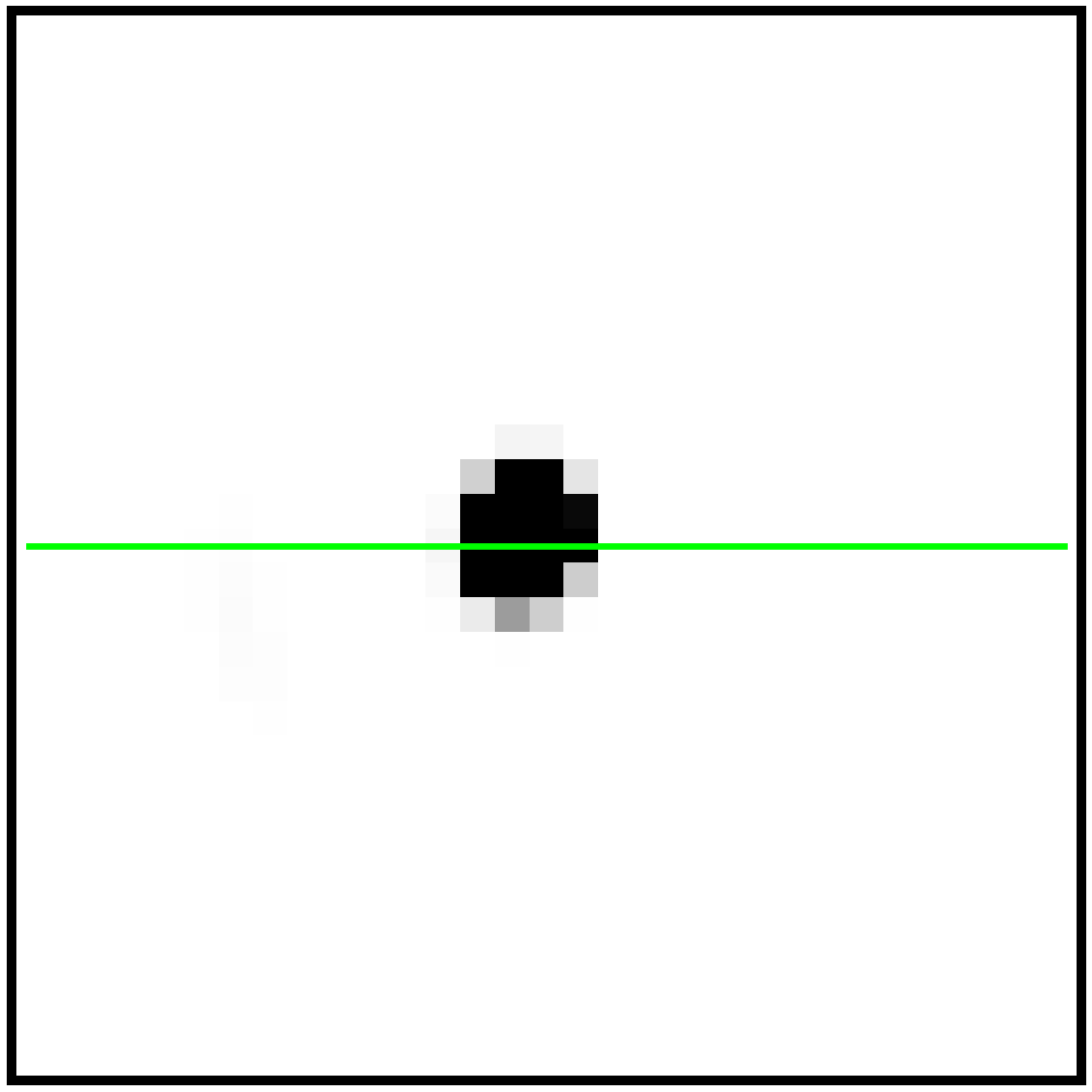}%
	}
	
	\caption{
		Positron range blurring comparison on real phantom data on the lesion located at the lung--soft tissue interface: 
		\protect\subref{subfig:ct_image1} \ac{CT} image showing the \ac{ROI} (green square) and the lesion's location (red spot), 
		\protect\subref{subfig:profiles1} intensity profiles for each method (cf. green line in reconstructed lesions), 
		\protect\subref{subfig:noprc1} no positron range correction, 
		\protect\subref{subfig:svtd1} \ac{SVTD}-reconstructed image, and 
		\protect\subref{subfig:ddconv1} \ac{DDConv}-blurred image.
	}
	\label{fig:ct_kernels1}
\end{figure}

\begin{figure}
	\subfloat[CT image with profile line \label{subfig:ct_image2}]{%
	\includegraphics[width=0.48\linewidth]{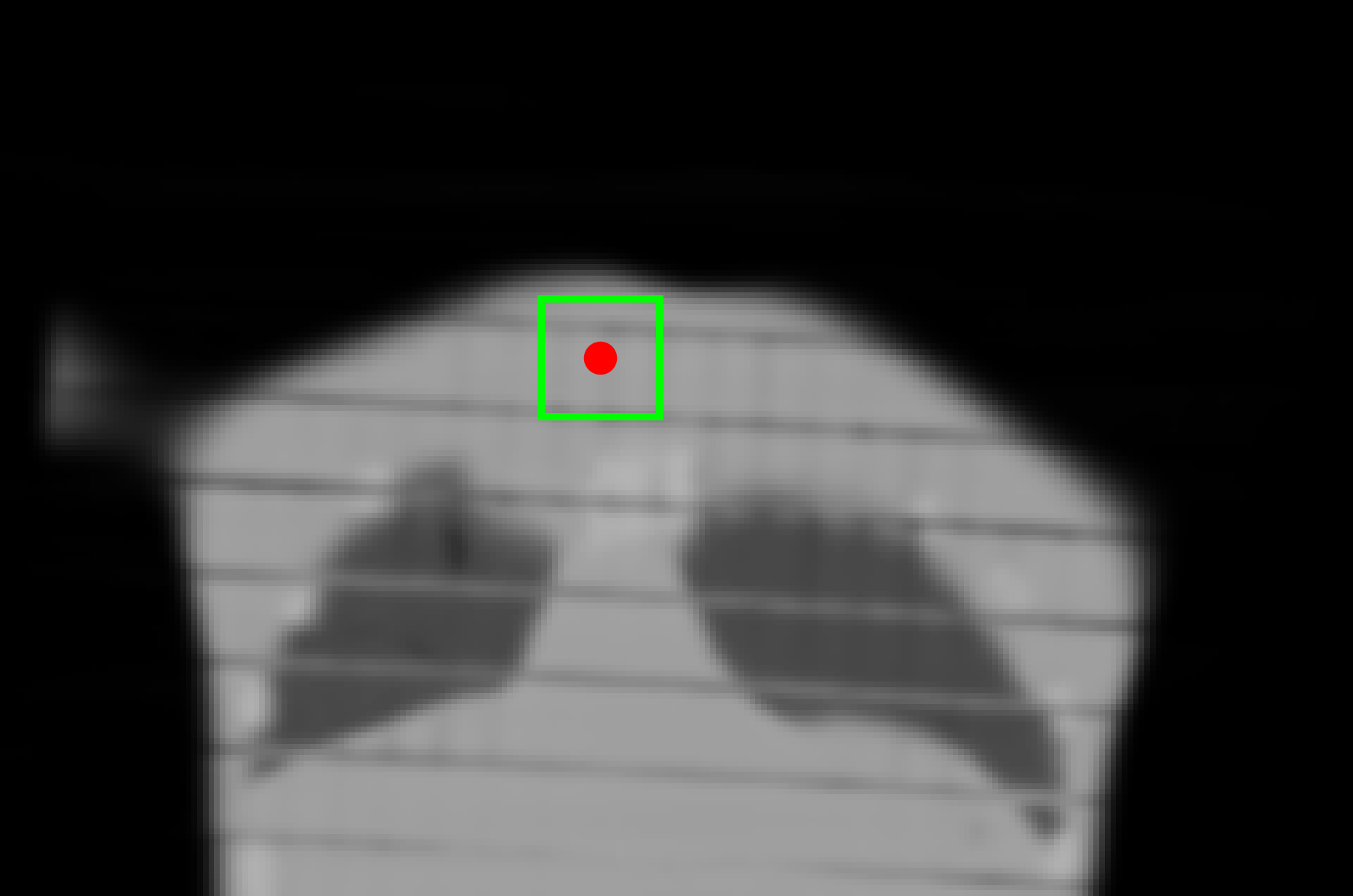}%
	}\hfill
	\subfloat[Line profiles across tumor region \label{subfig:profiles2}]{%
	\begin{tikzpicture}
		\begin{axis}[
			width=0.6\linewidth,
			height=0.45\linewidth,
			axis x line=bottom,
			axis y line=left,
			xmin=0, xmax=31,
			ymin=0, ymax=8229503,
			enlarge y limits={upper, value=0.05},
			clip marker paths=false,
		    grid=both,
			grid style={line width=.1pt, draw=gray!20},
			major grid style={line width=.2pt,draw=gray!20},
			minor tick num=1,
			tick label style={font=\fontsize{5}{5}\selectfont},
			xticklabel style={yshift=-1pt},
			xlabel={Position (voxel)},
			ylabel={Intensity},
			xlabel style={font=\fontsize{6}{6}\selectfont,yshift=3pt},
			ylabel style={
				font=\fontsize{6}{6}\selectfont,yshift=-20pt,
			},
			tick align=outside,
			tick pos=left,
			axis on top=true,
			legend style={
				at={(1,1)},
				anchor=north east,
				font=\fontsize{5}{5}\selectfont,
				row sep=0.5pt,
				cells={anchor=west},
				inner sep=1pt,
				outer sep=0pt
			},
			legend cell align={left},
			every axis plot/.append style={line width=1pt},
			]
	
		\fill[lightblue] (axis cs:0,0) rectangle (axis cs:31,8229503);      
		\node[anchor=north west, font=\scriptsize\bfseries, text=blue] at (axis cs:2,6229503) {Water};
		\addplot[black, solid, smooth] table [x expr=\coordindex, y index=0, col sep=comma] 
		    {images/RealPhantom/tumor2/window_tumor2_noprc.txt};
		\addplot[red, dashed, smooth] table [x expr=\coordindex, y index=0, col sep=comma] 
		    {images/RealPhantom/tumor2/window_tumor2_svtd.txt};
		\addplot[green!60!black, dotted, smooth] table [x expr=\coordindex, y index=0, col sep=comma] 
		    {images/RealPhantom/tumor2/window_tumor2_ddconv.txt};
		\legend{No \ac{PRC}, \ac{SVTD}, \ac{DDConv}}
		\end{axis}
	\end{tikzpicture}%
	}
	
	\subfloat[No \ac{PRC} \label{subfig:noprc2}]{%
		\includegraphics[width=0.31\linewidth]{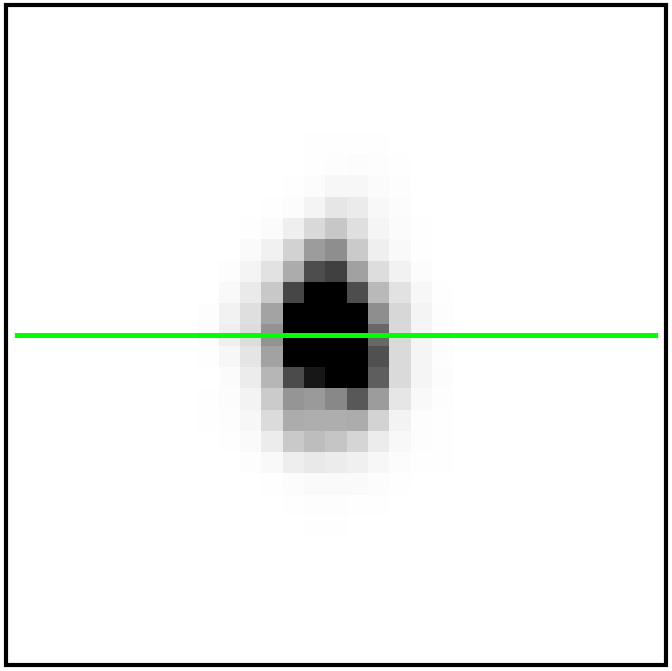}%
	}\hfill
	\subfloat[\ac{SVTD} \label{subfig:svtd2}]{%
		\includegraphics[width=0.31\linewidth]{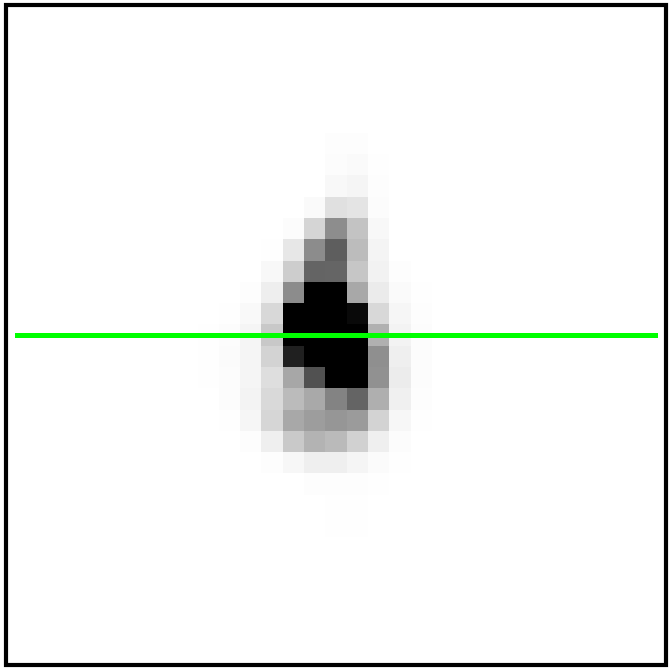}%
	}\hfill
	\subfloat[\ac{DDConv} \label{subfig:ddconv2}]{%
		\includegraphics[width=0.31\linewidth]{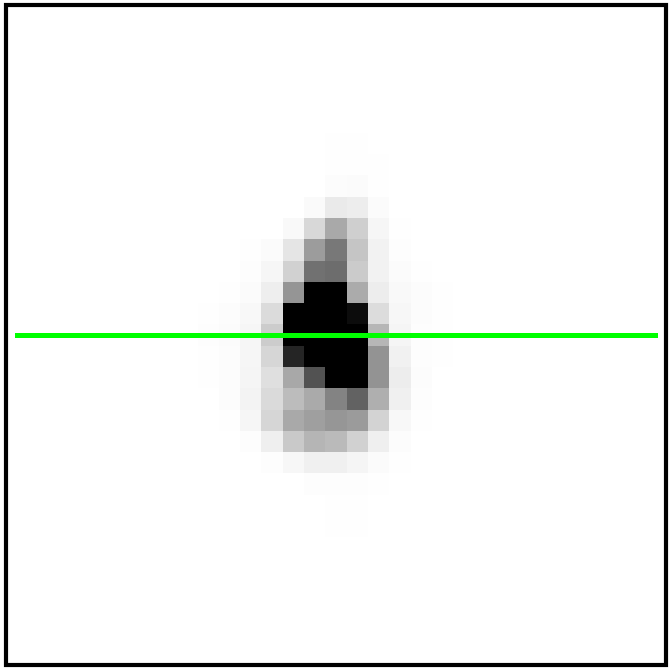}%
	}
	\caption{
		Positron range blurring comparison on real phantom data on the lesion located in soft tissues: 
		\protect\subref{subfig:ct_image2} \ac{CT} image showing the \ac{ROI} (green square) and the lesion's location (red spot), 
		\protect\subref{subfig:profiles2} intensity profiles for each method (cf. green line in reconstructed lesions), 
		\protect\subref{subfig:noprc2} no positron range correction, 
		\protect\subref{subfig:svtd2} \ac{SVTD}-reconstructed image, and 
		\protect\subref{subfig:ddconv2} \ac{DDConv}-blurred image.
	}\label{fig:ct_kernels2}
\end{figure}

We evaluated \ac{SVTD}, \ac{DDConv}, and no-\ac{PRC} reconstructions using real \ac{PET} data acquired on a Siemens Biograph Vision \ac{PET}/\ac{CT} scanner at Kuopio University Hospital (Kuopio, Finland), employing a physical phantom with injected activity to simulate lesions. We considered two lesions: (i) one located at the interface between the lung and soft tissues (Figure~\ref{subfig:ct_image1}) and (ii) one located in the soft tissues (Figure~\ref{subfig:ct_image2}). Profiles and magnified images of the reconstructed lesions are shown in Figures~\ref{subfig:profiles1}, \ref{subfig:noprc1}, \ref{subfig:svtd1}, and \ref{subfig:ddconv1} for lesion (i),  and in Figures~\ref{subfig:profiles2}, \ref{subfig:noprc2}, \ref{subfig:svtd2}, and \ref{subfig:ddconv2} for lesion (ii).

The results on lesion (i) (heterogeneous region) show that \ac{SVTD} recovers higher activity than \ac{DDConv}, while the results on lesion (ii) (homogeneous region) are similar. These results confirm the findings of the simulation experiments in Section~\ref{sec:exp1} and Section~\ref{sec:exp2}.

\section{Discussion}\label{sec:discussion}

A primary advantage of the proposed \ac{DDConv} approach is its ability to generate \ac{PR} blurring kernels with an accuracy comparable to that of \ac{MC} simulations while being several orders of magnitude faster. For example, in \ac{GATE}, the simulation of a single 31\texttimes{}31\texttimes{}31 kernel  using one million positron events takes approximately 1~min and 40~s, and this time increases proportionally with the number of simulated events. In contrast, the trained \ac{DDConv} model predicts the same kernel in about 162~ms, with a computation time that remains constant regardless of the number of positrons represented. For the full \ac{XCAT} dataset, which includes approximately $1.04\cdot 10^{10}$ positron events, the \ac{MC}-based generation required about 1,000 parallel \ac{GATE} simulations, each lasting 1 hour and 43 minutes on a computer cluster. The equivalent forward operator can be produced with the proposed \ac{DDConv} method in about 27 minutes (1,620~s, cf. Table~\ref{table:runtime}) using the full 31\texttimes{}31\texttimes{}31  kernel (cf. Appendix \ref{sec:runtime}), demonstrating a reduction in computation time while preserving physical accuracy.

From a reconstruction standpoint, this work addresses several limitations that have long hindered accurate \ac{PRC}. The proposed \ac{DDConv} framework enables the generation of physically realistic kernels that match the accuracy of \ac{MC} simulations while remaining computationally practical for iterative reconstruction. In addition, the explicit formulation of both forward and transposed operators ensures full mathematical consistency within the \ac{EM} algorithm, guaranteeing stable convergence and preserving the quantitative integrity of the reconstructed activity distribution. The selection of the kernel size is also critical for reconstruction efficiency and physical fidelity. In practice, the 11\texttimes{}11\texttimes{}11 neighborhood with a 2-mm voxel size was adopted as an optimal trade-off between accuracy and computational cost, following the recommendations of \citeauthor{kertesz2022implementation} \cite{kertesz2022implementation}. Analysis of \ac{MC}-derived reference kernels confirmed that this configuration retains about 84\% of the annihilation energy in lung and nearly 100\% in soft-tissue and bone, while larger 31\texttimes{}31\texttimes{}31 kernels provide negligible quantitative improvement but considerably increase computation time (cf.  Appendix~\ref{sec:kernelsize}).

Compared to prior \ac{PRC} methods, \ac{DDConv} offers substantial benefits in both precision and speed. Early approaches precomputed few generic kernels for different materials, or utilized simple deconvolutions; although computationally efficient, these approaches often fail at modeling \ac{PR} at lung--soft tissue or bone--soft tissue interfaces. Recent anisotropic spatially-variant kernels improve accuracy but still rely on combining multiple precomputed kernels, sometimes introducing trade-offs in accuracy or speed. In contrast, \ac{DDConv} predicts spatially variant \acp{PSF} in real time for each voxel neighborhood, thus maintaining \ac{MC}-like fidelity even in complex, inhomogeneous regions. The method's efficiency stems from its \ac{GPU}-based convolutional design: the heavy computation of blurring is delegated to highly optimized parallel operations, enabling fast kernel estimation across large images without sacrificing the high fidelity needed for accurate quantification (cf. Appendix \ref{sec:acceleration}). Notably, the full computation of \ac{SVTD} and \ac{DDConv} (accelerated version) for an entire \ac{XCAT} phantom volume takes approximately 18 seconds, demonstrating that the proposed approach remains practical for clinical applications with \ac{GPU} acceleration. 

Our preliminary results on real data suggest that \ac{DDConv} and \ac{SVTD} behave differently on heterogeneous regions and behave similarly on homogeneous regions, which confirms our findings on simulated data. 

From a clinical perspective, achieving accurate \ac{PRC}  can significantly improve image resolution and lesion detectability, particularly for higher-energy tracers such as \ac{Ga68}. The ability to correct for \ac{PR}-induced blurring in lung or bone interfaces offers more consistent quantitative accuracy across the \ac{FOV}. By delivering sharper images and preserving quantitative consistency for a wide array of positron emitters, \ac{DDConv} has the potential to improve \ac{PET} imaging standards and expand the use of isotopes previously considered too susceptible to range effects.
\section{Conclusion}\label{sec:conclusion}

In conclusion, this study introduced \ac{DDConv} as an efficient and accurate framework for positron range correction in PET imaging. By combining local attenuation maps with activity information, \ac{DDConv} dynamically estimates high-resolution blurring kernels, matching \ac{MC} accuracy at a fraction of the computational cost. Unlike previous methods that rely on precomputed or approximate models, \ac{DDConv}’s predictive approach integrates seamlessly into iterative reconstruction and preserves consistency between forward and backward operations. Demonstrations on digital phantoms and real phantom data confirm its ability to improve image resolution and quantitative accuracy, especially for high-energy positron emitters. These results underscore the clinical potential of \ac{DDConv} for routine \ac{PET}, enabling near-\ac{MC}-level corrections without prohibitive run times and thus contributing to more reliable disease detection and characterization.

\appendix
\subsection{Acceleration}\label{sec:acceleration}

The computation of $\boldB(\boldmu)\boldx$ and $\boldB(\boldmu)\transp\boldz$ can be accelerated by considering a single \ac{PR} \ac{PSF} for homogeneous regions in which the \ac{PSF} is independent of position.

\subsubsection{Homogeneity Map}

We considered a decomposition of the  $L=3$ materials (soft tissues, lungs and bones) which provides the binary images $\boldu_l \in \{0,1\}^J$, $l=1,\dots,L$, such that $\sum_{l=1}^L \boldu_l = \boldone$.

For each material $l$, a single \ac{PR} \ac{PSF}, which takes the form of an 11\texttimes{}11\texttimes{}11 image $\boldh_l \in \R^m$ ($m=11^3$), is generated from \ac{MC} simulations using a positron emission source in a homogeneous medium corresponding to material $l$; each of these simulation results in an isotropic Gaussian-like \ac{PSF}. For each region $l$, the blurred material images are computed, i.e.,
\begin{equation}
	\boldv_l = \boldu_l \ast \boldh_l
\end{equation}	
where `$\ast$' denotes the standard convolution with a position-independent kernel. Each image $\boldv_l$ ranges in $[0,1]$ and we define the subsets of indices
\begin{equation}
	\calS_{\mathrm{hom}}^l = \{ j\in \calS  , [\boldv_l]_j = 1  \} \, .
\end{equation}
The subset 	$\calS_{\mathrm{hom}}^l$ is the $l$th \emph{homogeneous} area, i.e., the area in material $l$ on which an emitted positron is certain to annihilate with an electron in the same material. Conversely, the set
\begin{equation}
	\calS_{\mathrm{het}} = \calS \setminus \bigcup_{l=1}^L\calS_{\mathrm{hom}}^l   
\end{equation} 	
is the \emph{heterogeneous} area.

\subsubsection{Forward Operator}

We first define the \emph{homogeneous} blurring operator $\boldB_{\mathrm{hom}}(\boldmu)$, which is computed by separately masking the entire activity image $\boldx$ with each region $\calS_{\mathrm{hom}}^l$ followed by convolution with the isotropic kernel $\boldh_l$, then summing over $l$, i.e.,   
\begin{equation}\label{eq:hom-forward}
	\boldB_{\mathrm{hom}}(\boldmu)\,\boldx \;=\; 
	\sum_{l=1}^L \left(\boldx\odot\,\mathbbm{1}_{\calS_{\mathrm{hom}}^l}\right)\,\ast \boldh_l,
\end{equation}
where `$\odot$' denotes the element-wise vector multiplication and $\mathbbm{1}_{\calS_{\mathrm{hom}}^l} \in \mathbb{R}^J$ denotes the indicator function of $\calS_{\mathrm{hom}}^l$.

For voxels in the heterogeneous subset $\calS_{\mathrm{het}} $, a dynamic kernel is needed. At each voxel $j$, the \ac{PR} predictor $\boldsymbol{G}_{\boldsymbol{\theta}}$ is used to compute a local \ac{PSF} $\boldw_j$ from its attenuation neighborhood~$\boldmu_{\mathcal{N}_j}$ and distance vector~$\boldd$. The \emph{heterogeneous} \ac{PR} blurring operator $\boldB_{\mathrm{het}}(\boldmu)$ is defined at each voxel $k$ as
\begin{equation}
	[\boldB_{\mathrm{het}}(\boldmu)\boldx]_{k} = \sum_{j\in\calN_k\cap \calS_{\mathrm{het}}}  w_{j\to k} \cdot x_j 
\end{equation}
which is computed by omitting voxels $j \notin \calS_{\mathrm{het}}$ in Algorithm~\ref{algo:pr}.

Finally, we have 
\begin{equation}
	\boldB(\boldmu) = \boldB_{\mathrm{hom}}(\boldmu) + \boldB_{\mathrm{het}}(\boldmu)
\end{equation}

\subsubsection{Transposed Operator}

The transposed homogeneous blurring operator $\boldB_{\mathrm{hom}}(\boldmu)\transp$ is obtained by interchanging the multiplication with the indicator function $\mathbbm{1}_{\calS_{\mathrm{hom}}^l}$ and the convolution with the isotropic kernel $\boldh_l$, i.e.,
\begin{equation}
	\boldB_{\mathrm{hom}}(\boldmu)\transp\,\boldz
	= \sum_{l=1}^L
	(\boldz \ast \boldh_l)\odot\,\mathbbm{1}_{\calS_{\mathrm{hom}}^l} \, ,
\end{equation}
while $\boldB_{\mathrm{het}}(\boldmu)\transp$ is defined as 
\begin{equation} 
	\left[\boldB_{\mathrm{het}}(\boldmu)\transp \boldz\right]_j= 
	\begin{cases}
		\sum_{k\in\calN_j}w_{j\to k}\cdot z_k  & \text{if}\ j \in \calS_{\mathrm{het}} , \\
		0 & \text{otherwise,}
	\end{cases}
\end{equation}
which is computed by omitting voxels $j \notin \calS_{\mathrm{het}}$ in Algorithm~\ref{algo:pr_t}.

Finally, we have 
\begin{equation}
	\boldB(\boldmu)\transp = \boldB_{\mathrm{hom}}(\boldmu)\transp + \boldB_{\mathrm{het}}(\boldmu)\transp.
\end{equation}

\subsection{Runtime Evaluation}\label{sec:runtime}

All experiments were carried out on a workstation equipped with an Intel Xeon E5-1650~v4 CPU (3.6\,GHz), 62\,GB RAM, and an NVIDIA GeForce RTX~3060 GPU (12\,GB VRAM) using PyTorch~2.5 with CUDA acceleration.  Unless stated otherwise, the runtime analysis was performed on the \ac{XCAT} volume of 
200\texttimes{}200\texttimes{}100 voxels with a batch size of 400. The results are summarized in Table~\ref{table:runtime}.

\begin{table}
	
	\label{tab:inference_times}
	\vspace{1mm}
	\setlength{\tabcolsep}{3pt} 
	\renewcommand{\arraystretch}{1.1} 
	\footnotesize
	\caption{Inference time comparison between \ac{SVTD}, \ac{DDConv}, and Accelerated \ac{DDConv} (cf. Appendix~\ref{sec:acceleration}).}\label{table:runtime}
		\begin{tabular}{p{1.5cm}p{1.5cm}p{2cm}p{2.2cm}}
			\hline
			\centering\textbf{Kernel} & 
			\centering\textbf{SVTD [s]} & 
			\centering\textbf{DDConv [s]} & 
			\centering\textbf{Acc.\ DDConv [s]} \tabularnewline
			\hline
			\centering $11^3$ & \centering \textbf{18} & \centering 74 & \centering \textbf{18} \tabularnewline
			\centering $21^3$ & \centering \textbf{40} & \centering 500 & \centering 138 \tabularnewline
			\centering $31^3$ & \centering \textbf{120} & \centering 1,620 & \centering 480 \tabularnewline
			\hline
		\end{tabular}
		
\end{table}

For the reference \ac{SVTD} method, the processing time increases almost linearly with the kernel size ($11^3 \rightarrow 21^3 \rightarrow 31^3$), from 18~s to 40~s and 120~s, respectively. This scaling occurs because GPU computations are limited mainly by memory bandwidth rather than by pure arithmetic throughput: larger kernels require transferring a larger ``halo'' region between memory and GPU cores,  while the compute units remain nearly saturated.  Consequently, the runtime follows the kernel volume ($k^3$) with good efficiency.

In contrast, the proposed \ac{DDConv} uses a per-voxel \ac{CNN} inference step  to predict a local \ac{PSF}, which constitutes the primary computational bottleneck.  The model inference takes on average 7.5~ms, 50~ms, and 162~ms per batch (400 voxels)  for kernel sizes of $11^3$, $21^3$, and $31^3$, respectively,  and dominates the total runtime when applied over the full image. The accelerated version (see Appendix~\ref{sec:acceleration}) mitigates this cost by using pre-computed homogeneous kernels for soft tissue, lung, and bone regions---computed with standard CUDA convolutions---and applying the learned model only in heterogeneous interface regions.  This hybrid strategy reduces the overall computation time  approximately by a factor of 3 while retaining the near \ac{MC} accuracy of the full \ac{DDConv}.  If the heterogeneous regions occupy most of the image, however, the runtime naturally approaches that  of the non-accelerated implementation.

\subsection{Analysis of Kernel Size}\label{sec:kernelsize}

The kernel side length fixes the spatial support over which \ac{PR} blurring is modeled and therefore sets the balance between physical fidelity and computational burden. To assess how much probability mass is lost when truncating the kernel, we started from an \ac{MC}-derived $31^3$ reference kernel and computed, for three representative materials, the fraction of its total energy contained in centered cubic crops of smaller sizes.

\begin{table}
	\renewcommand{\arraystretch}{1.05}
	\setlength{\tabcolsep}{3pt}
	\caption{Fraction of the kernel energy retained for different cubic support sizes, normalized to the full $31^3$ \ac{MC} reference kernel.}
	\label{tab:prc_energy}
	\centering
	\begin{tabular}{lccccc}
		\hline
		$\textbf{Material}$ & $\bm{31^3}$ & $\bm{22^3}$ & $\bm{11^3}$ & $\bm{9^3}$ & $\bm{7^3}$ \\
		\hline
		Lung  & 1.000 & 0.9999 & 0.837 & 0.722 & 0.574 \\
		Water & 1.000 & 1.0000 & 0.9999 & 0.9997 & 0.9982 \\
		Bone  & 1.000 & 1.0000 & 1.0000 & 0.9999 & 0.9999 \\
		\hline
	\end{tabular}
\end{table}

As summarized in Table~\ref{tab:prc_energy}, the $22^3$ crop already contains more than $99.99\%$ of the total kernel energy for all three materials, indicating that contributions outside this region are negligible. In contrast, reducing the support to $11^3$ voxels preserves only about $84\%$ of the energy in lung, while the loss in water and bone remains marginal (close to $100\%$ retention). Further shrinkage to $9^3$ or $7^3$ kernels leads to substantial truncation of the probability tail in low-density lung, whereas the effect in soft tissue and bone is minor. 

Based on this analysis, an $11^3$ neighborhood offers a practical compromise between accuracy and runtime for $2$-mm isotropic voxels: it is sufficiently large to capture most of the annihilation distribution in all materials, yet small enough to keep the cost of \ac{DDConv} manageable. For anisotropic voxel grids, the kernel dimensions should be scaled proportionally to the voxel spacing.

\section*{Acknowledgment}

All authors declare that they have no known conflicts of interest in terms of competing financial interests or personal relationships that could have an influence or are relevant to the work reported in this paper.

\AtNextBibliography{\footnotesize} 
\printbibliography
\end{document}